\documentclass[useAMS, usenatbib]{mn2e}
\usepackage{tablefootnote}

\makeatother
\usepackage[normalem]{ulem} 
\usepackage{epsfig,amssymb,amsmath,bm} 
\usepackage{graphicx,epstopdf}
\usepackage{amsfonts, cancel}
\usepackage{marvosym}
\usepackage{dsfont}
\usepackage{stmaryrd}
\usepackage{color}
\usepackage{subfigure}
\usepackage{verbatim}

\usepackage{latexsym}
\usepackage{subfigure}
\usepackage{mathtools}
\usepackage[english]{babel}
\usepackage{flushend}
\usepackage{balance}

\usepackage{upgreek}

\usepackage{caption}
\usepackage{dcolumn}
\newcolumntype{d}[1]{D{.}{.}{#1}}

\newcommand{\be}{\begin{equation}}
\newcommand{\ee}{\end{equation}}

\newcommand{\ball}{\begin{align}}
\newcommand{\eall}{\end{align}}

\title{Scattering analysis of LOFAR pulsar observations}
\author[M Geyer and A Karastergiou and PWG]{M. Geyer$^{1}$\thanks{marisa.geyer@physics.ox.ac.uk; aris.karastergiou@physics.ox.ac.uk}, 
A. Karastergiou$^{1,2,3 \star}$,
V.I. Kondratiev$^{4,5}$, K. Zagkouris$^{1}$,
\newauthor M. Kramer$^{6,7}$, B.W. Stappers$^{7}$, J.-M. Grie{\ss}meier$^{8,9}$, J.W.T. Hessels$^{4,10}$, 
\newauthor D. Michilli$^{4,10}$, M. Pilia$^{11}$ and C. Sobey$^{12,13}$\\
$^{1}$Astrophysics, University of Oxford, Denys Wilkinson Building, Keble Road, Oxford OX1 3RH, UK\\
$^{2}$Physics Department, University of the Western Cape, Cape Town 7535, South Africa\\ 
$^{3}$Department of Physics and Electronics, Rhodes University, PO Box 94, Grahamstown 6140, South Africa\\
$^{4}$ASTRON, The Netherlands Institute for Radio Astronomy, Postbus 2, 7990 AA Dwingeloo, The Netherlands\\
$^{5}$Astro Space Centre, Lebedev Physical Institute, Russian Academy of Sciences, Profsoyuznaya Str. 84/32, Moscow 117997, Russia\\
$^{6}$Max-Planck-Institut f\"{u}r Radioastronomie, Auf dem H\"{u}gel 69, 53121 Bonn, Germany\\
$^{7}$Jodrell Bank Centre for Astrophysics, School of Physics and Astronomy, The University of Manchester, Manchester M13 9PL, UK\\
$^{8}$LPC2E - Universit\'{e} d'Orl\'{e}ans /  CNRS, 45071 Orl\'{e}ans cedex 2, France\\
$^{9}$Station de Radioastronomie de Nan\c{c}ay, Observatoire de Paris, PSL Research University, CNRS, Univ. Orl\'{e}ans, OSUC, 18330 Nan\c{c}ay, France\\
$^{10}$Anton Pannekoek Institute for Astronomy, University of Amsterdam, Science Park 904, 1098 XH Amsterdam, The Netherlands\\
$^{11}$INAF - Osservatorio Astronomico di Cagliari, via della Scienza 5, I-09047 Selargius, Italy\\
$^{12}$International Centre for Radio Astronomy Research - Curtin University, GPO Box U1987, Perth, WA 6845, Australia\\
$^{13}$CSIRO Astronomy and Space Science, 26 Dick Perry Avenue, Kensington, WA 6151, Australia
}

\begin{document}

\date{Accepted 2017 May 09. Received 2017 May 04 ; in original form 2017 February 14}

\pagerange{\pageref{firstpage}--\pageref{lastpage}} \pubyear{2016}

\maketitle

\label{firstpage}

\begin{abstract} We measure the effects of interstellar
scattering on average pulse profiles from 13 radio pulsars with
simple pulse shapes. We use data from the LOFAR High Band Antennas,
at frequencies between 110 and 190~MHz. We apply a forward
fitting technique, and simultaneously determine the intrinsic pulse
shape, assuming single Gaussian component profiles. We find that the
constant $\tau$, associated with scattering by a single thin screen,
has a power-law dependence on frequency
$\tau \propto \nu^{-\alpha}$, with indices ranging from
$\alpha = 1.50$ to $4.0$, despite simplest theoretical models predicting $\alpha = 4.0$ or $4.4$.
Modelling the screen as an isotropic or extremely anisotropic
scatterer, we find anisotropic scattering fits lead to larger
power-law indices, often in better agreement with theoretically
expected values. We compare the scattering models based on the
inferred, frequency dependent parameters of the intrinsic pulse, and
the resulting correction to the dispersion measure (DM). We highlight the
cases in which fits of extreme anisotropic scattering are appealing,
while stressing that the data do not strictly favour either model for
any of the 13 pulsars. The pulsars show
anomalous scattering properties that are consistent with finite
scattering screens and/or anisotropy, but these data alone do not
provide the means for an unambiguous characterization of the screens. 
We revisit the empirical $\tau$ versus DM relation and consider how our results support a frequency dependence of $\alpha$. Very long baseline interferometry, and observations of the scattering and scintillation properties of these sources at higher frequencies, will provide further evidence. \end{abstract}


\section{Introduction}

Prominent evidence for radio wave scattering comes from observing that average pulsar profiles grow asymmetrically broader at low frequencies  (e.g. \citealt{Lohmer2001}). These observed \textit{scattering tails} represent the power of the pulsar delayed through multipath propagation in the ionized interstellar medium (ISM). The exponential broadening of the pulse profiles is parameterized by a characteristic scattering time-scale, $\tau$.  

From the variety of propagation effects associated with radio waves travelling through the ISM, interstellar scattering is expected to display the strongest frequency ($\nu$) dependence. 
The strong dependence on frequency is caused by inhomogeneities in the electron density gradients of the ISM along the line of sight to the pulsar \citep{Salpeter1967, Scheuer1968}. 

Different models of the ISM predict different frequency dependencies. Many theoretical models make use of the \textit{thin screen approximation}, by which it is assumed the scattering along the line of sight to the pulsar can be approximated by a single scattering surface approximately mid-way to pulsar, which is infinitely extended transverse to the line of sight and thin along the line of sight (e.g. \citealt{Williamson1972}).  Modelling the inhomogeneities of such a thin screen as a Kolmogorov turbulence in a cold plasma, leads to a dependence of scattering time-scales proportional to $\nu^{-4.4}$ \citep{LeeJokipii1976, Rickett1977}. A more simple approach, by which the scattering is assumed to be isotropic and described by a circularly symmetric Gaussian distribution in scattering angles, leads to a dependence of $\tau \propto \nu^{-4}$ \citep{Cronyn1970, Lang1971}.

Observations of scatter broadened pulsars have led to measurements of the power-law index $\alpha$ (where $\tau \propto \nu^{-\alpha}$) in agreement with the theoretical models, as well as to deviations of $\alpha$ from 4 or 4.4. Notably \citet{Lohmer2001} found a mean value of $\bar{\alpha} = 3.44 \pm 0.13$ from the scatter broadening measurements of nine pulsars, at frequencies between 600~MHz and 2.7~GHz. The pulsars in the study were selected to have large dispersion measure (DM) values, and as such the authors argue that the low $\alpha$ values could be due to multiple finite scattering screens along the long lines of sights to the sources. 

More recently, \citet{Lewandowski2013} measured $\alpha$ values in the range 2.77--4.59 (from 25 sources), and \citet{Lewandowski2015} published a similar range of 2.61--5.61 (based on 60 sources).  In both cases, even lower $\alpha$ values than these were found, but were considered spurious and were subsequently disregarded from the analysis. Their measurements were based on observations with the Giant Meterwave Radio Telescope (GMRT, Pune India) between 325~MHz and 1.2~GHz,  and with the 100-m Effelsberg Radiotelescope (2.6--8.35~GHz) as well as published scattering times from the literature.

Additionally, low frequency scattering and scintillation studies conducted at the Pushchino Radio Astronomy Observatory in Russia, have measured $\alpha$ values below the theoretically expected values \citep{Smirnova2008, Kuzmin2007}. The observatory hosts the Large Phased Array (LPA or BSA) and the DKR-1000 telescope operating at 111.9~MHz and 41, 62.4 and 88.6~MHz, respectively.  

Space-ground interferometry, which combined the efforts of the 10-m Space Radio Telescope aboard \textit{RadioAstron}, the Arecibo telescope and the Westerbork Synthesis Radio Telescope (WSRT), have led to observations with a 220 000 km projected baseline, and unprecedented resolution at metre wavelengths. 
Studying the scattering towards the nearby pulsar B0950+08 with this interferometer, \citet{Smirnova2014} inferred two independent scattering surfaces along the line of sight, and an $\alpha$ value of $3.00 \pm 0.08$.

Flatter $\tau$ spectra, i.e. spectra for which $\alpha < 4$, have been reasoned to be due to anomalous scattering mechanisms and geometries. This includes scattering by a finite (truncated) scattering screen \citep{CordesLazio2001}, the impact of an inner cut-off scale \citep{RickettJohnston2009} and anisotropic scattering mechanisms \citep{Stinebring2001, Tuntsov2012}.   Ionized gas clouds in the ISM, as small as an order of AU in transverse radius, have been promoted by e.g. \citet{Walker2001} to explain extreme scattering events (ESEs) observed in quasars. ESEs have also been observed in pulsars (e.g \citealt{Cognard1993, Coles2015}) and may be associated with the same ISM structures.

Evidence for anisotropic scattering has especially come from organized patterns in the dynamic spectra of pulsar observations \citep{Gupta1994}. These patterns translate to parabolic arc features in the secondary (power) spectra, which are considered to be tell-tale signs of anisotropic scattering \citep{Stinebring2001,Walker2004}.

In a previous paper, we have shown that fitting anisotropically scattered (simulated) data with an isotropic model can lead to $\alpha$ values less than the theoretically predicted value (\citealt{Geyer2016}, hereafter GK16). As an extension of \citet{CordesLazio2001}, we also showed how non-circular scattering screens at locations off-centred with respect to the direct line of sight, lead to low $\alpha$ values. In our chosen examples the theoretical setups lead to $\alpha$ values as low as 2.9. 

The proposed mechanisms for obtaining smaller $\alpha$ values have certainly led to an improved understanding of correlations between ISM structure and pulsar scattering, but the detailed structure of the ISM and the physical interpretation (e.g. the distribution and size of anomalous scattering clouds) remain unclear. It also remains to be investigated to what extent observations promote an evolution of $\alpha$ with frequency. 

Low frequency data sets, due to the strong expected dependence of scatter broadening on frequency,  are ideal for investigating scattering effects and ISM structure in depth. In this paper we analyse the scatter broadening of 13 pulsars using the core of the Low Frequency Array (LOFAR; \citealt{Stappers2011,vanHaarlem2013}). LOFAR provides not only access to low frequency data, but also a broad-band at these low frequencies. This allows us to channelize the data into several high signal-to-noise (S/N) average profiles within the band, such that a detailed $\tau$ spectrum can be calculated. We fit the scatter broadened profiles with two scattering models: an isotropic thin screen scattering model and an extremely anisotropic scattering model. The broadening functions associated with each of these models are described in Section \ref{sec:scatmodels}. 

The paper aims to answer the following questions:
\begin{enumerate}
\item Are anisotropic scattering mechanisms and/or finite scattering screens required to fit our data set, or are isotropic mechanisms sufficient?
\item What are the mean values of $\alpha$ obtained for this set of pulsars, and how do the values differ with scattering models?
\item What additional information regarding the profile evolution and DM can we derive from the fits for scattering?
\item Do we have any evidence for correlations between flux density and scattering from the data?
\end{enumerate}

The paper is organized as follows: in Section \ref{sec:obsdata}, we describe the LOFAR data sets and subsequent data reduction. Thereafter, in Section \ref{sec:scatmodels}, we discuss the fitting methods used to extract $\tau$ values from the scattered profiles.  In Section \ref{sec:results}, we present a scattering analysis for each pulsar independently,  supplemented by Appendix \ref{app:profiles}. Thereafter in Section \ref{sec:disc}, we discuss our results, returning to the questions posed above.

\section{Observations and Data reduction}\label{sec:obsdata}

An overview of the 13 chosen sources and their associated parameters is given in Table \ref{table:one}. Three different LOFAR data sets are used, all of which were recorded using the LOFAR core with the High Band Antennas (HBAs, 110--190~MHz).  The data sets are LOFAR Commissioning Data, HBA Census Data \citep{Bilous2016} and Cycle 5 timing data (project code: LT5\_003; PI: Verbiest), discussed in more detail below. The pulsars were selected on the basis of being bright and scattered in this frequency band, as well as exhibiting simple average profile shapes, as inferred from less scattered profiles at higher frequencies. Depending on the data set and data quality, 8 or 16 profiles over the HBA band were analysed. The number of frequency channels was typically reduced for average pulse profiles with a peak S/N$ < 2.7$. In some cases individual low S/N frequency channels were removed from the analysis, as discussed on a pulsar-by-pulsar basis in Section \ref{sec:results}.

\subsection{Commissioning data}
Data recorded using the LOFAR HBA Core (19--23 stations), during the pre-Cycle 0 and Cycle 0 period (ending in November 2013), are here collectively called Commissioning data. The data are of similar quality as the data presented in \citet{Pilia2016}.

The Commissioning data are typically split into 6400 frequency channels across the HBA band, and have a phase bin resolution of 1024 bins across the pulse period. The data are incoherently dedispersed, with the exception of PSR~J1913$-$0440, for which the data are coherently dedispersed. The largest per channel DM smearing of 11~ms is for PSR~J1909+1102 at 110~MHz (4\% of its pulse period).

From the Commissioning data, we present nine sources that exhibit clear scatter broadening. We mostly form eight average profiles across the observing bandwidth, to which our scattering models are fit independently.

\subsection{Census data}

The original LOFAR HBA Census data set \citep{Bilous2016} includes 194 pulsars, observed from February to May 2014, at declinations  \mbox{$\delta> 8^{\circ}$}, and galactic latitudes $|b| > 3^{\circ}$. These specifications were chosen to maximize the telescope sensitivity (which is reduced with increasing zenith angle) and to avoid higher background sky temperatures towards the Galactic plane, and are as such not necessarily highly scattered pulsars.  From this set of pulsars we picked the ones that have a high peak S/N, exhibit clear exponential scattering and have simple single component profiles. All of  the pulsars were observed for  $> 1000$ rotations or at least 20 min using the full LOFAR HBA Core. 

The recorded bandwidth was split into 400 sub-bands, with either 64 or 128 channels per sub-band. The phase bin resolution varies from 128 to 1024 bins per pulse period. In this paper the Census data are typically presented as 16 average profiles across the HBA band. More detail on the observing strategy and data acquisition can be found in \citet{Bilous2016}.

\subsection{Cycle 5 data}

As part of an ongoing timing programme with LOFAR  (Cycle 5, project code: LT5\_003), two of the pulsars that overlap with the Commissioning data subset are continuously monitored at HBA frequencies. The data are recorded using the full LOFAR Core (23 stations), producing 10 min observations with 400 frequency channels, and then coherently dedispersed. We use Cycle 5 timing data for the two overlapping pulsars, and pick the observations with the highest S/N for each pulsar (ObsIDs L424139 and L423987).  The data are averaged to form 16 profiles across the band. We compare the outcomes to the lower S/N Commissioning data.

\subsection{Data Reduction}

All the observations are pre-processed using the standard LOFAR Pulsar Pipeline (PulP, \citealt{Kondratiev2016}). The complex-voltage data from individual stations are summed coherently, after which the data are dedispersed and folded offline.  Radio frequency interference (RFI) was removed using the {\tt clean.py} tool from the {\textsc CoastGuard}\footnote{https://github.com/plazar/coast\_guard} package \citep{Lazarus2016}. Profiles were flux density calibrated in the same way as described in \citet{Kondratiev2016} resulting in an initial conservative error estimate of 50\% \citep{Bilous2016}. Thereafter, we calculate a corrected flux density value for a given averaged pulse profile, which compensates for flux density losses due to the effects of extreme scattering.  
The correction is determined from the fitting parameter obtained by the scattering model, as described in more detail in the next section. 


\section{Analysis and Fitting techniques}\label{sec:scatmodels}

\begin{table*}
\centering
\begin{tabular}{p{1.8cm} p{1.7cm} p{1.0cm} p{1.5cm} p{1.6cm} p{1.2cm} p{1.2cm} p{1.2cm} p{1.2cm}}
\hline
Pulsar \mbox{J-name} & Pulsar \mbox{B-name} & Period (s) & DM \mbox{(pc cm$^{-3}$)} &$D$ \,\, \mbox{(kpc)} &Flux density$^{\dagger}$ (mJy)& Flux spectral index$^{\dagger}$, $\beta$ & Data & MJD \\ [0.5ex] 
\hline
\hline
J0040+5716   &   B0037+56   &   1.12    &   92.5146  & 2.42 (2.99)  &   5.00$^{*}$  & 1.8  & Census & 56753\\
J0117+5914   &   B0114+58  &   0.10    &   49.4210   & 1.77 (2.22)  &   43.40$^{*}$& 2.4& Comm.& 56518\\
   		     &     			    &     		       &   49.4207   & &   & & Census & 56781\\
J0543+2329  &   B0540+23      &   0.25    &   77.7026   & 1.56 (2.06)  &   29.00& 0.7& Census & 56780\\
J0614+2229\textsuperscript{k}    &  B0611+22   &   0.33    &   96.9100   & 1.74 (2.08) &  29.00&2.1 & Cycle 5&57391\\
    &    &       &   96.9030   &  &  & & Comm. &56384\\
J0742$-$2822\textsuperscript{k,l}    & B0740-28    &   0.17    &   73.7950   &  2.00$^{\ddagger}$  & 296.00& 2.0& Comm.&56603\\
J1851+1259   &   B1848+12   &   1.21    &   70.6333   & 2.64 (3.50)  &   8.00 & 1.8 &Census&56687\\
J1909+1102\textsuperscript{k,l}    &  B1907+10   &   0.28    &   150.0050  & 4.80$^{\ddagger}$  &   50.00 &2.5 &Comm.&56388\\
J1913$-$0440\textsuperscript{k,l}    &  B1911-04   &   0.83    &   89.3700  &  4.04 (2.79) & 118.00 & 2.6&Comm.&56259\\
   &    &      &   89.3850 &  &  & &Cycle 5&57391\\
J1917+1353\textsuperscript{k,l}   &  B1915+13   &   0.19   &   94.6580   &  5.00$^{\ddagger}$  & 43.00 & 1.8&Comm.&56525\\
J1922+2110\textsuperscript{k,l}   &  B1920+21  &   1.08    &   217.0220  &  4.00$^{\ddagger}$  & 30.00 & 2.4&Comm.&56388\\
J1935+1616\textsuperscript{k,l}    &  B1933+16 &   0.36    &   158.6210  &   3.70$^{\ddagger}$  &242.00 & 1.4&Comm.&56607 \\
J2257+5909&   B2255+58     &   0.37    &   151.1330  &  3.00$^{\ddagger}$  & 251.90$^{*}$ & 0.8&Comm. &56518\\
J2305+3100\textsuperscript{l}    &   B2303+30  &   1.58    &   49.5845 & 25.00 (3.76) &  24.00 & 2.3&Census&56773\\ [1ex]
\hline
\end{tabular}
\caption{The list of sources analysed in this paper. The periods are given to two decimal values. The DM values, to four decimals, are the values with which the data files are dedispersed. The quoted distance values ($D$) are obtained from the ATNF pulsar catalog\protect\footnotemark  \citep{ATNF2005}, and are mostly computed using the updated YMW16 electron density distribution model \citep{Yao2017}. Bracketed values are older estimates based on the NE2001 \citep{Cordes2002} electron density model. The superscripts indicate, $^{\ddagger}$: distances derived independently from DM values, e.g. from parallax measurements or the association with objects, such as supernova remnants. These are typically more reliable than values calculated from electron density distribution models; k: sources appearing in \citet{Krishnakumar2015}; l: sources in \citet{Lewandowski2015}; $^{*}$: flux density values (at 350~MHz) from \citet{Stovall2014}; $^{\dagger}$: flux density values (at 408~MHz) and flux density spectral indices, $\beta$, taken from \citet{Lorimer1995}, where $S_{\nu}\sim \nu^{-\beta}$, with $S$ the flux density and $\nu$ the frequency.}
\label{table:one}
\end{table*}

\begin{figure}
\centering
\includegraphics[width=\columnwidth]{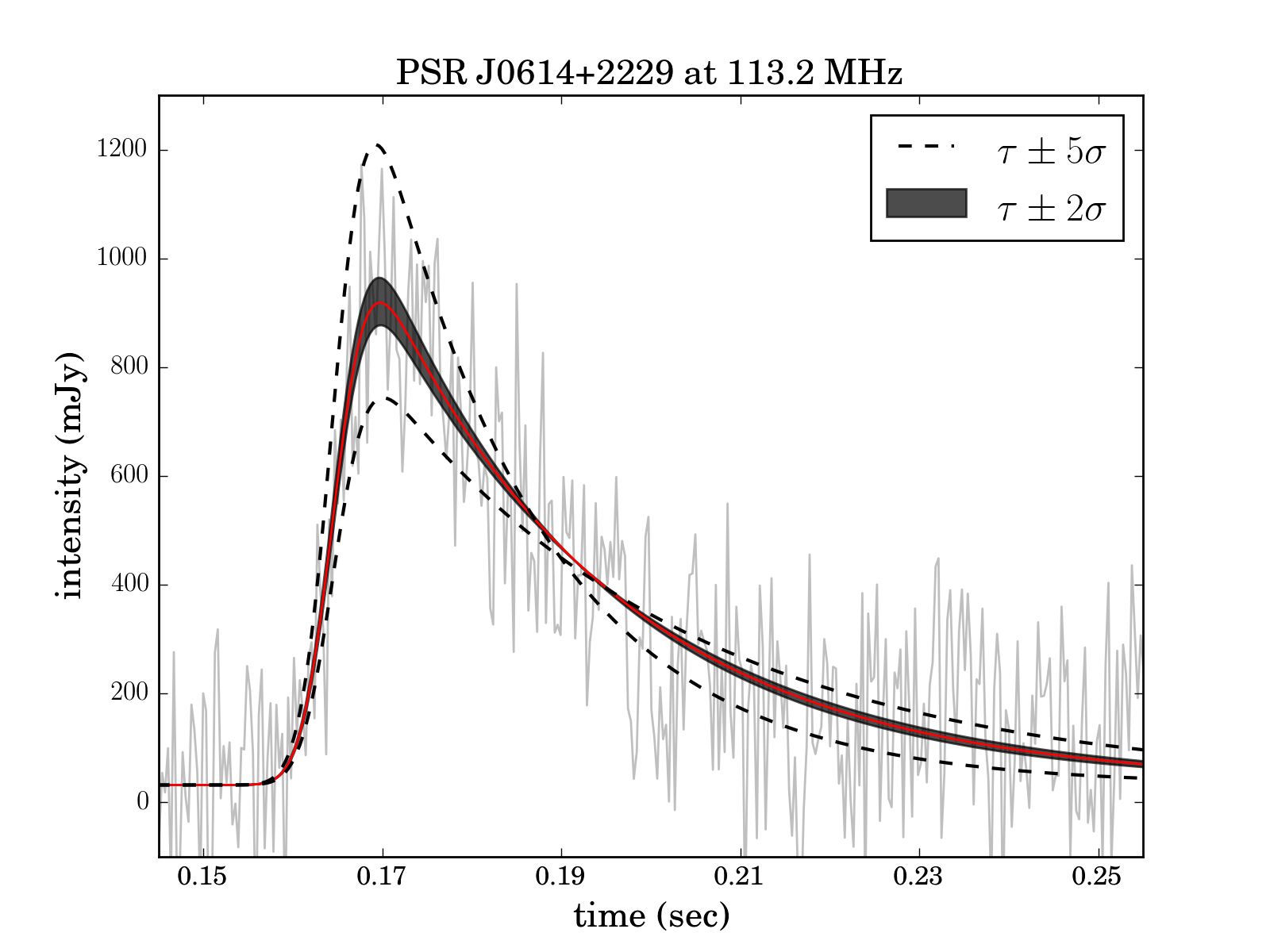}
\caption{The range of pulse shapes for  PSR J0614+2229 resulting from the best fit $\tau$ value (solid red line) and the best fit value with an added 2$\sigma$ (shaded region) or 5$\sigma$ error range (dashed lines). To ensure that the error ranges are visible, only 10~s of the pulse period is shown.}
\label{fig:25sig}
\end{figure}

After the data are reduced and flux density calibrated, as discussed in Section \ref{sec:obsdata}, we write them out to ascii format (using \texttt{pdv} in \textsc{psrchive}; \citealt{vanStraten2012}). These ascii files are subsequently analysed by our \textsc{python} scattering code, which fits each channelized profile and produces diagnostic plots, as well as $\tau$ and scatter-corrected flux density spectra. The scattering method used is the \textit{train + DC} method, described in detail in GK16. This method produces a fit to the scattered profile assuming a Gaussian intrinsic pulse profile and an approximately zero off-pulse baseline. The five fitting parameters are the amplitude ($A$), the width ($\sigma$) and the mean (centroid, $\mu$) of the intrinsic pulse, as well as the characteristic scattering time ($\tau$) imparted by the ISM and a DC fluctuation (on the order of the noise of the data) in the baseline away from zero. 

The \textit{train + DC} method also models the profile shapes that result when individual pulses in the pulse train overlap due to high levels of scattering. Under these circumstances, the off-pulse baseline is raised significantly, such that the mean flux density calculated from zero baselined data would lead to an underestimation of the flux density (for more details see GK16). For a given scattering time-scale and intrinsic pulsar parameters, we calculate the associated raised baseline level and use this value to convert the observed \textit{uncorrected flux density} to the \textit{corrected flux density}. Values quoted in the Section \ref{sec:results} are these corrected flux density measurements and their associated errors. We note that from the initial flux calibration conducted during the data reduction stage, these errors should have a minimum value of 50\%, as discussed in \citet{Bilous2016}.

The underlying Gaussian fitting parameters also allow us to obtain a DM-type value, by fitting

\begin{align}
\Delta \mu = \Delta \rm{DM} (\frac{1}{\nu_i^2} - \frac{1}{\nu_H^2}),
\label{eq:DM}
\end{align}
 
\noindent where $\Delta \mu = \mu_H - \mu_{i}$ is the shift in the intrinsic pulse centroid value between a given frequency channel ($\nu_i$) and the highest frequency channel ($\nu_H$). We label this proportionality constant as $\Delta$DM, since it represents a change in the DM value obtained from dedispersing the scattered pulse profiles during the initial PulP analysis.

\footnotetext{http://www.atnf.csiro.au/research/pulsar/psrcat}

We consider two independent scattering mechanisms, in each case using the same fitting procedure as described above. In the first instance, we consider an isotropic scattering model (IM). This model assumes that a single thin scattering screen, which scatters radio waves isotropically (following a circularly symmetric Gaussian distribution), is responsible for the broadened pulse profile. The temporal broadening function associated with the ISM takes the form, 

\begin{align}
{f_{t}} &= \tau^{-1}e^{-t/\tau}{U(t)}\label{eq:ftiso}
\end{align}

\noindent The unit step function, ${U(t)}$, ensures that we only consider time $t >0$. Equation
\eqref{eq:ftiso} is valid assuming the scattering screen is infinite transverse to the line of sight. 
For this model the theoretically expected frequency dependence is $\tau \propto \nu^{-4}$. 

The second scattering model is an extremely anisotropic scattering model (AM). In the limiting case, where the anisotropy~$\rightarrow \infty$, such a scattering screen will only scatter radio waves along a single dimension. The associated broadening function is then, 

\begin{align}
f_{t} &= \frac{e^{-t/\tau}}{\sqrt{\pi t \tau}} U(t)\label{eq:ft1D},
\end{align}
 
\noindent with again  $\tau \propto \nu^{-4}$. These scattering mechanisms are described in more detail in \citet{CordesLazio2001} and GK16.

In order to account for the frequency integration of the channelized data, (e.g. integrating over an 8~MHz band when HBA data are averaged to 16 frequency channels), we make use of equation (6) in GK16 to extract the monochromatic frequency associated with an obtained $\tau$ value. The impact of this effect is discussed in detail in GK16. It should be noted that for the LOFAR data in this paper the correction in $\alpha$ is less than 0.5\%.

After the $\tau$ value and the corresponding frequency value for each channelized profile is extracted,  a power-law fit (weighted inversely with the square of the errors) to the $\tau$ spectrum is obtained. For both the profiles and $\tau$ spectra we report a 1$\sigma$ error. A fit to the average profile of PSR~J0614+2229 in Fig. \ref{fig:25sig} shows the typical impact of $\tau$ errors on the scatter broadened shape. The dark grey shaded region around the best-fit $\tau$ value (red solid line) represents a 2$\sigma$ error, and the dashed line shows the  5$\sigma$ error margin. 

To evaluate the goodness of the model fits we make use of two standard metrics, the reduced Chi-squared ($\chi^2_{\rm{red}}$) value and the Kolmogorov--Smirnov (KS) test. The KS test is applied to the residuals (data $-$ model) to test the Gaussianity of the residuals. This test provides the probability that the residuals follow a Gaussian distribution. The main objective is to identify examples where the residuals are severely non-Gaussian. 
We compare the outcomes of these metrics for the two models (isotropic and extremely anisotropic) for all the sources.


\section{Results}\label{sec:results}

\subsection{Scattering and flux density analysis for each pulsar}\label{sec:pulsarbypulsar}

Here, we discuss our results for each pulsar individually. For convenience, we
provide sub-headings for each pulsar, summarizing the period,
scattering time and DM, with no errors, as well as the time and frequency resolution of the data ($\delta t$ and $\delta \nu$). Data sets are abbreviated as Co, Ce and Cy for Commissioning, Census and Cycle 5 data, respectively. 

Thereafter, we show the outcomes of our fitting models and consider the goodness of the fits (detailed results can be found in Appendix \ref{app:profiles}). We present the fitted scattering values ($\tau$), along with the computed $\tau$ spectra, providing comparisons to values from
the literature. We then discuss the flux density spectra that result from the
scattering analysis, and consider how these values compare to published
studies. The basic pulsar properties of the set are summarized in
Table~\ref{table:one}.

\begin{figure*}
\centering
\includegraphics[width=\textwidth]{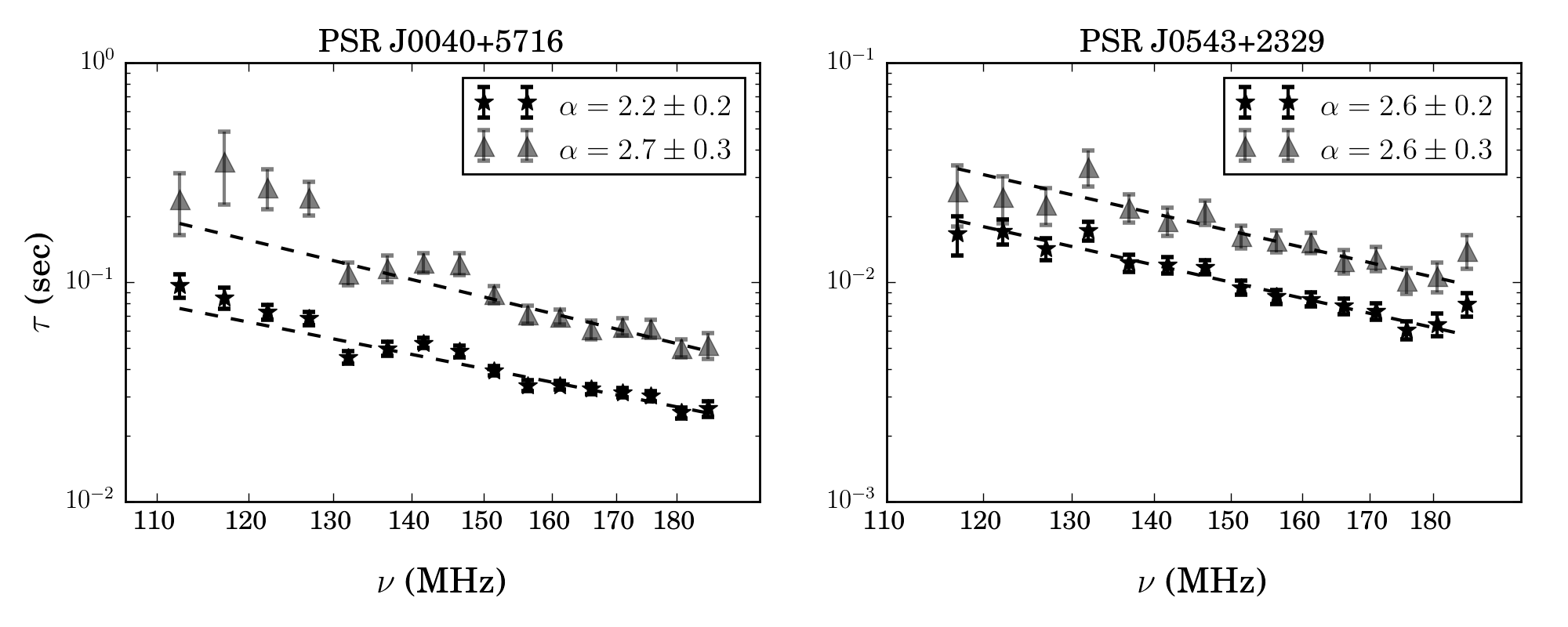}
\caption{The $\tau$ spectra for PSRs J0040+5716 and J0543+2329. IM fits are shown in black (stars) and AM fits in grey (triangles).}
\label{fig:tau1}
\end{figure*}

\subsubsection{PSR~J0040+5716} 
{\footnotesize{$\rm{P}=1.12$ s, $\tau_{150}=40$~ms, $\rm{DM}=92.5$ pc cm$^{-3}$, $\delta t=2.2$~ms, $\delta \nu=1.5$ kHz}}\\

\vspace{3mm}

PSR~J0040+5716 was discovered in a search for low-luminosity pulsars
at 390~MHz using the 92-m transit telescope at Green Bank (92m-GBT) in West Virginia, US \citep{Dewey1985}, and has the lowest tabulated flux density value for our list
of sources. The distance estimate of this pulsar has changed from 4.48 kpc \citep{Taylor1993} to 2.99 kpc (NE2001 model, \citealt{Cordes2002}) to more recently 2.42 kpc \citep{Yao2017}. Its DM value of 92.5~pc~cm$^{-3}$ is close to the
mean of the set DM values.

We use LOFAR HBA Census data for this pulsar. From the European Pulsar Network (EPN) data base\footnote{http://www.epta.eu.org/epndb/} \citep{Lorimer1998}, we
find that it is a single component pulsar at 408~MHz
\citep{GouldLyne1998} and, as it exhibits no clear secondary
components in our data set, it is likely a single component pulsar down
to low frequencies. There are currently no scattering measurements
published for this pulsar.

The profile fits over all 16 channels (for which the lowest peak S/N
is equal to 4.4) produce a $\tau$ spectrum with spectral index
$\alpha = 2.2 \pm 0.2$ for the IM and $\alpha = 2.7 \pm 0.3$ for the AM fits (Fig. \ref{fig:tau1}, left panel). The
error bars on $\tau$ and the spread in $\tau$ values obtained from the
IM (black, stars) are typically smaller than for the AM (grey,
triangles). The goodness of fit parameters, shown in Table \ref{table:app} in Appendix \ref{app:profiles}, do not favour a particular scattering model.

The $\tau$ spectrum appears to have three segments with breaks around
130~MHz and 150~MHz. The residuals of the profile at 136~MHz are less
Gaussian than for the other frequency
channels. 

Splitting the bandwidth into only four channels to increase the S/N and
refitting, leads to a similar $\alpha$ value of $2.1 \pm 0.3$. We
also find that the $\sigma$ (unscattered pulse width) versus frequency
relationship for this pulsar follows a rough power-law dependence (see Section 5.3).
Fixing the $\sigma$ values to this power-law dependence and refitting,
increases the $\alpha$ value to 2.3, which is within the errors of
the original value. We conclude that the obtained $\alpha$ value is
robust, and not critically sensitive to these tested changes in
fitting method.

The $\Delta$DM values obtained from equation \eqref{eq:DM}, are presented in Table \ref{table:two}.

The flux density values inferred from our fitting models show no clear
dependence on frequency (and therefore scattering). \citet{Bilous2016}
measured a mean flux density at HBA frequencies of $33 \pm 17$~mJy. Similarly,
we find a frequency-averaged mean (corrected) flux density of 31.4~mJy (IM), and 35.8~mJy (AM). 
At 151.3~MHz, we measure flux density values of $34 \pm 30$~mJy and $37 \pm 22$~mJy (IM and AM). \citet{Stovall2014} published a value of 4.7--5.0~mJy at 350~MHz.
Using a simple power-law spectrum and the spectral index from Table \ref{table:one}, the \citet{Stovall2014}
result implies a flux density of $21.9 \pm 0.7$~mJy at 151.3~MHz, well within
our flux density error margins.

\subsubsection{PSR~J0117+5914} 

\begin{figure*}
\centering
\includegraphics[width=\textwidth]{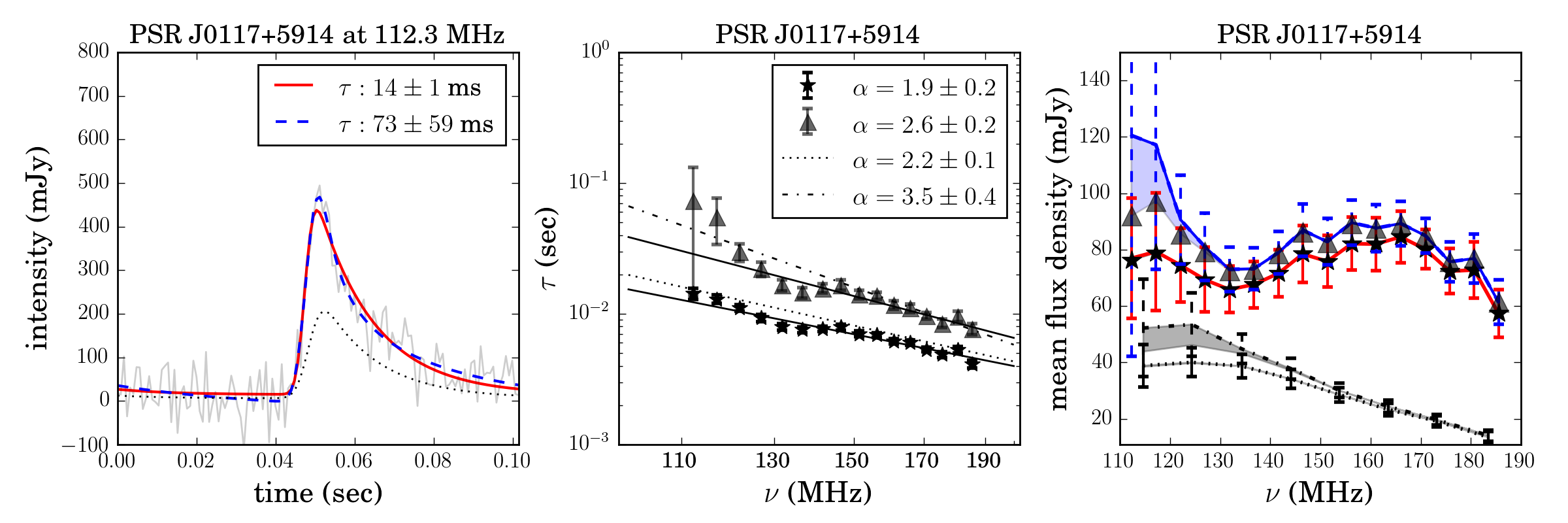}
\caption{PSR~J0117+5914. \textit{Left:} average Census profile with the IM fit in red (solid line) and the AM fit in blue (dashed line). An IM fit to the Commissioning profile at 114.6~MHz is also shown (dotted line). \textit{Middle:} the associated $\tau$ spectra. It shows $\tau$ values obtained from the IM (Census data: stars with solid line fit, Commissioning data: dotted line) and from the AM (Census data: triangles with solid line fit and Commissioning data: dash--dotted line). \textit{Right:} flux spectra calculated from scattered profiles. Uncorrected flux density values are shown as stars (IM) and triangles (AM) for the Census data, with corrected flux density values as red and blue solid lines, respectively. For the IM, the uncorrected and corrected values are near equal and for the AM, the shaded region represents the increase in flux density when applying a scattering correction. The error bars associated with the corrected flux density values are shown in solid (red, IM) and dashed (blue, AM). The flux density spectrum for the Commissioning data appears at lower mean flux density values (dotted line for corrected IM flux density and dash--dotted line for corrected AM flux density).}
\label{fig:tau0117}
\end{figure*}

{\footnotesize{$\rm{P} = 0.10$ s, $\tau_{150} = 7$~ms (Ce) 8~ms (Co), $\rm{DM}= 49.4$ pc cm$^{-3}$, $\delta t= 0.8$~ms (Ce) 0.1~ms (Co), $\delta \nu=3.1$ kHz (Ce) 12.2 kHz (Co)}}\\
\vspace{3mm}

PSR~J0117+5914 is the fastest rotating pulsar in the set and was
discovered in a 92m-GBT survey for short-period pulsars
\citep{StokesTaylor1985}. It is one of the closest pulsars in the set (1.77~kpc) and has the lowest DM value (49.4 pc cm$^{-3}$) of the studied pulsars. This is the only pulsar for which we have both Commissioning and
Census data. We use eight frequency channels across the band for the
Commissioning data, and 16 for the Census data.

Fig. \ref{fig:tau0117} shows the results from both data sets jointly.
The left-hand panel shows the average pulse profile of the Census data
at 115~MHz, with the IM fit to the data in red (solid), and the AM fit
in blue (dashed). These profiles have lower phase resolution than the
other data sets, with only 128 bins across the pulse profile. We note
that the $\tau$ value obtained using the AM is more than four times that
obtained using the IM. The AM seems to fit pulse peak better. 

The middle panel shows the $\tau$ spectra for both
data sets and models, with Census data as solid lines (stars (IM) and triangles (AM) data points)
and Commissioning data in dotted (IM) and dash--dotted (AM) lines.

The data sets show similarities, and in the case of the IM, the power
$\alpha$ values agree within 1$\sigma$. The IM fit
produces one of the lowest $\alpha$ values of our set,
$\alpha = 1.9 \pm 0.2$ for the Census and
$\alpha = 2.2 \pm 0.1$ for the Commissioning data. The high S/N of the
Census data constrains the goodness of fit of the two models well (Appendix \ref{app:profiles}, Table \ref{table:app}).

The right-hand side of Fig. \ref{fig:tau0117} shows the flux density
measurements from the scattering fits. The shaded regions towards low frequency indicate the flux density corrections that account for scattering effects. Corrected flux density values with error bars, range from approx. 50 to 100~mJy (IM) or 40 to 200~mJy (AM)
for the Census data. In contrast, the Commissioning data flux density values are much lower.
\cite{Stovall2014} recently measured the flux density to be 43.4~mJy at 350
MHz. Using the spectral index from Table~\ref{table:one}, this flux density is
translated to approx. 330~mJy at 150~MHz -- higher than the
measured values between 30 and 90~mJy (Co and Ce) at 150~MHz here, even with an additional 50\%
error range. However, the flux densities obtained from the Census data, do
agree well with the mean flux density value of $S_{\rm{HBA}} = 79 \pm 39$~mJy,
published in \cite{Bilous2016}, where flux density values were not corrected for scattering as in this paper.

For both data sets, the flux density values obtained from the AM are higher
than the corresponding IM values. The high quality Census data for
this pulsar are shown in Appendix \ref{app:profiles}. Below 135~MHz, the data suggest that scattering tails are `wrapping
around', i.e. the scattering tail stretches beyond the pulse period.
This could be correlated with turnover in the flux density spectrum of the
Commissioning data. However, the Census data do not show a clear
turnover at these frequencies. 

In Table A2 in the Appendix \ref{app:profiles}, we compare DM corrections for pulsars for which we have more than one data set.  We note that applying the $\Delta$DM correction from the AM for PSR~J0117+5914, results in DM values that remain more similar between the two
observing epochs.

\subsubsection{PSR~J0543+2329} 

{\footnotesize{$\rm{P}= 0.25$ s, $\tau_{150} = 10$~ms, $\rm{DM}= 77.7$ pc cm$^{-3}$, $\delta t=1.0$~ms, $\delta \nu=1.5$ kHz}}\\
\vspace{3mm}

This pulsar is associated with the supernova remnant IC~443. It was
discovered in one of the earliest pulsar searches with the Lovell
telescope at Jodrell Bank, at 408~MHz \citep{Davies1972}. The Census
data were channelized to 16 average pulse profiles, of which the
lowest channel is excluded from the analysis for having a low peak S/N
value. As seen in Fig. \ref{fig:tau1}, the $\tau$ spectra for both
scattering models have similar power-law indices.

\cite{Cordes1986} gives $\tau = 2.69$ $\rm{\mu}$s at 1~GHz. Our
measurement of $\tau = 0.01$~s at 150~MHz and $\alpha = 
2.61$, would translate to $\tau = 66$ $\rm{\mu}$s at 1~GHz. The required
spectral index to link our 186~MHz observation to the 1~GHz
measurement, is $\alpha = 4.6$, close to the Kolmogorov value of
4.4. \cite{Kuzmin2007} published $\tau = 15 \pm 5$~ms at 111~MHz,
using data from the Large Phased Array Radio Telescope (BSA) at the
Pushchino Radio Astronomy Observatory in Russia. Again using our obtained
IM power-law index we find $\tau = 22 \pm 6$~ms at 111~MHz, which lies
just outside their error bounds.

The flux density measurements show no clear frequency dependence. At 151.4~MHz, the computed flux density values are  $46 \pm 15$~mJy (IM) and  $47 \pm 12$~mJy (AM). To incorporate the uncertainty in the initial flux calibration during the data reduction stage, we augment these values to $46 \pm 23$~mJy (IM) and  $47 \pm 24$~mJy (AM).
\citet{Bilous2016} published a similar mean flux density of $36 \pm 18$~mJy. The measured flux density of $S_{408} = 28.9 \pm 1.3$~mJy at 408~MHz
\citep{Lorimer1995} leads to an expected flux density of around 58~mJy at 150~
MHz, using a simple power-law spectrum with spectral index 0.7 (Table \ref{table:one}).  This value lies within 50\% of our HBA flux density measurement.

\subsubsection{PSR~J0614+2229} 
{\footnotesize{$\rm{P}= 0.33$ s, $\tau_{150} = 15$~ms (Co \& Cy) , $\rm{DM}= 96.9$ pc cm$^{-3}$, $\delta t=0.3$~ms (Co \& Cy), $\delta \nu=195$ kHz, coherently dedispersed (Cy), 12.2 kHz (Co)}}\\
\vspace{3mm}

\begin{table*}
\centering
\begin{tabular}{p{1.7cm} p{1.8cm} p{1.7cm} p{1.7cm} p{1.8cm} p{1.8cm} p{1.6cm}}
\hline
Reference	&	\textit{Kuzmin 2007}	&	\textit{This paper }& \textit{Slee 1980}	&	\textit{Alurkar 1986}	&	\textit{L\"{o}hmer 2004}	&	\textit{K15}	\\
\hline
Freq (MHz)	&	102 /111			&	150  		& 160 				&	160 			&	243 			&	327 	\\
\hline
J0614+2229	&	$40 \pm 10$		&$15 \pm 0 $		&					&				&				&	$1.74 \pm 0.03$	\\
J0724$-$2822	&	$22 \pm 5.3^{*}$	& $20 \pm 2$		&					&$24.5 \pm 2.8$	&				&	$0.71 \pm 0.01$	\\
J1909+1102	&					& $42 \pm 3$		& $27 \pm 7$			&$26.5 \pm 8.1$	&				&	$1.35 \pm 0.02$	\\
J1913$-$0440	&	$35 \pm 15^{*}$	& $7 \pm 0$		&$32 \pm 5$			&$16.7 \pm 1.8$	&				&	$0.19 \pm 0.01$	\\
J1917+1353	&	$40 \pm 20^{*}$	&$11 \pm 1$		&$12 \pm 3$			&$11.7 \pm 1.9$	&				&	$0.36 \pm 0.01$	\\
J1922+2110	&					&$42 \pm 2$		&					&$96.8 \pm 50$	&$4.4 \pm 1.5 $	&	$2.3 \pm 0.1$	\\
J1935+1616	&	$50 \pm 15$		&$20\pm1$		&$25 \pm 4$			&$21.7 \pm 1.6$	&$4.6 \pm 0.2$	&	$3.21 \pm 0.02$	\\
J2303+3100	&	$13 \pm 3$		&$9 \pm0$		&					&$9.9 \pm 3.6$	&				&					\\
\hline
\end{tabular}
\caption{Low frequency characteristic scattering time values ($\tau$, in~ms) from the literature for pulsars with two or more of these values quoted in text. $^{*}$Values at 102~MHz, the rest of the \citet{Kuzmin2007} values are at 111~MHz. See text for full references.}
\label{table:lit}
\end{table*}

\begin{figure*}
\centering
\includegraphics[width=\textwidth]{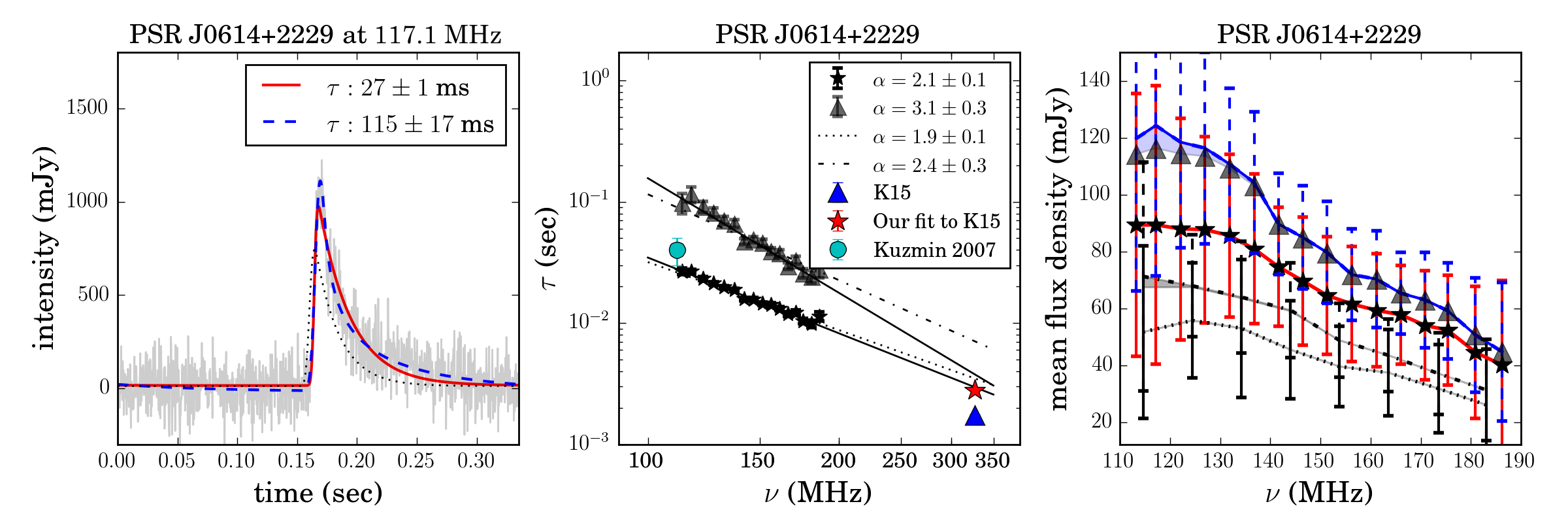}
\caption{\textit{Left:} the average profile of PSR~J0614+2229 at 117.1~MHz (grey), and the IM (red, solid line) and AM fits (blue, dashed line). The dotted line shows the Commissioning data average profile at 124.4 MHz. \textit{Middle:} associated $\tau$ spectrum for the IM (Cycle 5 data: stars with solid line fit and Commissioning data: dotted line) and for the AM (Cycle 5: triangles with solid line fit and Commissioning data: dash--dotted line) with added data points from K15 and \citet{Kuzmin2007}. \textit{Right:} associated flux density spectra, with Cycle 5 data at the top and Commissioning data at lower flux density values. The markers are similar to Fig. \ref{fig:tau0117}.}
\label{fig:tau0614}
\end{figure*}

PSR~J0614+2229 was discovered in the same survey as PSR~J0543+2329.
For this pulsar we have LOFAR Commissioning and Cycle~5 data. The Cycle 5 data lead
to a $\tau$ spectrum for which $\alpha = 2.1 \pm 0.1$ (IM), and $\alpha = 3.1 \pm 0.3$ (AM),
as shown in Fig.~\ref{fig:tau0614}. The figure also shows $\alpha$
values $1.9 \pm 0.1$ (IM) and $2.4 \pm 0.3$ (AM) from the Commissioning data, such that 
especially for the IM the values are in good agreement.  
 
The left-hand panel of Fig. \ref{fig:tau0614} shows the comparison of profile fits, between the IM (red, solid line) and the AM (blue, dashed line) at 117.1~MHz.  The computed $\tau$ values differ by a factor of approx. 4, and the AM fit follows the tip of the profile peak more closely. The computed $\chi_{\rm{red}}^2$ values are consistently lower for the AM,
with mean values quoted in Table \ref{table:app} in Appendix \ref{app:profiles}.

This pulsar lies at a distance of 1.74~kpc (previously estimated to be at 4.74~kpc, see Table~\ref{table:one}) and has a DM value of
96.9~pc~cm$^{-3}$, which is similar to PSR~J0040+5716. The $\alpha$
values obtained using the IM for these two pulsars are also comparable.

This pulsar is the first in our source list to overlap with an Ooty
Radio Telescope (ORT) data set at 327~MHz, for which characteristic
scattering times were recently published (\citealt{Krishnakumar2015},
hereafter K15\footnote{The data used in \citet{Krishnakumar2015} are
  available from http://rac.ncra.tifr.res.in/da/pulsar/pulsar.html.}).
Their data and methods are different to ours in several ways: (1) the
average pulse profiles are far less scatter broadened than our sample
set. (2) They follow the method described in \citet{Lohmer2001}, in
which a high frequency pulsar is used to estimate and fix the width of
a Gaussian template used in the fitting method. (3) Wrap around
scattering is not modelled (and not required at this frequency for
these pulsars) and (4) for all pulsars marked as `double-component'
profiles, only the trailing component of the pulse profile is fitted
for.

The K15 characteristic scattering time for this pulsar is
$\tau = 1.74 \pm 0.03$~ms at 327~MHz. Refitting their data with our
techniques gives $\tau = 2.80 \pm 0.02$~ms. The main difference seems
to come from their choosing a fixed width. A fit with our code,
specifying a fixed width as inferred from a high frequency profile,
leads to a more similar $\tau$ value of 1.80~ms, although we note that
the fit to the amplitude of the profile becomes considerably worse. We
also note that the pulse profile is not highly scattered at a
frequency of 327~MHz, which increases the uncertainty in the obtained
$\tau$ values.  \cite{Kuzmin2007} published a characteristic scattering time of
$\tau = 40 \pm 10$~ms at 111~MHz. At 113.2~MHz, we find
$\tau = 26.7 \pm 1.6$~ms, which translates to $28.1 \pm 2.7$~ms at
111~MHz when using our obtained IM $\alpha$ value. The literature values are shown in the middle panel of Fig.~\ref{fig:tau0614}. The extrapolations of the Cycle 5 data fits 
are in good agreement with our (IM) fit to the K15 data at 327~MHz. Low frequency scattering time results from the literature for this pulsar and subsequent
ones are summarized in Table \ref{table:lit}.

The flux density and spectral index (see Table \ref{table:one})
suggests a mean flux density of 237~mJy at 150~MHz, much larger than our measured values of $65 \pm 21$~mJy (IM) or $80 \pm 18$~mJy (AM). Again these error bars should be augmented to a minimum value of 50\%, such that at 150~MHz they are $65 \pm 33$~mJy (IM) or $80 \pm 40$~mJy (AM). The flux density spectrum associated with the IM flattens out towards the lowest frequencies, with no obvious turnover. The AM shows a more clear turnover towards lowest frequencies, in
agreement with both the raised baseline measurements (represented by the shaded area in the flux density spectrum) and the wrapped scattering tails seen in the AM profile fits (Appendix \ref{app:profiles}). 

\subsubsection{PSR~J0742$-$2822}\label{sec:J0742} 

{\footnotesize{$\rm{P}= 0.17$ s, $\tau_{150} = 20$~ms, $\rm{DM}= 73.8$ pc cm$^{-3}$, $\delta t=0.2$~ms, $\delta \nu=12.2$ kHz}}\\
\vspace{3mm}

\begin{figure*}
\centering
\includegraphics[width=\textwidth]{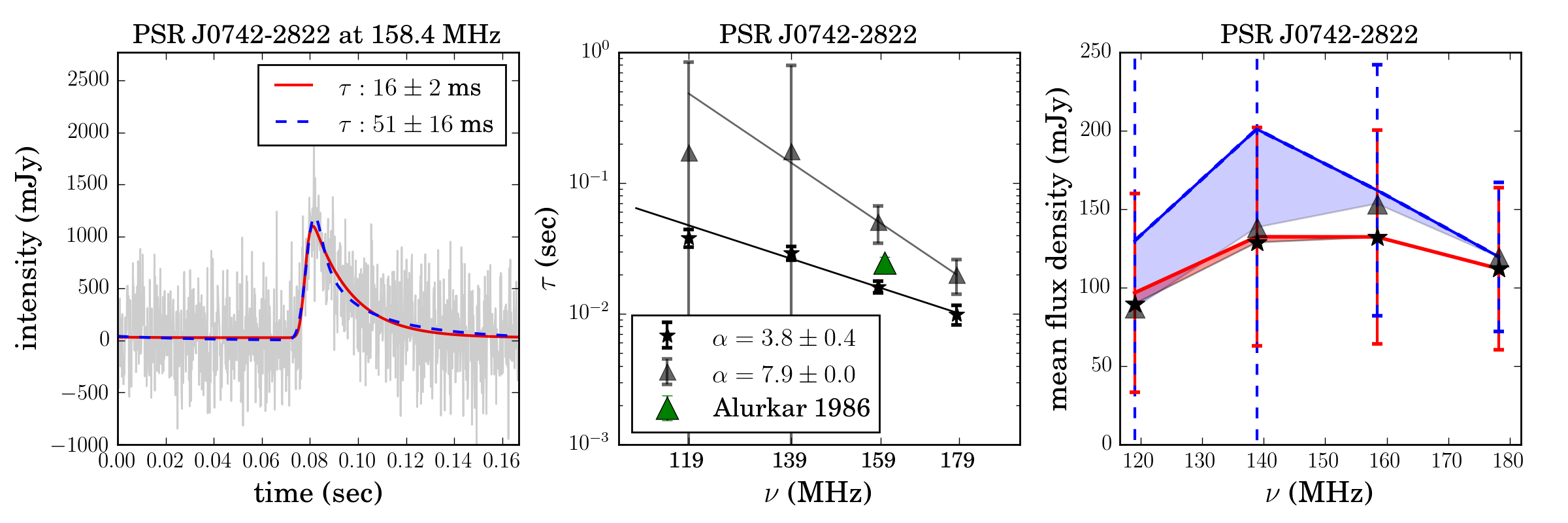}
\caption{\textit{Left:} low S/N fitted profile from Commissioning data. \textit{Middle:} associated $\tau$ spectrum with an added data point from \citet{Alurkar1986}. \textit{Right:} flux spectrum showing uncorrected and corrected flux density values. All markers and colours as in previous figures.}
\label{fig:tau0742}
\end{figure*}

PSR~J0742$-$2822 is a young pulsar associated with an H\textsc{i} shell in Puppis
\citep{StacyJackson1982}.

The Commissioning data we have for this pulsar have
relatively low peak S/N values. Using 16 frequency channels across the band
results in profiles for which all have S/N$ < 2.7$. We therefore use four
channels, which still yield a maximum S/N of only 4.5. Using four
channels, we find $\alpha = 3.8 \pm 0.4$ for the IM. This is the
first fit for which the theoretical value of $\alpha = 4$ lies within
the error bars of the measured $\alpha$ value. Furthermore, the AM
fails to produce convincing fits, as can be seen in Fig.
\ref{fig:tau0742}, resulting in a large $\alpha$ value with a tight error
margin, $\alpha = 7.9 \pm 0.0$. The error bars on $\tau$ for the
first two channels are large ($> 350\%$), such that the power-law fit
weighted by the inverse of the errors squared, is dominated by the two
higher frequency $\tau$ values only, leading to an inaccurate
spectrum.

A scattering measurement of $\tau = 24.5 \pm 2.8$~ms at 160~MHz was
obtained by averaging nine observations using the Culgoora circular array
\citep{Alurkar1986}. This data point is shown in Fig. \ref{fig:tau0742}. We find a $\tau$ measurement at 159.6~MHz of
$16.1 \pm 1.6$~ms. Since \citet{Alurkar1986} used a fixed width model
(obtained from profiles at 408~MHz), a difference is expected. K15 find
$\tau = 0.71 \pm 0.01$~ms at 327~MHz. In comparison, our spectral
index of $\alpha = 3.8 \pm 0.4$ implies $\tau = 1.02 \pm 0.25$~ms at
327~MHz. Fitting the K15 data with our code, gives
$\tau = 1.5 \pm 0.1$~ms, however, at 327~MHz their data show that the
top peak of the pulse profile has split into two components. In these
cases, they fit only the secondary component (as shown in Fig. 1 in
K15), whereas we fit the whole profile. It is worth noting that at 327~MHz, the profile is not highly scattered. Lastly, \citet{Kuzmin2007} published a value of $22 \pm 5.3$~ms at 102~MHz. This is much lower
than predicted by our spectral index. A summary of the relevant
literature values is given in Table~\ref{table:lit}.

The low S/N data make definitive statements about
this pulsar hard. A fit using the AM fails, making it an unlikely model. The trend in $\Delta\mu$ versus frequency is very weak, providing a $\Delta$DM value with large error bars (Table \ref{table:two}).

The flux spectral index in Table \ref{table:one} suggest a mean flux density of over
2~Jy at 150~MHz. The low S/N data lead to corrected flux density values with large error bars.  At 140~MHz, the measured values are  $133 \pm 70$~mJy and $201 \pm 476$~mJy for the IM and AM, respectively. 

For both models, the corrections to the flux density values due to scattering are substantial (Fig. \ref{fig:tau0742}, shaded), and represent a maximum increase in the best fit flux density of 8\% (IM) and 50\% (AM).  The flux density spectrum for this pulsar shows a turnover around 140~MHz, and the profile fits suggest that this could be ascribed to wrap around scattered pulses. 

\newpage
\subsubsection{PSR~J1851+1259} 

{\footnotesize{$\rm{P}= 1.21$ s, $\tau_{150} = 6$~ms, $\rm{DM}= 70.6$ pc cm$^{-3}$, $\delta t=1.2$~ms, $\delta \nu=3.1$ kHz}}\\
\vspace{3mm}

\begin{figure}
\centering
\includegraphics[width=\columnwidth]{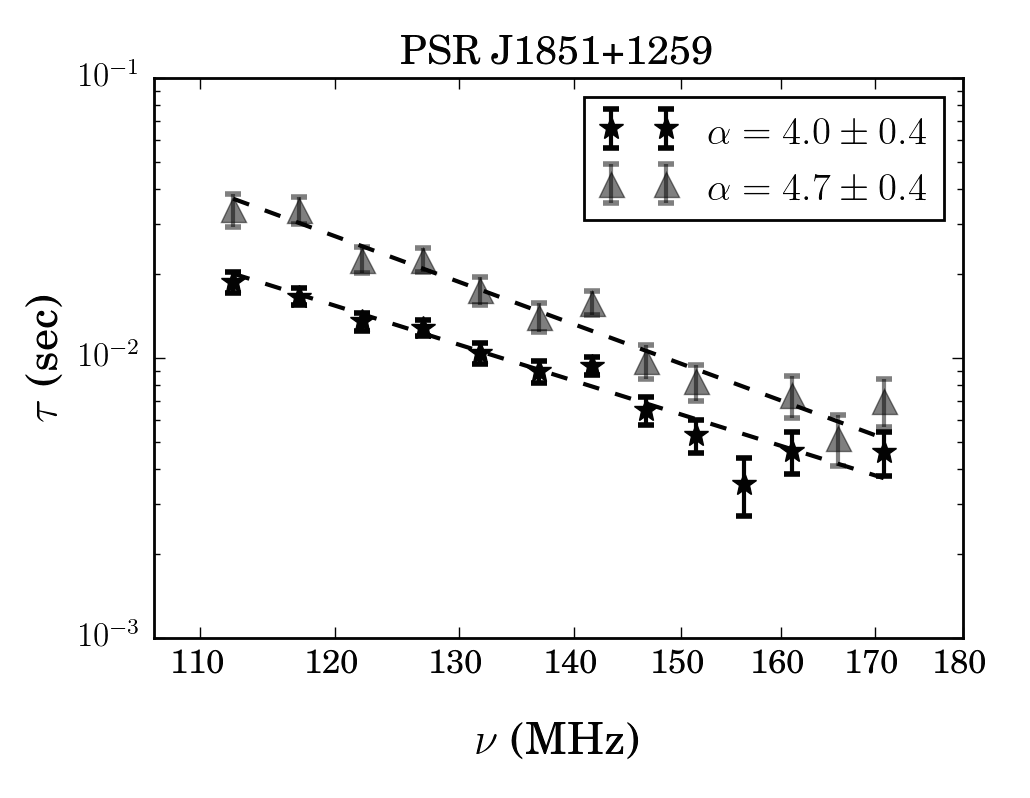}
\caption{The $\tau$ spectra for PSR J1851+1259. IM fits are in black (stars) and AM fits in grey (triangles).}
\label{fig:tau2} 
\end{figure}

This pulsar was discovered in the same survey as PSR~J0117+5914
\citep{StokesTaylor1985}. Here, we fit Census data only. Towards the
higher end of the HBA band, the profiles appear marginally scattered.
For this reason, four of the higher frequency channels were not used. The
remaining 12 channels were all fitted with $\tau$ values of which the
1$\sigma$ error bars were less than 23\%.

The IM fits to the data result in an $\alpha$ value as
predicted by a Gaussian isotropic scattering screen, namely
$\alpha = 4.0 \pm 0.4$. For the AM, $\alpha = 4.7 \pm 0.4$, as
shown in Fig. \ref{fig:tau2}. This is the second pulsar for which the
obtained values match the theoretically predicted spectral index. 

The mean Census flux density measurement for this pulsar is equal to
$37 \pm 19$~mJy \citep{Bilous2016}. The expected flux density measurement at
150~MHz using the values in Table \ref{table:one} is 48.5~mJy. Both
these figures are in good agreement with our results, namely
$43 \pm 29$~mJy (IM) and $43 \pm 33$~mJy (AM) at 151.4~MHz. The flux density spectrum is shown in Appendix \ref{app:flux}.

\subsubsection{PSR~J1909+1102} 

{\footnotesize{$\rm{P}= 0.28$ s, $\tau_{150} = 42$~ms, $\rm{DM}= 150.0$ pc cm$^{-3}$, $\delta t=0.3$~ms, $\delta \nu=12.2$ kHz}}\\
\vspace{3mm}

\begin{figure*}
\centering
\includegraphics[width=\textwidth]{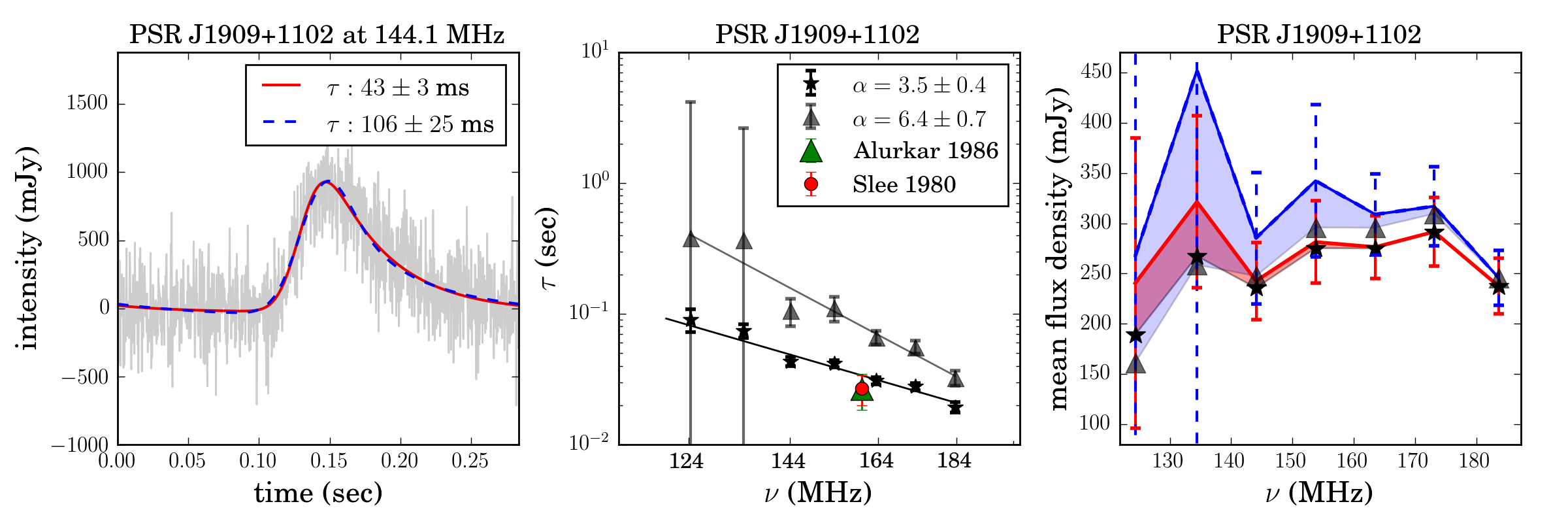}
\caption{\textit{Left:} scattering fits to the profile of PSR~J1909+1102 at 144.1~MHz. IM is shown in red (solid) lines and AM is in blue (dashed) lines. \textit{Middle}: the $\tau$ spectra associated with the scattering fits. IM in black (stars) and AM in grey (triangles). \textit{Right:} the uncorrected flux density values (IM, stars and AM, triangles) and corrected flux densities (IM, red line and AM, blue line). The flux density corrections shifts the turnover in the spectrum from higher towards lower frequencies.}
\label{fig:tau1909}
\end{figure*}

PSR~J1909+1102 was discovered in a low-latitude pulsar survey with the
Lovell Telescope at Jodrell Bank in 1973 \citep{Davies1973}. This
pulsar has one of the higher DM values in our observing set and is
highly scattered. In Section \ref{sec:disc} (Fig. \ref{fig:DistDM}), we
show the DM versus distance values for our set of sources. PSR~J1909+1102 is one of the outliers on this
plot.

Since the pulsar is at a low latitude, we only have Commissioning data.
To increase the S/N, we split the HBA band into eight frequency channels. 
The peak S/N values for all but the lowest frequency channel (which we exclude from the analysis) range from 3.5 to 9.1. The $\tau$ spectrum resulting from these measurements is shown
in Fig. \ref{fig:tau1909} (middle panel). 

Two similar $\tau$ measurements at 160~MHz were obtained by
\citet{Alurkar1986} and \citet{Slee1980} using the Culgoora circular
array, namely $26.5 \pm 8.1$ and $27 \pm 7$~ms (Table \ref{table:lit}). At 163.5~MHz, we obtain
$30.7 \pm 1.9$~ms, within their estimated error at 160~MHz.

\citet[hereafter L15]{Lewandowski2015} published a spectral index
value of $\alpha = 3.61 \pm 0.03$, based on literature values and
a fit to a higher frequency profile from the
EPN data base. L15 follow the fitting approach as described in
\citet{Lohmer2001}. L15 do, however, allow for the width of the
unscattered profile to change. Many of the error bars on $\alpha$ values (and some of the $\alpha$ values itself) in L15 were updated in \citet[hereafter L15b]{Lewandowski2015b}. For PSR~J1909+1102, the updated value is $\alpha = 3.61^{+0.79}_{-0.74}$. Their $\alpha$ value lies well within
the error bars of our spectral index measurement of $\alpha = 3.5 \pm 0.4$ (IM).

K15 published a value of \mbox{$\tau = 1.35 \pm 0.02$~ms} at 327~MHz. Fitting the K15 data ourselves, we find $\tau = 2.4$~ms at 327~MHz. As with PSR~J0614+2229, the main difference seems to be coming from whether or not a model uses a fixed width intrinsic profile. Fitting
their data with a fixed width template we find $\tau = 1.10 \pm 0.1$~ms, in closer agreement to their published value.

The flux density value and flux spectral index in Table \ref{table:one} implies a flux density of 610~mJy at 150~MHz (using a simple power-law model) which is
more than twice our measured IM value of  $281 \pm 41$~mJy at 154~MHz. At this frequency, the AM predicts $342 \pm 76$~mJy. To adhere to 50\% error margins, we change these to $281 \pm 141$~mJy (IM) and $342 \pm 171$~mJy (AM). 
The flux density spectrum of PSR~J1909+1102 (Fig.
\ref{fig:tau1909}, right-hand panel), shows significant observed flux density loss due to scattering (shaded areas) that we correct for. The corrected spectrum shows a turnover at around 140~MHz. We note that this pulsar has the largest ratio
of $\tau$ at 150~MHz to pulse period, namely $\tau_{150}/P = 0.15$,
and the scattering tails wrap around the full rotational phase.

\subsubsection{PSR~J1913$-$0440}

{\footnotesize{$\rm{P}= 0.83$ s, $\tau_{150} = 7$ms (Cy) 9ms (Co), $\rm{DM} = 89.4$pc cm$^{-3}$, $\delta t=0.8$~ms (Cy \& Co), $\delta \nu=195$ kHz (Cy \& Co), coherently dedispersed}}\\
\vspace{3mm}

\begin{figure*}
\centering
\includegraphics[width=\textwidth]{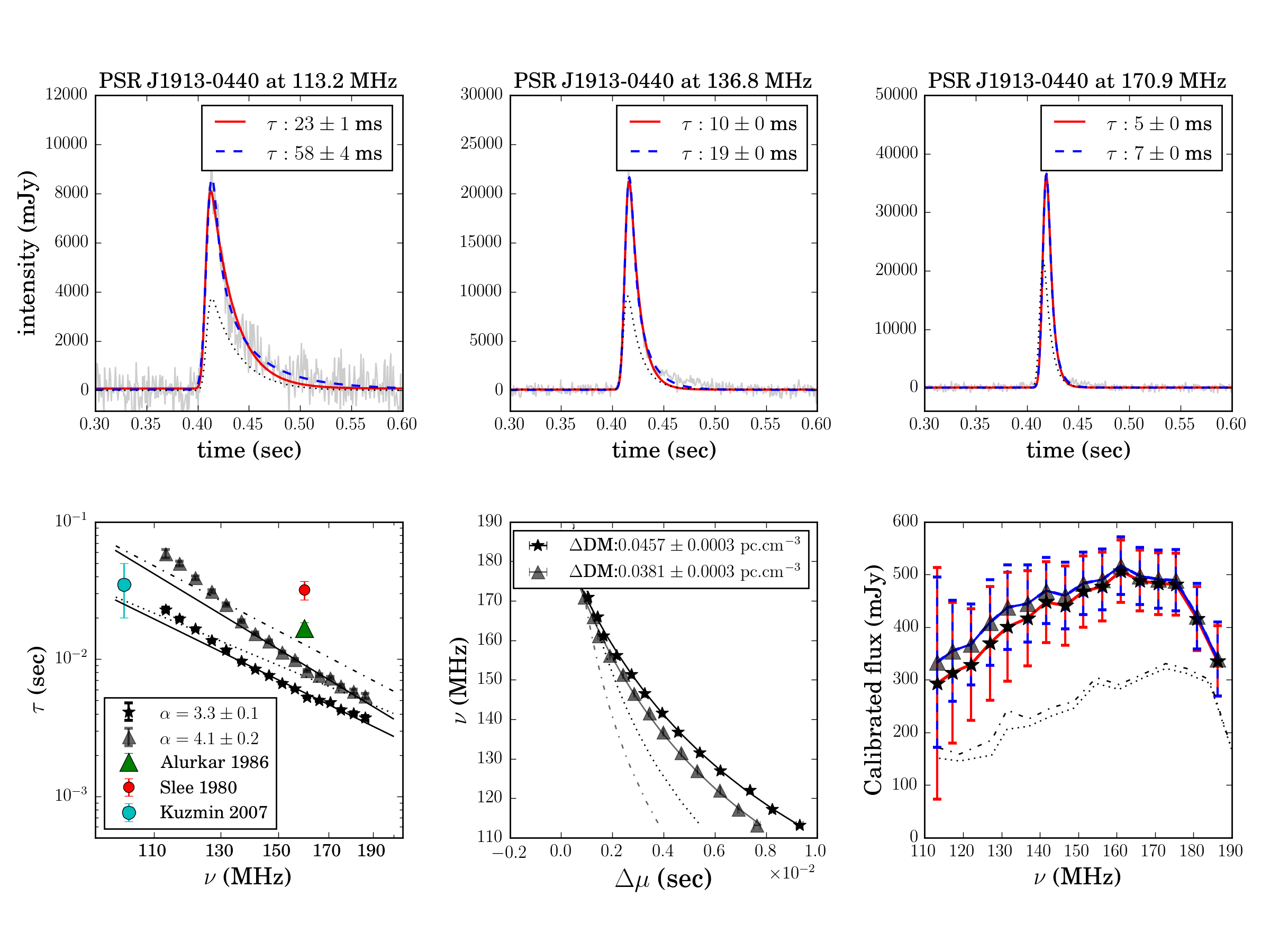}
\caption{\textit{Top:} scattered broadened profiles of PSR~J1913$-$0440 at three different frequencies in the HBA band (Cycle 5 data: IM fit in solid red and AM fit in blue dashed line. Commissioning data: IM fits at similar frequencies to Cycle 5 profiles in dotted lines). A secondary feature can be made out in the middle and right-hand panels. The profiles are enlarged to show 0.3 s of the 0.83 s pulse period. \textit{Bottom, left:} the $\tau$ and flux density spectra for PSR~J1913$-$0440 (IM, stars and AM triangles), along with data points from the literature. The Commissioning data fits are also shown as a dotted line (IM, $\alpha = 2.7 \pm 0.2$) and dash--dotted line (AM, $\alpha = 3.5 \pm 0.3$). \textit{Bottom, middle:} fits to the obtained excess DM values. Markers are as in the previous panel. \textit{Bottom, right:} the flux density spectra for IM (stars: uncorrected flux density and red solid line: corrected flux density) and AM (triangles: uncorrected flux density and blue (top) solid line: corrected flux density). The AM of the lowest two frequency channels is associated with large $\tau$ error bars (dashed), leading to flux density errors of several 100\%. Flux spectra for the Commissioning data are shown as dotted (IM) and dash--dotted (AM) lines.}
\label{fig:tau1913}
\end{figure*}

PSR~J1913$-$0440 features in both the Commissioning and the
Cycle 5 data set. From the latter, we obtain \mbox{$\alpha = 3.3 \pm 0.1$} (IM) and $\alpha = 4.1 \pm 0.2$ (AM). The Commissioning data lead to somewhat lower spectral indices with larger error bars,
namely $\alpha = 2.7 \pm 0.2$ (IM) and $\alpha = 3.5 \pm 0.3$ (AM).

In the following, we concentrate on the Cycle 5 data, which, with 16
frequency channels, have higher S/N (each profile has peak
S/N$ > 30$). In Fig. \ref{fig:tau1913}, we show the profile fits to
three frequency channels, along with the $\tau$ and flux density spectra and
the $\Delta$DM trends. The full set of profiles is shown in Appendix \ref{app:profiles}.

We note a small flat feature appearing at frequencies above 130~MHz
(see the top middle and right-hand panels of Fig. \ref{fig:tau1913}). At
low frequencies, the feature is fit together with the primary component
leading to an overestimation in $\tau$. This is visible as a deviation
in the fitted $\tau$ spectrum. Fitting only components above 130~MHz
(IM) would lead to a slightly lower spectral index of
$\alpha = 3.2 \pm 0.1$.

Previous scattering measurements for this pulsar exist at 102~MHz
\citep{Kuzmin2007} and 160~MHz (\citealt{Alurkar1986, Slee1980}; see
Table \ref{table:lit}). Using these values as well as their own fits to
EPN profiles at higher frequencies (up to 408~MHz), L15 published an
$\alpha$ value equal to $2.62 \pm 0.86$. We note (from e.g. the K15 data) that the profile is nearly unscattered at 327~MHz, such that we do not value the inclusion of ever higher frequency profiles by L15. The L15 value was changed to $\alpha = 4.18 \pm 0.44$ in L15b, after the inclusion of measurements with the GMRT at 150 and 235~MHz.

Using $\alpha  = 3.3 \pm 0.1$ we extrapolate to a
$\tau$ value of $25 \pm 2.2$~ms at 102~MHz, which lies within the
error bars of the \citet{Kuzmin2007} published value, $35 \pm 15$~ms.
Our $\tau$ value at 161.1~MHz is $5.3 \pm 0.1$~ms, much lower than the
values at 160~MHz of $16.7 \pm 1.8$and $32 \pm 5$~ms in
\citet{Alurkar1986} and \citet{Slee1980}, respectively. The difference
is likely due to their using lower S/N data, in which the secondary
component at these frequencies can not be isolated. K15 find
$\tau = 0.19 \pm 0.01$~ms at 327~MHz. Extrapolating our data would
lead to a value of around $\tau \sim 54$~ms. We note that K15 have
labelled this pulsar as a double-component pulsar, in which case they
fit only the secondary component, different from us. However, at
327~MHz, the pulsar is very weakly scattered.

Due to the presence of the low frequency feature, our fits provide
weak average goodness of fit statistics (Table \ref{table:app}, Appendix \ref{app:profiles}). The shape of the AM
profiles fit the secondary feature well, leading to residuals that are
much more Gaussian, especially at lower frequencies, than the IM. The
deviation from a simple power-law in the $\tau$ spectrum, seen in Fig.~\ref{fig:tau1913}, therefore also starts at higher frequencies, than for the AM. These better fits lead to an $\alpha$ value close to the
theoretically expected value of 4. The secondary feature disappears at
high frequencies, i.e. at low scattering. If in fact this low
frequency feature is due to scattering alone it would provide strong
support for an extremely anisotropic scattering mechanism along this
line of sight. However, we can not rule out that the feature is
intrinsic to the pulsar.

We see a clear $\Delta$DM trend in both the Commissioning
Cycle 5 data for the IM and AM fits (Fig. \ref{fig:tau1913}, lower middle panel). Introducing the
$\Delta$DM corrections lead to the improved DM values in Table
\ref{table:app2} in Appendix \ref{app:profiles}. The corrections show that the DM values associated with the two data sets become more equal after the corrections are applied, both for the IM and AM. In this respect, the AM performs more favourably.
 
The flux density associated with our scattering fits shows a clear turnover at
160~MHz (Fig. \ref{fig:tau1913}, lower right panel). This can not be understood through long scattering
tails, since the scattering values w.r.t the pulse period are not large
enough, and the corrected flux density differs negligibly from the uncorrected flux density. The flux density and spectral index values from Table \ref{table:one}
suggest a flux density of approx. 1590~mJy at 150~MHz. In comparison our measured values at 
151~MHz are $468 \pm 68$~mJy (IM) and $483 \pm 59$~mJy (AM), which becomes $468 \pm 234$~mJy (IM) and $483 \pm 242$~mJy (AM) with increased error margins. The Commissioning data for this pulsar result in even lower values of $246 \pm 123$~mJy (IM) and $259 \pm 130$ (AM)~mJy at 149~MHz (with increased error margins).

\subsubsection{PSR~J1917+1353} 

{\footnotesize{$\rm{P}= 0.19$ s, $\tau_{150} = 11$~ms, $\rm{DM}= 94.7$ pc cm$^{-3}$, $\delta t=0.2$~ms, $\delta \nu=12.2$ kHz}}\\
\vspace{3mm}

PSR~J1917+1353 was discovered by \citet{Swarup1971}. It is one of the most distant (5.00 kpc) pulsars in the set.  The EPN data base suggests this pulsar has a single component up to 1.4~GHz and the fits are therefore unlikely to be biased by the presence of secondary
components.

\begin{figure*}
\centering
\includegraphics[width=\textwidth]{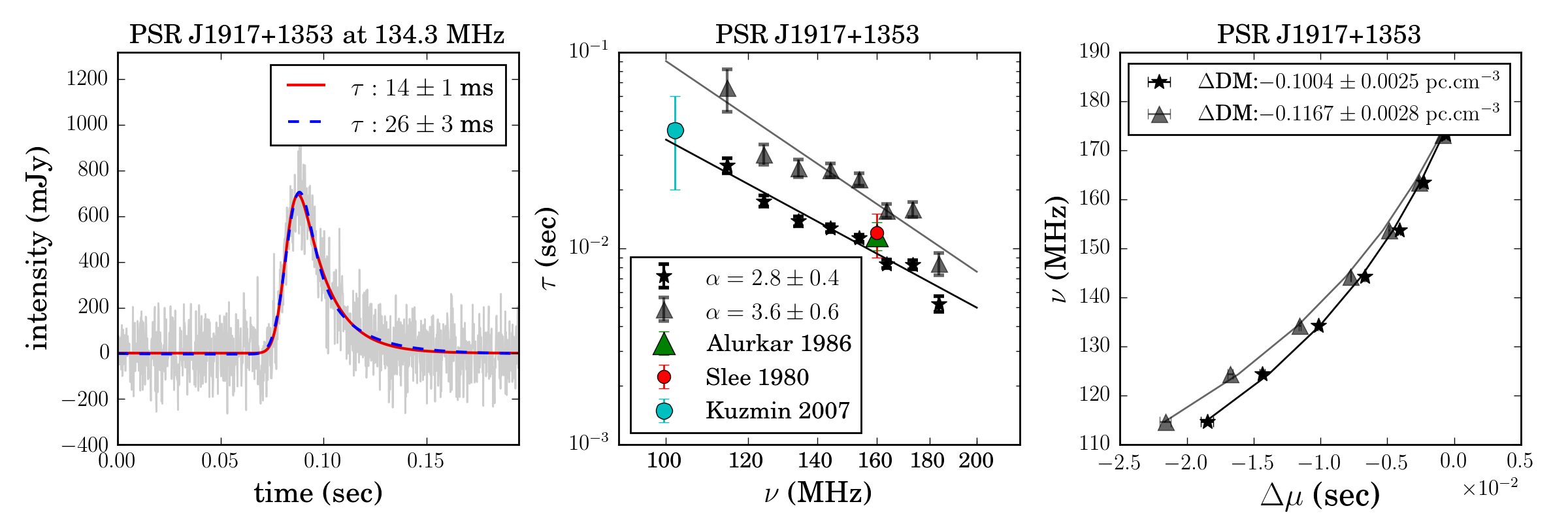}
\caption{\textit{Left:} the scattering fits to a profile shape of PSR~J1917+1352 (IM, red solid and AM, blue dashed). \textit{Middle:} the associated $\tau$ spectra along data points from the literature. \textit{Right:} the obtained $\Delta$DM fits for both models (as before: IM, stars and AM, triangles).}
\label{fig:tau1917}
\end{figure*}

We channelize the Commissioning data for this pulsar into eight frequency
channels, which provides peak S/N values between 5.2 and 12.6. The
$\alpha$ values obtained are $2.8 \pm 0.4$ (IM) and
$3.6 \pm 0.6$ (AM, see Fig.~\ref{fig:tau1917}, middle panel), the latter of which lies close to theoretical
predictions. Scattering values from the literature exist at 102, 160
 and 327~MHz (see Table \ref{table:lit}).

The K15 data point again underestimates $\tau$ according to our own
measurements, whereas the data point from \cite{Kuzmin2007} seems in
good agreement. 

The expected flux density at 150~MHz using the input of Table \ref{table:one} and a simple power-law, is approx. 288~mJy. Our fits suggest flux density values that are similar for both models.  At 154~MHz, we find $80 \pm 11$ ~mJy (IM) and $86 \pm 9$~mJy (AM), or $80 \pm 40$~mJy (IM) and $86 \pm 43$~mJy (AM) using 50\% error margins. Over the HBA band, the flux density changes monotonically in the range 48--126~mJy. At the lowest
frequency channel the profile shows scatter broadening that stretches
across the pulse period. This can be seen most clearly for the AM (Appendix \ref{app:flux}, shaded region). The flux density spectrum shows no turnover at low frequencies.

\newpage
\subsubsection{PSR~J1922+2110} 

{\footnotesize{$\rm{P} = 1.08$ s, $\tau_{150} = 42$~ms, $\rm{DM} = 217.0$ pc cm$^{-3}$, $\delta t=1.1$~ms, $\delta \nu=12.2$ kHz}}\\
\vspace{3mm}

\begin{figure*}
\centering
\includegraphics[width=\textwidth]{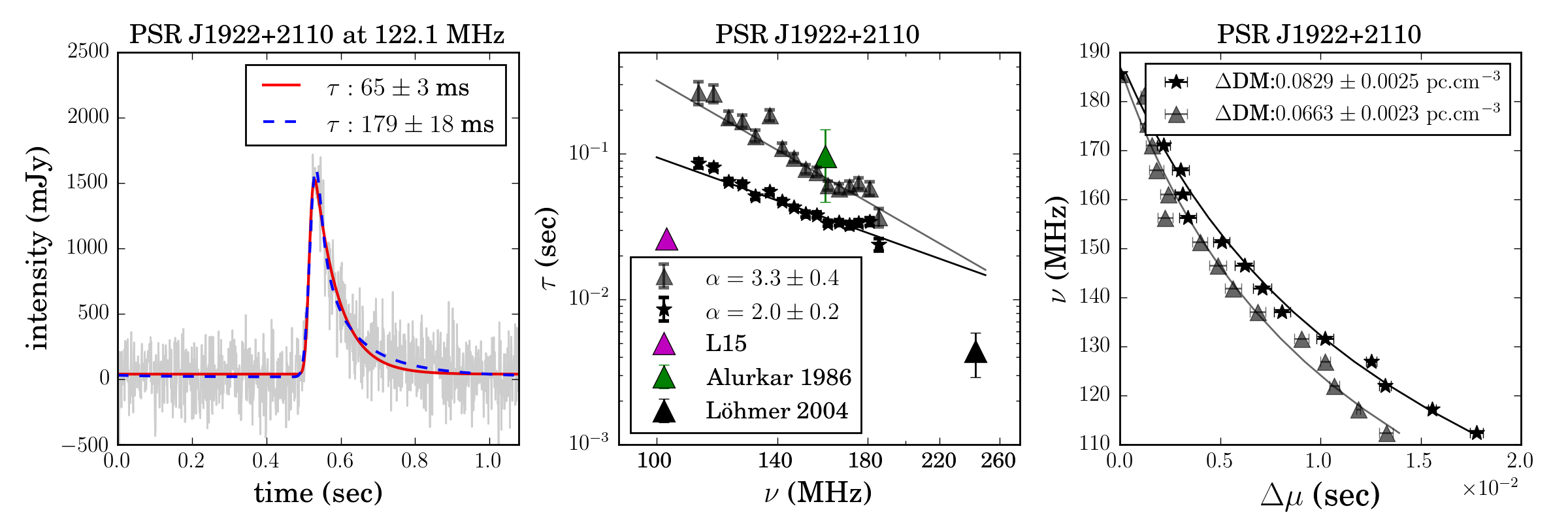}
\caption{Similar to Fig. 10, for PSR~J1922+2110.}
\label{fig:tau1922}
\end{figure*}

PSR~J1922+2110 has the highest DM value in our data set, and is seen
to be an outlier in the DM versus distance plot of Fig.~\ref{fig:DistDM}.
It was discovered in the same low-latitude pulsar survey as PSR J1909+1102
\citep{Davies1973}. We analyse Commissioning data for this pulsar. The
high S/N allows us to channelize the band into 16 average profiles.
Fitting the IM leads to $\alpha = 2.0 \pm 0.2$, and
$\alpha = 3.3 \pm 0.4$ for the AM (Fig.
\ref{fig:tau1922}). 

Several scattering measurements are available for this pulsar (Table \ref{table:lit}).
The literature values do not agree well with our estimations. L15 obtained $\alpha =  0.94 \pm 0.88$ for this pulsar, but omitted the value from their subsequent analyses for
being `suspiciously low'. They measured a $\tau$ value at 102.75~MHz
of roughly 25~ms, more than three times lower than our data point of
$86.1 \pm 6$~ms at 112.5~MHz. Their value comes from fitting an EPN
data base profile at 102.75~MHz \citep{Kassim1999}, for which the phase resolution is poor.
L15 further claim that PSR~J1922+2110 has an asymmetric profile, such
that scattering fits to this pulsar could be fitting profile
evolution as well. We note that above 410~MHz, the pulsar has a secondary
component. \citet{Lohmer2004} published a $\tau$ value of $4.5 \pm 1.5$~ms at 243~
MHz. This was the only frequency at which the authors present an obtained
$\tau$ value with a standard error. For the higher frequencies, they obtained upper limits on $\tau$ only.

We conclude that the LOFAR data set is likely the best data for
measuring the scattering parameters, since it provides high peak S/N
values (between 6.8 and 16.4) and good phase resolution (1024 bins per
pulse period).

As seen from the profile in the left-hand panel of Fig.
\ref{fig:tau1922}, the IM and AM produce very similar fits to the
data. This is true for all the frequency channels. The main difference
can be seen in the fit of the scattering tail, where the AM follows a
flatter trend. Clear and well-fitted $\Delta$DM trends are shown in the right-hand panel
of Fig. \ref{fig:tau1922}.

The flux density and a power-law
spectral index values of Table \ref{table:one}, leads to an expected flux density of 316~mJy at 150~MHz. The flux density values estimated from our scattered profile fits at 151.4
~MHz are again much lower: $89 \pm 35$~mJy (IM) and $96 \pm 30$~mJy (AM), or $89 \pm 45$~mJy (IM) and $96 \pm 48$~mJy (AM) with increased error margins. The spectra have relatively simple structure, decreasing with frequency. In the AM spectrum, small contributions due to the corrected flux density calculation can be seen, concurrent with the onset of a wrap around profile shape (Appendices \ref{app:profiles} and \ref{app:flux}).

\newpage
\subsubsection{PSR~J1935+1616} \label{sec:J1935}

{\footnotesize{$\rm{P}= 0.36$ s, $\tau_{150} = 20$~ms, $\rm{DM} = 158.6$ pc cm$^{-3}$, $\delta t=0.4$~ms, $\delta \nu=12.2$ kHz}}\\
\vspace{3mm}

\begin{figure*}
\centering
\includegraphics[width=\textwidth]{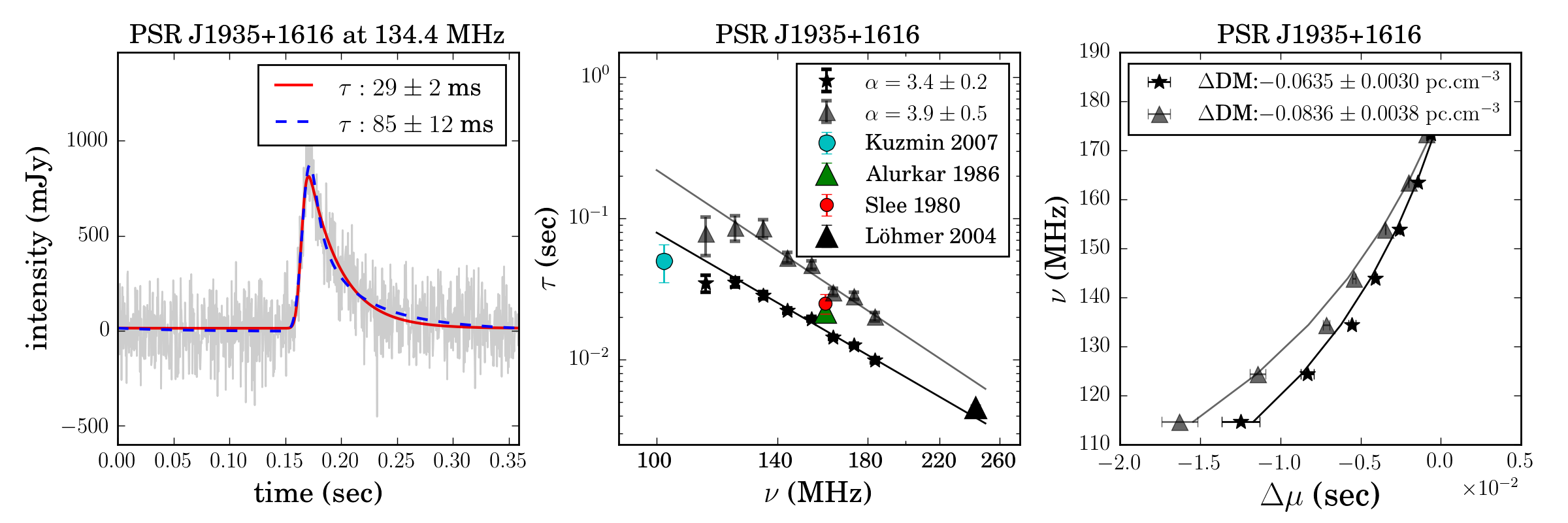}
\caption{Similar to Fig. 10, for PSR~J1935+1616.}
\label{fig:tau1935}
\end{figure*}

This pulsar has one of the highest flux density values in the set, having been
discovered in a single pulse search using the Lovell Telescope
at Jodrell Bank \citep{Davies1970}. In Fig. \ref{fig:DistDM} we see that this
pulsar, at a distance of 3.7~kpc, has a higher DM than other pulsars at similar distances. 

The Commissioning data for this pulsar are split into eight frequency
channels, leading to peak S/N values between 3.6 and 21.2. We obtain
an $\alpha$ value of $3.4 \pm 0.2$ using the IM. This value is in
good agreement with the values obtained by \citet{Lohmer2004}, $\alpha = 3.4 \pm 0.2$,
and L15b, $\alpha = 3.35^{+0.36}_{-0.41}$. Both these literature values of $\alpha$ are determined using other published scattering measurements between 110 and 250~MHz \citep{Rickett1977,Slee1980,Alurkar1986,Lohmer2001,Kuzmin2007}, and therefore serve as an independent check of our measurements. Our AM leads to $\alpha = 3.9 \pm 0.5$, in agreement with theoretical values of 4 or 4.4. The fits for both models look alike, with the AM reaching slightly higher into the
peaks of the pulse profile and having a different slope in
the scattering tail. 

Published $\tau$ measurements from the literature include $50 \pm 15$~ms at 111~MHz \citep{Kuzmin2007} and two sets of measurements at 160~MHz of $21.7 \pm 1.6$~ms \citep{Alurkar1986} and $25 \pm 4$~ms
\citep{Slee1980}, as shown in Fig. \ref{fig:tau1935} and Table
\ref{table:lit}. At 327~MHz, K15 published a $\tau$ value of
$3.21 \pm 0.02$~ms. Our refitting of the data (which also fits for the
underlying width of the intrinsic pulse) leads to a value of
$1.8 \pm 0.1$~ms, whereas keeping the value of $\sigma$ fixed (as
obtained from a 600~MHz template taken from the EPN data base), we find
a value of 3.3~ms, in closer agreement to theirs. We note that the K15
profiles at 327~MHz that overlap with our set of pulsars, are
generally not very scattered, and that a fixed width method often
leads to a fit that does not model the peak of the scattered profile
well. Extrapolating to 327~MHz using our obtained isotropic spectral
index value, leads to a $\tau$ of $1.42 \pm 0.18$~ms instead, in close
agreement to our fit of their data at 327~MHz. 

The flux density measurements obtained from our scattering fits (IM and AM) show a
turnover at around 170~MHz, i.e. towards the highest observed frequency (Appendix \ref{app:flux}). This turnover is not purely associated with a wrap around scattering tail, as scattering corrections to the flux density only become significant below 145~MHz. The calculated flux density values at 154.1~MHz are $95 \pm 27$~mJy (IM) and $105 \pm 22$~mJy (AM). Again, these errors are increased to 50\% to reflect the initial flux calibration uncertainties. However, using a simple power-law and the flux parameters of Table \ref{table:one} implies a flux density of approx. 955~mJy at 150~MHz, roughly 10 times larger than our values. 

Several authors have noted that the line of sight to PSR~J1935+1616 is
puzzling, e.g \citet{Lohmer2004}. In their study of nine pulsars with
intermediate DM values (150--400 pc cm$^{-3}$), only PSR~J1935+1616
had an $\alpha$ value inconsistent with the expected Kolmogorov
prediction. The authors suggested the low $\alpha$ value could be due
to multiple scattering screens of finite size or varying scattering
strengths. Our data certainly point towards anomalous scattering, such
as the discussed AM. Alternatively, a truncated scattering screen
could provide a basis to interpret the data, as it would lead to both
a low $\alpha$ value and a decrease in flux towards low frequencies. A
turnover in the flux spectrum (unrelated to a long scattering tail) is
seen for PSR~J1935+1616. However, as shown in \citet{Geyer2016}, such a
scenario would involve an observable change in the pulse profile
shapes at low frequencies, which is not evident in the data. Soft
edges of the screen, multiple screens and low S/N profiles can render
these particular effects difficult to discern. The questions,
therefore, surrounding the nature of the scatterer(s) towards this
pulsar remain open.

\subsubsection{PSR~J2257+5909} 

{\footnotesize{$\rm{P}= 0.37$ s, $\tau_{150} = 31$~ms, $\rm{DM}= 151.1$ pc cm$^{-3}$, $\delta t=0.4$~ms, $\delta \nu=12.2$ kHz}}\\
\vspace{3mm}

\begin{figure*}
\centering
\includegraphics[width=\textwidth]{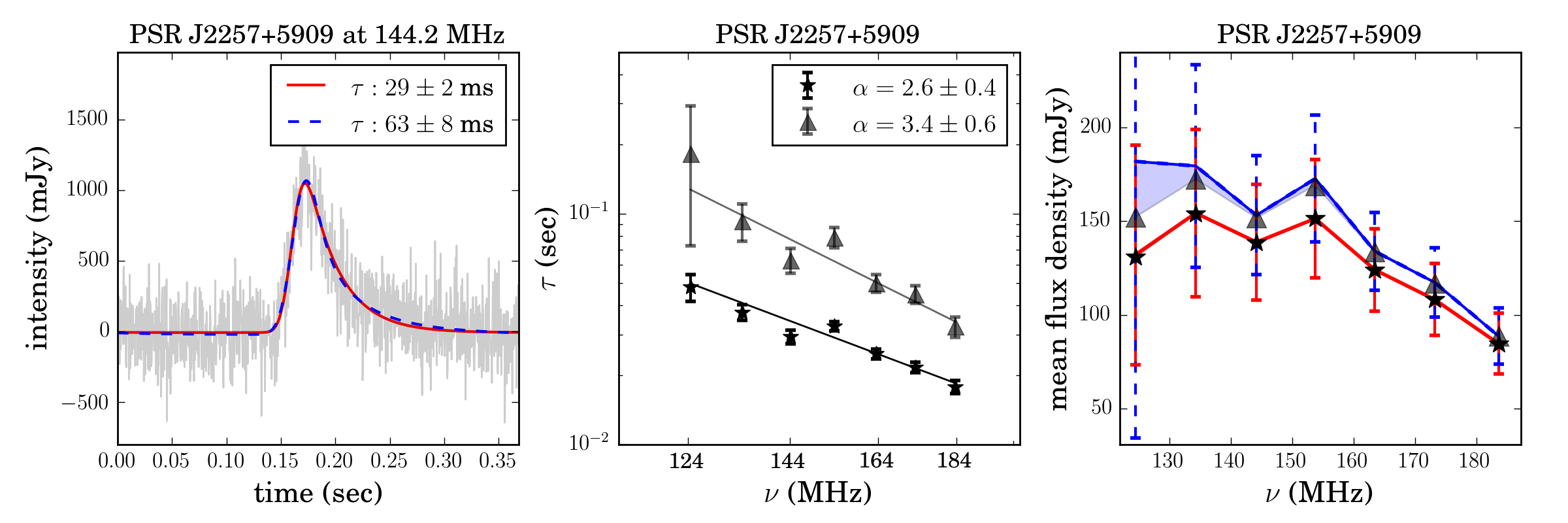}
\caption{Similar to Fig. 8, for PSR~J2257+5909.}
\label{fig:tau2257}
\end{figure*}

PSR~J2257+5909 is a bright pulsar discovered in the same survey as
B0540+23 and B0611+22 \citep{Davies1972}. Similar to PSR~J1935+1616,
it is an outlier on the DM versus distance plot (see Fig.
\ref{fig:DistDM}). We could not find previously published scattering times for this pulsar.
We split the Commissioning data for this pulsar into eight frequency
channels, and excluded the lowest frequency channel for which the peak
S/N is only 2.34. Fitting the remaining seven scattered
profiles, we obtain $\alpha = 2.6 \pm 0.4$ (IM) and
$\alpha = 3.4 \pm 0.6$ (AM), which approaches the theoretical value of
$\alpha = 4$.

The profile fits obtained by these models are very similar (Fig. \ref{fig:tau2257}, left-hand panel).
The $\chi_{\rm{red}}^2$ values are near equal, with the AM value consistently smaller than the
IM value (Table \ref{table:app}, Appendix \ref{app:profiles}). The low $p$-values from the KS test are dominated by two frequency
channels. Omitting these two (out of seven) channels, increases the $p$-values to 85.6 \% (IM) and 93.8\% (AM). The $\Delta$DM trends are less well fit for this pulsar and both models show negative $\Delta$DM fits.

Our flux density spectra, as calculated from the both models, shows the onset of turning over towards lower frequencies. In the case of the AM, the corrected flux density spectrum is straightened out w.r.t the uncorrected flux density spectrum (shaded region, right-hand panel of Fig. \ref{fig:tau2257}). The associated profile fits of especially the AM show related pulse wrap around at frequencies below 150~MHz (see Appendix \ref{app:profiles}). The calculated flux density values are $151 \pm 32$~mJy (IM) and $173 \pm 34$~mJy (AM) at 153.7~MHz, or $151 \pm 76$~mJy (IM) and $173 \pm 87$~mJy (AM) with 50\% error margins. The Table~\ref{table:one} flux parameters, implies a much larger flux density value of approx. 500~mJy at 150~MHz, using a simple power-law.

\subsubsection{PSR~J2305+3100} 

{\footnotesize{$\rm{P}= 1.58$ s, $\tau_{150} = 9$~ms, $\rm{DM}= 49.6$ pc cm$^{-3}$, $\delta t=1.5$~ms, $\delta \nu=3.1$ kHz}}\\
\vspace{3mm}

\begin{figure*}
\centering
\includegraphics[width=\textwidth]{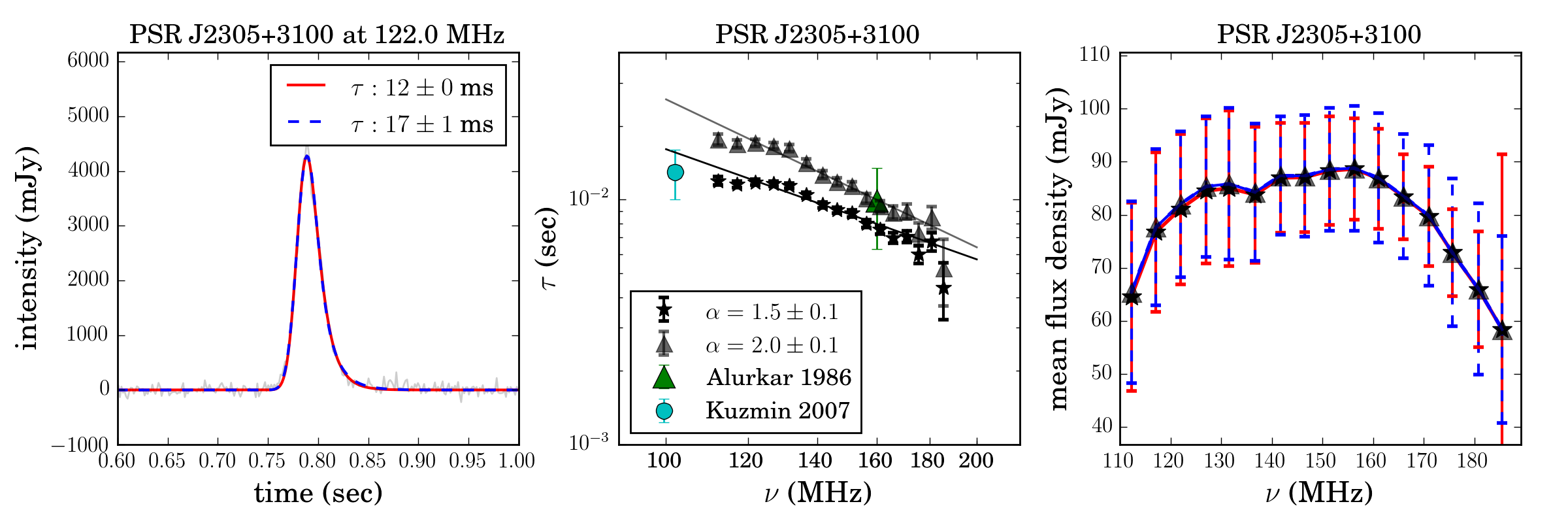}
\caption{Similar to Fig. 8, for PSR~J2305+3100.}
\label{fig:tau2303}
\end{figure*}

This pulsar was discovered using Arecibo in 1969 \citep{Lang1969}. It has
one of the lowest DM values in the set, but in contrast its distance is estimated to have a lower limit of 25~kpc \citep{Yao2017}. It is the pulsar with the longest pulse period (1.58 s) in
the set. We analyse Census data for this pulsar, split into 16
channels. All across the HBA band, the pulse remains
only slightly scattered. We find $\alpha = 1.5 \pm 0.1$ (IM)
and $2.0 \pm 0.1$ (AM). An example pulse along with the $\tau$
and flux density spectra are shown in Fig.~\ref{fig:tau2303}. The pulse shape
is enlarged, showing the pulse phase from 0.6
to 1.0 s.

An $\alpha$ value of $3.42 \pm 0.26$ was recently published by
L15. They tabulate their own measurements using
the GMRT (at 410~MHz to 1.4~GHz), as well as the results from
\citet{Kuzmin2007} (at 44, 63 and 111~MHz) and \citet{Alurkar1986} at
160~MHz (see Table \ref{table:lit} for literature values). The $\tau$ values
published by \cite{Kuzmin2007} are $\tau_{44} = 300 \pm 100$~ms,
$\tau_{63} = 110 \pm 20$~ms and $\tau_{111} = 13 \pm 3$~ms. A fit
across these three values leads to a spectral index of
$\alpha = 3.8 \pm 0.1$. \citet{Alurkar1986} published a value of
$\tau_{160} = 9.9 \pm 3.6$~ms. Our frequency channel closest to an observing frequency of 
\cite{Kuzmin2007} is 112.3~MHz. For this channel, we find
$\tau_{112.3} = 11.9 \pm 0.5$~ms in good agreement with their value at
111~MHz.  Extrapolating our $\tau$ values to 44~MHz, leads to a $\tau$ value of
around 55~ms (using the IM), much lower than calculated by \citet{Kuzmin2007}.

The $p$-values for both models are low (Table \ref{table:app}, Appendix \ref{app:profiles}), but significantly increased, to 79.6\% (IM) and 80.8\% (AM), when only the first five channels with more notable scattering, are considered.

The near identical scattering fits of the two models and the low levels of scattering, lead to near identical flux density spectra. We compute $88 \pm 10$~mJy (IM) and $89 \pm 12$~mJy (AM) at 151.3~MHz. These values are amended to have error bars 44 and 45~mJy, respectively, in line with the original flux calibration uncertainties. The mean published Census flux density value is $70 \pm 35$~mJy
\citep{Bilous2016}, which agrees well with our result. We see a turnover in the flux density spectra at around 150~MHz, however, due to the low $\tau$ values compared to the
pulse period, this is unlikely due to scattering effects.

\section{Discussion}\label{sec:disc}

We are now in a position to summarize our results and return to the questions posed in the Introduction, addressing each of them.  
In order, we first assess the profile fits and address whether AMs are required to fit this LOFAR data set. Thereafter, we consider the distributions of spectral indices obtained by the two models.  Next, we investigate what information is gained regarding the profile evolution and DM from the profile fits. This includes analysing the width evolution of the intrinsic profiles.  In Section \ref{sec:multi}, we briefly discuss the impact of finite scattering screens, and how this relates to our measurements.
Lastly, we return to the well-studied $\tau$ versus DM relation and consider how our data correspond to published trends.

\subsection{$\tau$ measurements using two models}

\begin{figure}
\centering
\includegraphics[width=0.9\columnwidth]{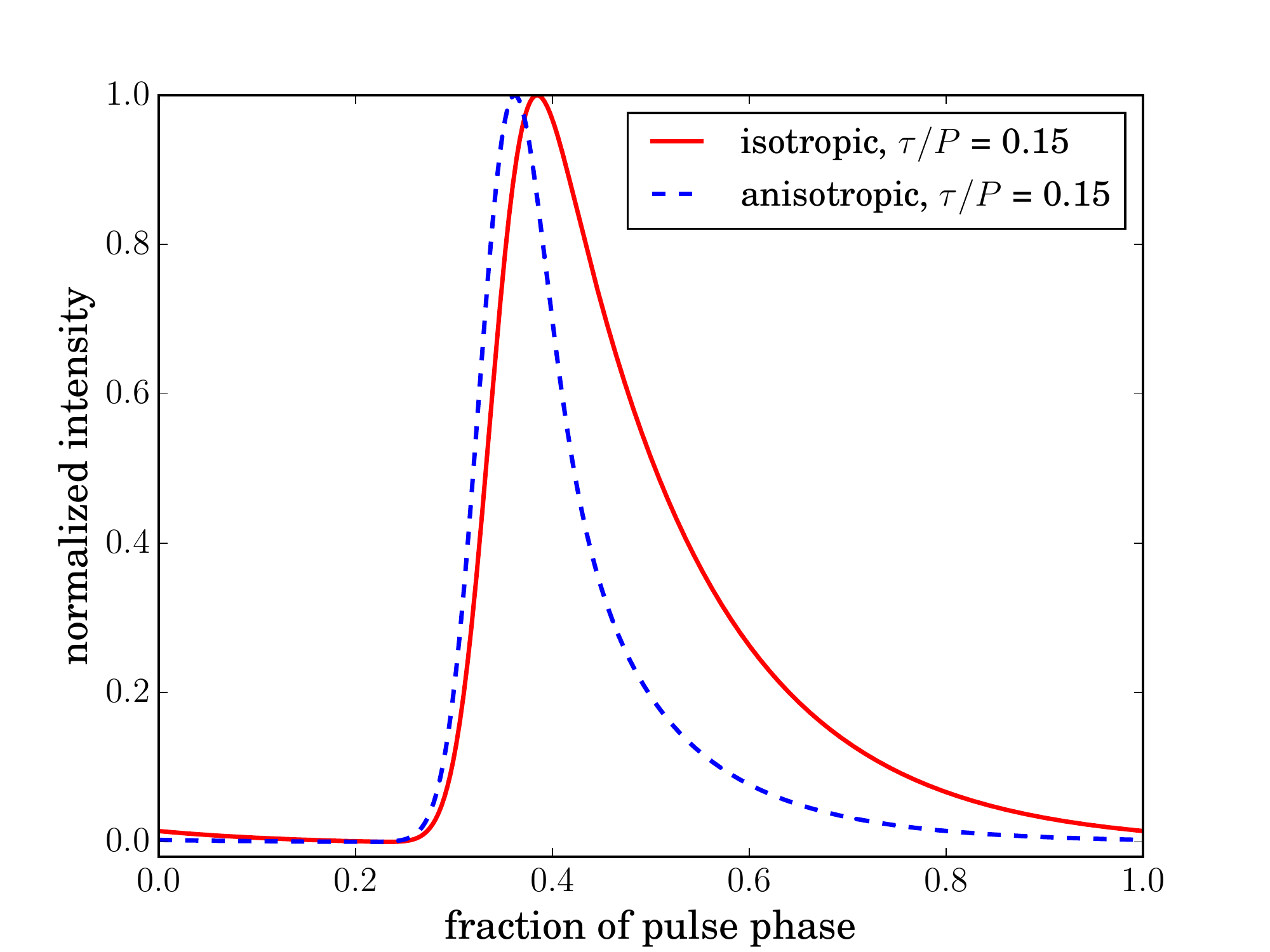}
\caption{A comparison of modelled isotropic and anisotropic pulse shapes with equal characteristic scattering times. The profiles are plotted to have a peak value equal to unity. All other fitting parameters ($\sigma$, $\mu$, A and DC) are equal for both profiles as well.}
\label{fig:shapes}
\end{figure}

\begin{figure}
\centering
\includegraphics[width=\columnwidth]{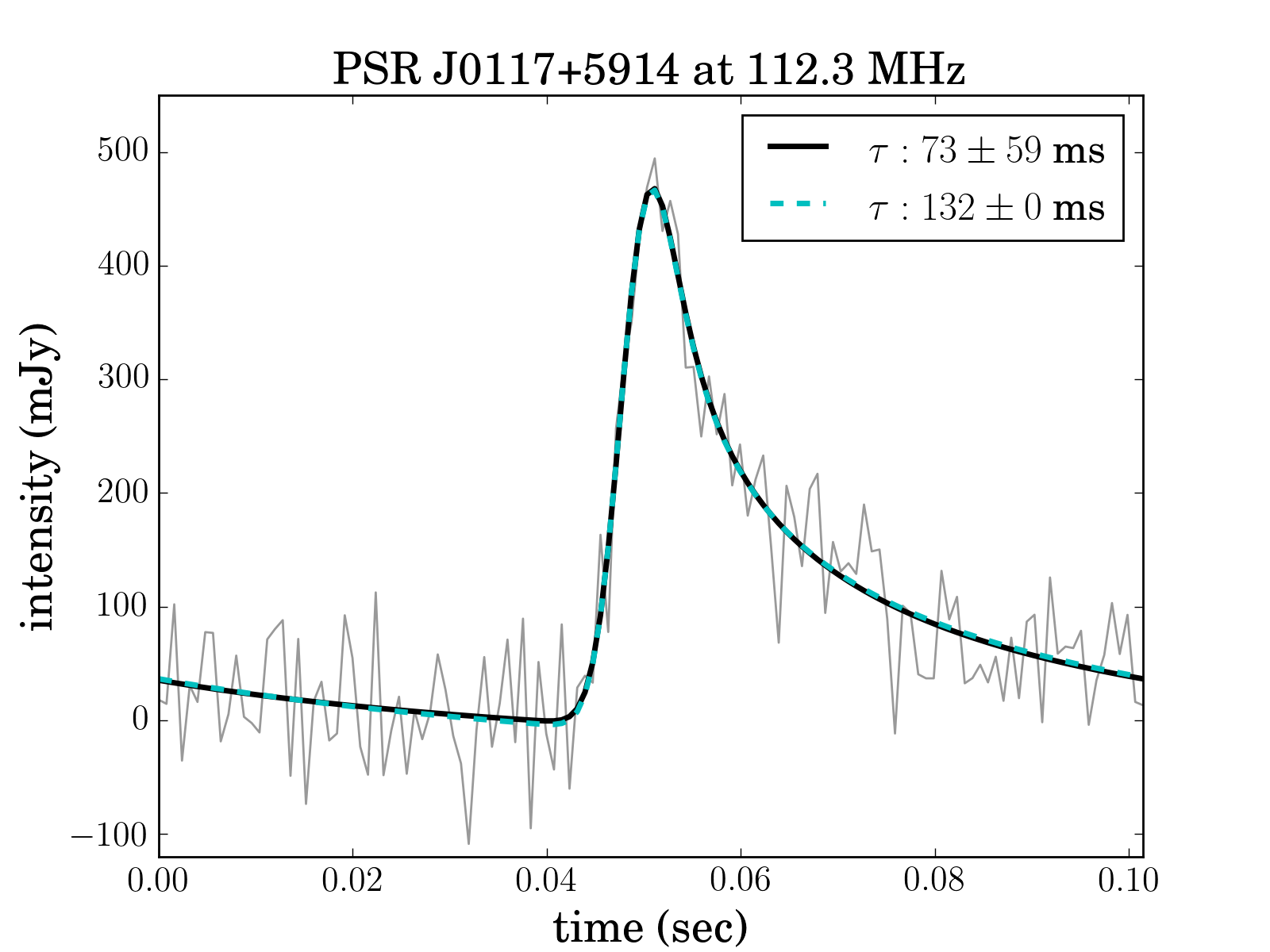}
\caption{PSR J0117+5914 Census data at the lowest frequency channel of 112.3~MHz. Resulting profile shapes for anisotropic model fits: (1) obtained by calculating the best fit $\tau$ value (black, solid line), (2) obtained by fixing the $\tau$ value to the upper value of the error margin on the best fit $\tau$ (0.1322 s), and fitting for the rest of the parameters (cyan, dashed). These two model fits are indistinguishable.}
\label{fig:anisotau}
\end{figure}

To consider whether the data necessitate AMs, we return to our fits of the scatter broadened data, and thereafter consider the impact on secondary parameters, such as the scattering spectral index $\alpha$.

We have used $\chi^2_{\rm{red}}$ values and the KS test to investigate the goodness of fits of the scattering models (Table \ref{table:app}, Appendix \ref{app:profiles}). Judging from the outcomes of these tests, we do not find any pulsar for which the AM is a requirement. All data are well fitted with the IM. In one case (PSR~J0742$-$2822), the anisotropic result is unreliable, as discussed in Section \ref{sec:J0742}, producing a spectral index of $\alpha \approx 8$.

We find, as expected, that the fitted $\tau$ value for a given profile is larger when applying the AM compared to the IM.  Fig.~\ref{fig:shapes} compares the profile shapes modelled with the same $\tau$ value. The shapes agree with the intuition that an isotropic scattering medium will scatter more pulsar radiation into the observer's line of sight, than an equivalent extremely anisotropic scatterer.  

We also note that the errors in $\tau$ are larger for the extreme AM than for the IM, especially at low frequencies. Fig.~\ref{fig:anisotau} shows that the best fit profile shape produced by the AM fitting code, (black, solid line), is indistinguishable from the shape produced when the $\tau$ value is fixed at the best fit value plus a 1$\sigma$ error, and the remaining parameters are refit (cyan, dashed line). The large error bars in $\tau$ stem from the anisotropically scattered shape being closer in shape to the unscattered Gaussian, which makes it more insensitive to changes in $\tau$ at low frequencies.  Larger errors in $\tau$ lead to larger errors in the spectral indices, as indicated in Table \ref{table:two}.

We note that for four of the pulsars there is some evidence for extreme anisotropy, although it is far from conclusive. In the case of PSRs J0117+5914 and J0614+2229, we find that the $\chi_{\rm{red}}^2$ value at all frequencies is closer to 1 for the AM and that similarly the $p$-values from the KS test are higher for the AM. The differences in these values are larger than for most other pulsars in the set.  We also find that the $\alpha$ values associated with the two models are well separated for both these pulsars. The other two possible candidates for anisotropy are PSRs J1913$-$0440 and J2257+5909. In the case of PSR~J1913$-$0440, we see a secondary feature that is well fit by the AM, leading to $\chi_{\rm{red}}^2$  values closer to 1, and higher p-values in the KS test. The $\Delta$DM correction also minimizes the difference in DM values of the two data sets for this pulsar (Table \ref{table:app2}, Appendix \ref{app:profiles}). Again, the $\alpha$ values obtained by the different models seem to be well separated, with the anisotropic case in agreement with theoretical values. PSR~J2257+5909 shows the widest separation in obtained $p$-values (85.6\% versus 93.8\%). The $\alpha$ values for the two models are also well separated and the anisotropic value is in agreement with theoretical values.  

More detailed tests of anisotropy in the temporal domain would require much higher S/N data, to be able to tell the model profile shapes apart. We have only made use of extreme (1D) anisotropic and isotropic models. It would be even more difficult to distinguish between models if lower degrees of anisotropy were included in the scattering models. 

Combining temporal analysis with secondary (power) spectra analysis (e.g. \citealt{Stinebring2001}) and the imagining of pulsars where possible, will increase the efficiency of tests for anisotropy. The very-long-baseline interferometry (VLBI) constructed image of PSR B0834+06 \citep{Brisken2010} already provides concrete evidence for highly anisotropic scattering surfaces in the ISM.

\subsection{Scattering spectral index distribution}\label{sec:spindx}

\begin{table*}
\centering
\begin{tabular}{p{2.4cm} p{1.3cm} p{1.3cm} p{2.2cm} p{1.3cm} p{1.3cm} p{2.2cm}}
\hline
Pulsar &  \multicolumn{3}{>{\centering\arraybackslash}l}{Isotropic Scattering } &  \multicolumn{3}{>{\centering\arraybackslash}l}{Extreme (1D) Anisotropic Scattering} \\
\hline
           &  $\tau_{150}$ (ms) & $\alpha$  & $\Delta$DM (pc cm$^{-3}$) & $\tau_{150}$ (ms) & $\alpha$  & $\Delta$DM (pc cm$^{-3}$) \\ [0.5ex] 
\hline
\hline

J0040+5716  &$ 40 \pm 2 $&   $2.2    \pm   0.2$   & $0.0378 \pm 0.0024$  &$86 \pm 8$&   $2.7 \pm  0.3$ & $0.0143 \pm 0.0022$ \\
J0117+5914 (Co)  & $7 \pm 0$&   $2.2 \pm  0.1$    & $0.0082 \pm 0.0009$ &$14 \pm 1$&  $3.5 \pm  0.4$ &$0.0041 \pm 0.0011$ \\
J0117+5914  (Ce) & $8 \pm 1$&   $1.9   \pm  0.2$ & $0.0064 \pm 0.0006$    &$16 \pm 2$&   $2.6 \pm 0.2$ &$0.0038 \pm 0.0006$ \\
J0543+2329  &$10 \pm 1$&   $2.6  \pm   0.2$    &   $0.0155 \pm 0.0020$ &$17 \pm 2$& $2.7  \pm  0.3$ &$0.0031 \pm 0.0020$ \\
J0614+2229  (Co) &  $15 \pm 1$ &$1.9   \pm   0.1$    & $0.0030 \pm 0.0007$& $44 \pm 4$  & $2.4  \pm   0.3$ & $-0.0033 \pm 0.0006$ \\
J0614+2229  (Cy) &$15 \pm 0$&   $2.1  \pm  0.1$  &  $-0.0053 \pm 0.0006$ &$44 \pm 3$& $3.1  \pm  0.3$ & $-0.0109 \pm 0.0008$\\
J0742$-$2822  &$20 \pm 2$&   $3.8  \pm   0.4$    &  $0.0013 \pm 0.0027$ &&  ... & ... \\
J1851+1259  &$6 \pm 1$&   $4.0  \pm   0.4$    & $0.0264 \pm 0.0022$ &$10 \pm 1$&  $4.7  \pm  0.4$& $0.0158 \pm 0.0017$\\
J1909+1102  &$42 \pm 3$&   $3.5  \pm   0.4$ & $0.0351 \pm 0.0085$   &$120 \pm 27$&   $6.4  \pm  0.7$ & $-0.0276 \pm 0.0077$ \\
J1913$-$0440  (Co) &$9 \pm 0$&   $2.7  \pm   0.2$    &   $0.0240 \pm 0.0009$ &$16 \pm 1$& $3.5 \pm  0.3$ & $0.0161 \pm 0.0011$\\
J1913$-$0440 (Cy) &$7 \pm 0$&   $3.3 \pm  0.1$    &  $0.0457 \pm 0.0003$ &$12 \pm 0$& $4.1 \pm  0.2$ & $0.0381 \pm 0.0003$\\
J1917+1353  &$11 \pm 1$&   $2.8   \pm   0.4$    & $-0.1004 \pm 0.0025$ &$21 \pm 2$&  $3.6  \pm   0.6$ & $-0.1167 \pm 0.0028$ \\
J1922+2110  &$42 \pm 2$&   $2.0    \pm   0.2$    &  $0.0829 \pm 0.0025$ &$85 \pm 6$& $3.3   \pm   0.4$ & $0.0663 \pm 0.0023$\\
J1935+1616  &$20 \pm 1$&   $3.4    \pm   0.2$    & $-0.0635 \pm 0.0030$ &$46 \pm 4$&   $3.9   \pm   0.5$ & $-0.0836 \pm 0.0038$\\
J2257+5909  &$31 \pm 2$&   $2.6    \pm   0.4$    & $-0.0317 \pm 0.0058$ &$68 \pm 9$&  $3.4   \pm   0.6$ & $-0.0530 \pm 0.0050$\\
J2305+3100  &$9 \pm 0$&   $1.5    \pm   0.1$  &$0.0184 \pm 0.0035$  &$11 \pm 0$&  $2.0  \pm  0.1$ & $0.0144 \pm 0.0023$\\

\hline
\hline
$\langle\alpha\rangle$       &&   $2.7  \pm   0.2$    &  && $3.5  \pm   0.4$ &\\[1ex]
\hline
\end{tabular}
\caption{List of obtained $\tau$ values, spectral indices and $\Delta$DM values, using two models.}
\label{table:two}
\end{table*}

\begin{figure}
\centering
\includegraphics[width=\columnwidth]{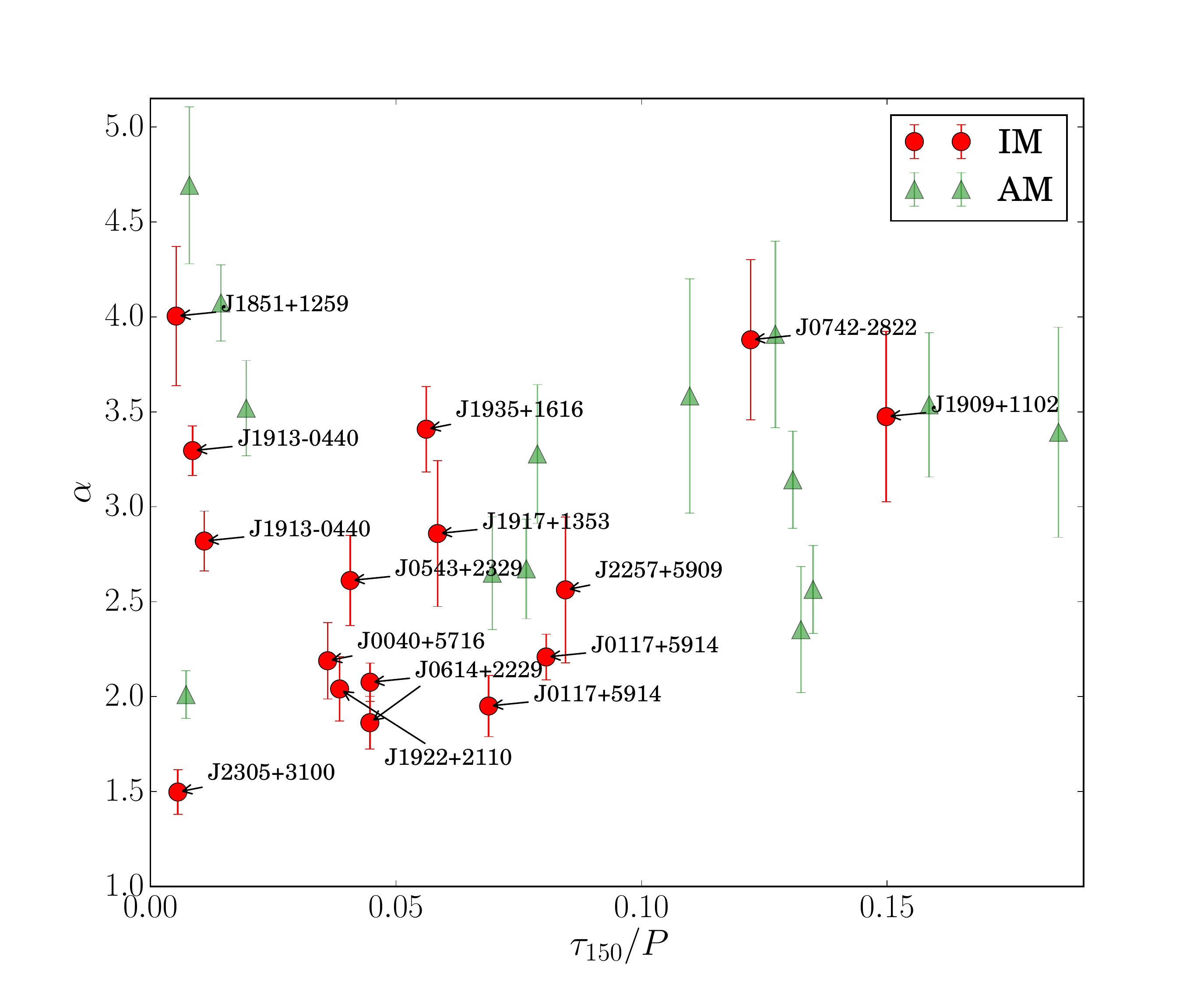}
\caption{Spectral indices $\alpha$ plotted versus $\tau$ values at 150~MHz in units of the pulse period . These $\tau$ values are obtained from the best fit power-law to the $\tau$ spectrum of each pulsar. Isotropic values are shown in red (circles), and anisotropic values in green (triangles). Isotropic data points are labelled by the corresponding pulsar names. The larger the $\tau_{150}/P$ values, the more the scattering tail of the average pulse profile will wrap around the full rotational phase.}
\label{fig:alphatauP}
\end{figure}

In Table \ref{table:two}, we summarize the obtained spectral indices for each pulsar for a given scattering model.  The average spectral index, using the IM, is $2.7 \pm 0.2$. This is much lower than the theoretically predicted values of $\alpha = 4$ or 4.4.

Using the IM, the vast majority of pulsars (11 out of 13) have $\alpha$ values smaller than 3.8. Not a single pulsar is measured to have an $\alpha$ value larger than 4.0. Only three pulsars have spectral indices in close agreement with the theoretically predicted values (viz. PSRs J0742$-$2822, J1851+12 and J1909+1102).

Similar low frequency spectral indices have been reported in L15 and \citet[hereafter L13]{Lewandowski2013}, although many of the values were dropped from the subsequent analysis in the papers for being \textit{suspiciously low}. Four of the pulsars that were excluded in L15 for which we also have measurements, include: PSRs  J0614+2229 (L15: $\alpha = 1.73 \pm 0.5$, our $\alpha = 2.1 \pm 0.1$), J0742$-$2822 (L15: $2.5 \pm 0.3$, our value: $3.9 \pm 0.4$), J1917+1353 (L15: $2.11 \pm 0.05$, our value: $2.9 \pm 0.4$) and J1922+2110 (L15: $0.98 \pm 0.88$, our value: $2.0 \pm 0.2$).

In GK16, at the hand of simulated data, we discuss how the accuracy of $\tau$ and consequently $\alpha$ measurements,  depend on the relative value of $\tau$ to the pulse period ($P$). Large $\tau/P$ values lead to scattering tails that wrap around the full rotational phase, and can in extreme cases lead to less accurate estimates of $\tau$ values. Fig. \ref{fig:alphatauP} shows the $\alpha$ values as a function of the degree of scattering relative to the pulse period, characterized by the fraction $\tau_{150}/P$, with $\tau_{150}$ the scattering time value at 150~MHz. We see that for all pulsars $\tau_{150}/P < 0.15$, using the IM. There is no clear dependence of $\alpha$ on these values. For a given $\tau_{150}/P$ value, we find a range of $\alpha$ values. This is reassuring, as it shows no sign of systematic offsets in $\alpha$ values obtained from our code. 

In the previous section we have discussed screen anisotropy. We note that the mean spectral index for the AM is equal to $3.5 \pm 0.4$, higher than for the IM and much closer to the theoretically expected values. In 10 out of the 13 objects, the increase in $\alpha$ using an AM is larger than 20\%, and for all but 1 the increase is larger than 10\%. However, for five objects, the $\alpha$ values associated with the IM and AM lie within the error bars of each other, making the outcomes of the models less distinct. The anisotropic $\alpha$ value for the majority of pulsars (9/13) has error bars reaching to the theoretically expected values. However, as also noted in the previous section, the error bars on the anisotropic spectral indices are typically larger than for the IM. 

In GK16, we have shown that fitting simulated anisotropic pulsar profiles with an IM, leads to incorrect $\tau$ values and lower spectral indices. The model-dependent $\alpha$ values obtained here, for which anisotropic values are often closer to theoretically predicted values, can be considered potential evidence for anisotropic scattering. Other possibilities for lower spectral indices include screens that are truncated, that have extreme scattering properties or multiple screens along a given line of sight. In Section~\ref{sec:multi} we discuss whether we find evidence for finite scattering screens. 

\subsection{Profile evolution and DM corrections}

\subsubsection{Profile evolution}

\begin{figure*}
\includegraphics[width=\textwidth]{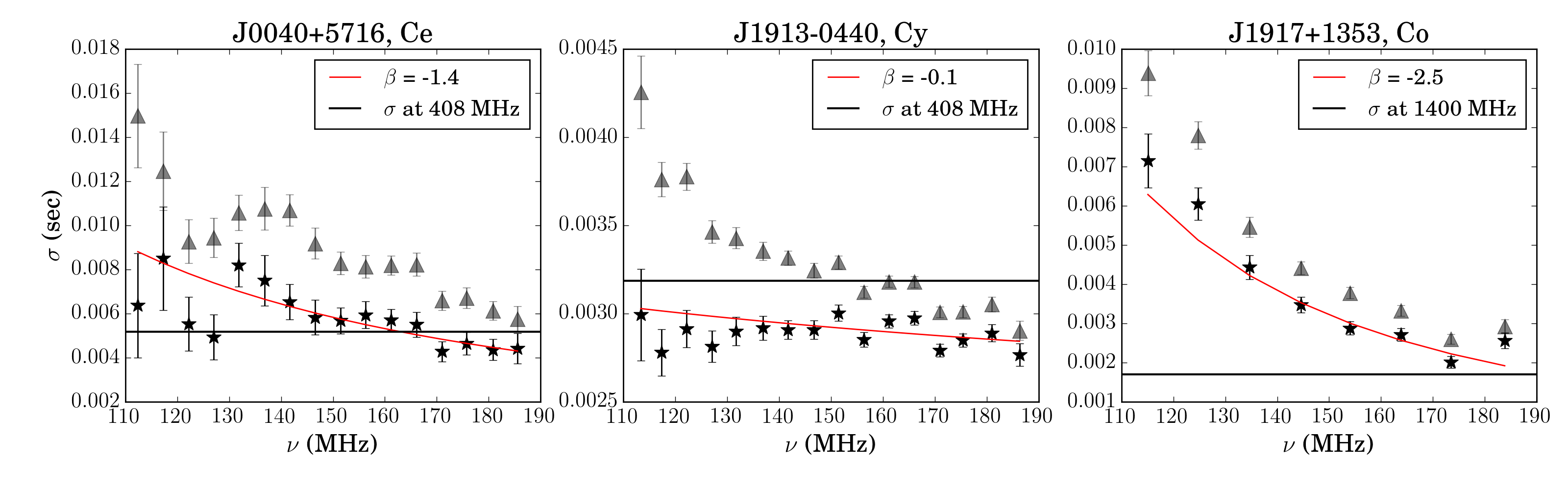}
\caption{The evolution of intrinsic pulsar widths (represented by $\sigma$) with frequency, for a subset of the pulsars. The width evolution is shown for both the isotropic model (stars) and extremely anisotropic model (circles). The high frequency widths as published in \citet[at 408~MHz]{Lorimer1995} and \citet[at 1.4 GHz]{Hobbs2004} are shown as horizontal solid lines. We have transformed their full width at half-maximum values (w${}_{50}$) to $\sigma$ values. The spectral indices of a power-law fit to the isotropic data ($\sigma \propto \nu^{\rm{pow}}$) are shown in the legends. Plots for the rest of the source list can be found in Appendix \ref{app:widths}.}
\label{fig:widths}
\end{figure*}

\begin{figure}
\includegraphics[width=\columnwidth]{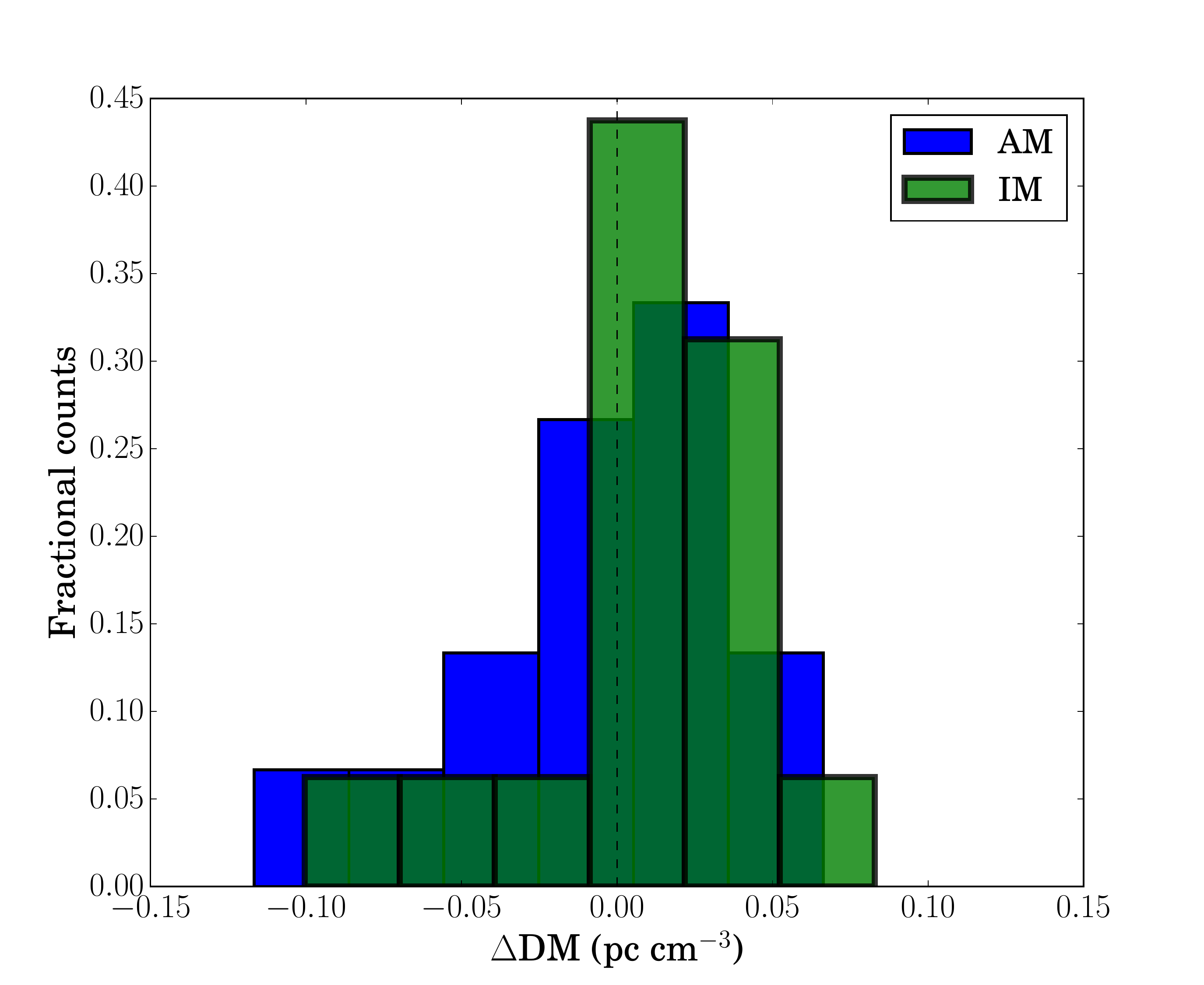}
\caption{Spread in $\Delta$DM values obtained for our set of pulsars.}
\label{fig:deltadm2}
\end{figure}

The simplified canonical pulsar emits a cone-shaped beam along its magnetic axis. Observational evidence shows that pulse profiles are often broader at low frequencies, and that multi component separation decreases with frequency. Physically this is understood to mean that high frequency radiation is emitted closer to the neutron surface and low frequencies higher up in the magnetosphere \citep{Thorsett1991}, leading to the phenomenon known as \textit{radius-to-frequency mapping} (RFM, \citealt{Cordes1978}).

Apart from this expected evolution of the pulsar width, a wide variety of intrinsic profile changes with frequency have been observed (see e.g. \citealt{LyneMan1988} and \citealt{MitraRankin2002}). \citet{Hassall2012} analysed the profile evolution of four pulsars at LOFAR frequencies, and found that none of them are well explained by RFM. 

An optimal de-scattering technique would remove the effects of scattering, revealing the intrinsic evolution of the pulsar with frequency. We have picked the sources in this paper not only based on the scattering tails they exhibit, but also for having simple single component profile shapes that do not show dramatic profile evolution. However, when scattering dominates the observed profile shape, as is the case for most of our sources at HBA frequencies, it is difficult to decouple scattering effects from profile evolution. Ideally, an exact understanding of the scattering mechanism and its dependence on frequency, would allow us to disentangle scattering and profile shape changes. As the evidence for anomalous scattering and deviations from expected scattering trends  (e.g. $\tau \propto \nu^{-4}$) increases, this becomes less straightforward. 

Our model assumes an intrinsic Gaussian-shaped profile and fits for the underlying Gaussian components in each frequency channel. The obtained standard deviation of the Gaussian ($\sigma$) is used as a proxy for pulse width. We analyse the evolution of $\sigma$  with frequency.  We compare the obtained $\sigma$ values at HBA frequencies to the pulsar widths measured at higher frequencies by \citet[at 408~MHz]{Lorimer1995} and \citet[at 1.4~GHz]{Hobbs2004}. The results for a subset of pulsars are shown in Fig. \ref{fig:widths}, the rest of which appear in Appendix \ref{app:widths}. 

We find that for the majority of the pulsars, the widths correspond well to the higher frequency results. In most cases, the widths measured at the high-frequency end of the HBA band are close to the corresponding widths at higher frequencies, as can be seen for PSR~J0040+5716 in the left-hand panel of Fig. \ref{fig:widths}. The cases for which the high frequency widths are greater than the widths obtained by the IM, include PSRs J0117+5914, J0614+2229, J1913$-$0440 and J1935+1616. In the case of PSR~J0117+5914, as was seen in its flux density spectra as well, outcomes differ significantly between the Commissioning and Census data. PSR~J1913$-$0440 (middle panel of Fig. \ref{fig:widths}) has been discussed as a pulsar with possibly a secondary component, evidence for which is seen not only in the profile shapes, but also in the deviations of the $\tau$ spectrum at low frequencies. PSR~J1913$-$0440 has also been considered a candidate for evidence for anisotropy. Here, we see that the width evolution of the intrinsic pulse modelled by the IM and AM is significantly different (middle panel of Fig. \ref{fig:widths}). The AM shows a clear decrease in width with frequency and is in closer agreement to the high frequency width obtained by \citet{Lorimer1995}. The pulsar which shows the most well-defined width evolution is PSR~J1917+1353, as depicted in the right-hand panel of Fig. \ref{fig:widths}. The isotropic case is fitted with a power-law, $\sigma \propto \nu^{-2.5}$. \citet{Hobbs2004} estimated w${}_{50} = 4.0$~ms, which translates to $\sigma =1.7$~ms. Our lowest measured width value for PSR~J1917+1353 is $2.0 \pm 0.15$~ms.

\subsubsection{DM corrections}

As seen throughout Section \ref{sec:results}, by fitting for the centroid values ($\mu$) of the intrinsic Gaussian components, we can estimate small corrections to the DM values, that we have labelled as $\Delta$DM.
Fig. \ref{fig:deltadm2} shows the distribution in $\Delta$DM values for both models. 
Although the histogram represents only a small number of sources, it appears that positive $\Delta$DM values are obtained more frequently than negative values, for both fitting models. A positive value here refers to an overestimation in the original data DM value by an amount of $\Delta$DM. Rerunning the data reduction with a value set to $\rm{DM} - \Delta$DM removes this dependence of $\Delta\mu$ on frequency.

From Fig. \ref{fig:deltadm2} we conclude that small errors in DM values are likely to be introduced by the traditional method in which DM values are obtained. This relies on finding the DM value for which the S/N of the sum of the channelized average pulse profiles is maximized, and is therefore sensitive to the location of the peaks of the channelized data in general. In the case of highly scattered profiles, the true location of the centroid of the intrinsic pulse lies earlier in time to when the peak of the signal is observed. Furthermore, this change is larger at low frequencies and smaller at high frequencies, such that, typically, the DM value required to maximize the S/N of the sum of the intrinsic pulse profiles is lower than the value required to maximize the sum of the scattered pulses. 

We also have instances in which the obtained $\Delta$DM is negative, and therefore the original DM was an underestimation. The asymmetric effect of scattering on pulse shapes tends to drag the profile centroids to larger $\tau$ values. However, effects such as the intrinsic profile evolution described previously, can cause shifts in either direction. It is therefore possible, through such counteracting impacts, to obtain DM values that are over or underestimated. 

DM corrections are critical to improve pulsar timing models. However, due to both intrinsic pulse evolution and the fact that at high frequencies (where pulsar timing observations are typically conducted), a smaller angular scale of the ISM is sampled than at lower frequencies \citep{CordesShannon2016}, the DM corrections obtained here, can not straightforwardly be extrapolated to timing observations.

\subsection{Finite scattering screens}\label{sec:multi}

As discussed in Section \ref{sec:spindx}, and in more detail in \citet{CordesLazio2001} and GK16, lower $\alpha$ values can result from finite scattering screens. 

If indeed some of the lower $\alpha$ values obtained at low frequencies here, are the result of finite scattering screens in the ISM, we can estimate the scattering screen size for each pulsar at which deviations from theoretical $\alpha$ values (associated with infinite screens) will become observable. For a mid-way screen, $\tau$ can be expressed as

\begin{equation}
\tau = \frac{D \sigma_{\theta}^2}{c},
\end{equation}

\noindent with $D$ the distance from the pulsar to the observer, $\sigma_\theta$ the standard deviation of the distribution in observing angles and $c$ the speed of light.  Using the $\tau$ values at 150~MHz for each pulsar (obtained from the $\tau$ spectra power-law fits), we calculate the maximum scattering screen size at 150~MHz at which observations become sensitive to the finite nature of the screen. Table \ref{table:blines} shows the associated scattering screen sizes for six pulsars as well as the required observing baseline to be able to resolve them. We have chosen $\sigma_\theta$ as a representation of the radial screen size (such that the diameter is 2$\sigma_\theta$).

\begin{table}
\centering
\small
\begin{tabular}{p{2.6cm} p{2cm} p{2cm}}
\hline
Pulsar & \mbox{Screen size}  \mbox{(2$\sigma_\theta$, mas)} & Baseline \qquad (km)\\[0.5ex]
\hline
\hline
J0040+5716	&	165	&	2485	\\
J1922+2110	&	130	&	3149	\\
J2257+5909	&	130	&	3150	\\
J1851+1259	&	62	&	6571	\\
J1917+1353	&	61	&	6726	\\
J1913$-$0440	&	53	&	7665	\\
\hline
\hline
\end{tabular}
\caption{The angular size of mid-way screens at 150~MHz, for six pulsars, based on their obtained $\tau$ values at 150~MHz and the distance to the source. The six shown, are the ones with the three largest and three smallest associated scattering screens (excluding PSR~J2305+3100 at 25~kpc). The last column provides the required baseline ($B = \lambda/2\sigma_\theta$) to resolve such screens with a low frequency VLBI network, where $\lambda = 2$ m is the wavelength corresponding to 150~MHz.}
\label{table:blines}
\end{table}

To show conclusively that the observed low $\alpha$ values are correlated with finite scattering screens, would therefore likely require a low frequency interferometer with large baselines, as given in Table \ref{table:blines}. The current version of the Low Frequency VLBI Network (LFVN) mostly includes 18 cm and 92 cm receivers in e.g Russia, India and China. 
An alternative approach, as mentioned in the Introduction, is to use the space-based observatory \textit{RadioAstron}, jointly with ground-based instruments to form extreme baseline interferometers \citep{Smirnova2014}.

A dependence of $\tau$ on frequency that shows a break in its power-law spectrum, can too be evidence for truncated screens, as flux density is lost at longer wavelengths where observations become sensitive to the size of the screen, altering the shape of the observed profiles (GK16). None of the pulsars in our set show clear evidence for broken power-law $\tau$ spectra across the HBA frequency band. Broad-band observations at higher frequencies are required to investigate whether these breaks are observed. However, it is likely that small scattering screens will not have physically sharp edges, and therefore could feasibly result in increasingly flatter frequency dependencies on $\tau$ for a given range of frequencies.

\begin{figure}
\includegraphics[width=1.1\columnwidth]{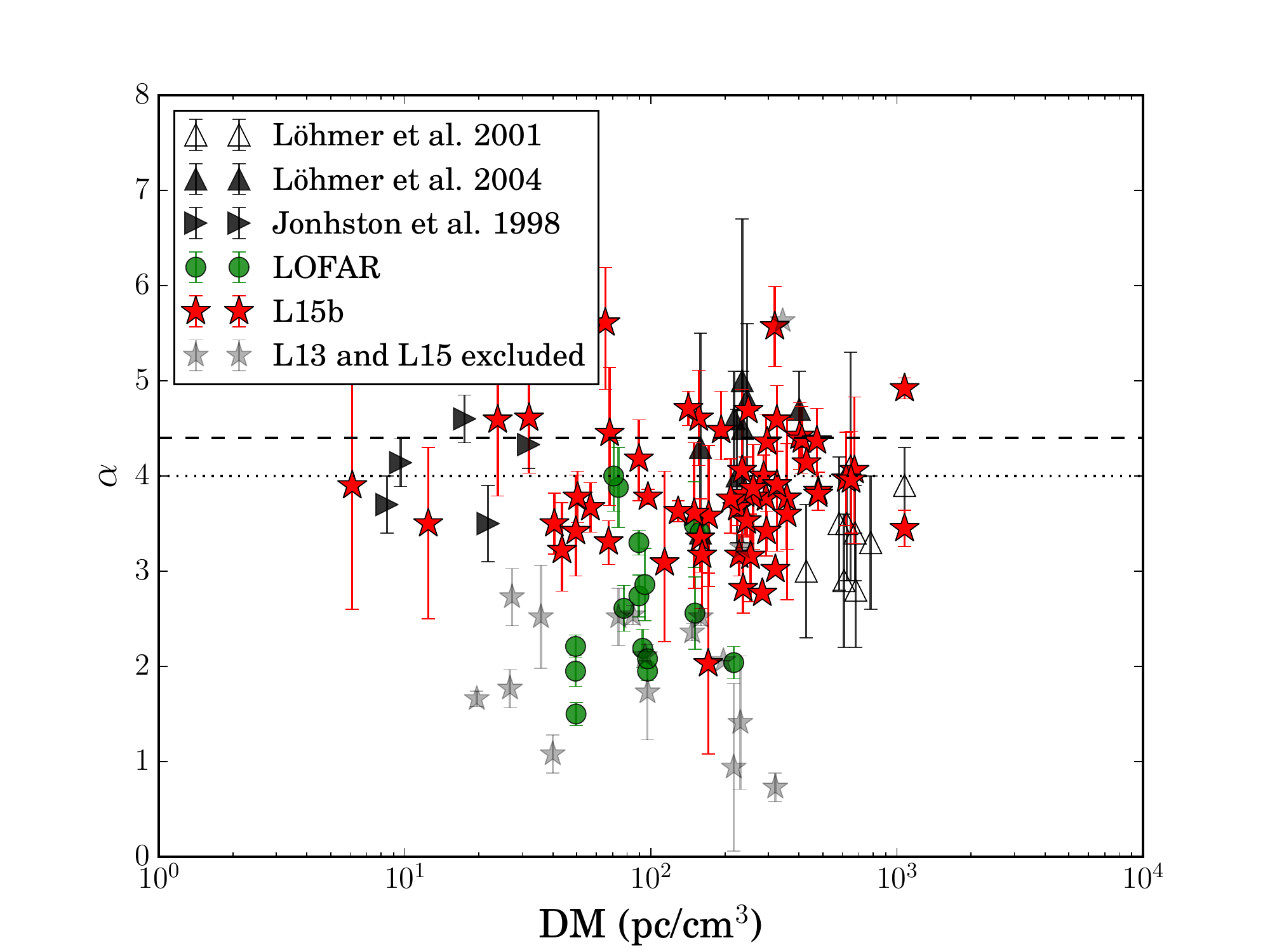}
\caption{Spectral indices ($\alpha$) are plotted against the corresponding DM values. The LOFAR data set of this paper is shown in green (dark circles), with other data points from the literature as indicated in the legend. We show the updated L13 and L15 values as given in L15b (red, dark stars), along with the lower $\alpha$ values from these papers, that were excluded from further analyses (grey, lighter stars). }
\label{fig:alfDM}
\end{figure}

A correlation between low $\alpha$ values and high DM values, has been promoted by data sets such as in \citet{Lohmer2001}. In Fig. \ref{fig:alfDM}, we show the dependence of $\alpha$ with DM. This figure includes data points from several other papers, as described in the legend and the caption of the figure. 

\citet{Lohmer2001} suggested that beyond a given DM threshold ($\rm{DM} > 300$ pc cm$^{-3}$), $\alpha$ values deviated from the theoretically expected values. This threshold has been revisited by L13 (amongst others), who suggested that deviations in $\alpha$ start occurring at lower DM values, around $\rm{DM} = 230$--250 pc cm$^{-3}$. In their more recent paper (L15), however, additional $\alpha$ measurements have led them to conclude that the previously postulated DM threshold was based on data biased by a small number of $\tau$ measurements.   

\begin{figure}
\includegraphics[width=\columnwidth]{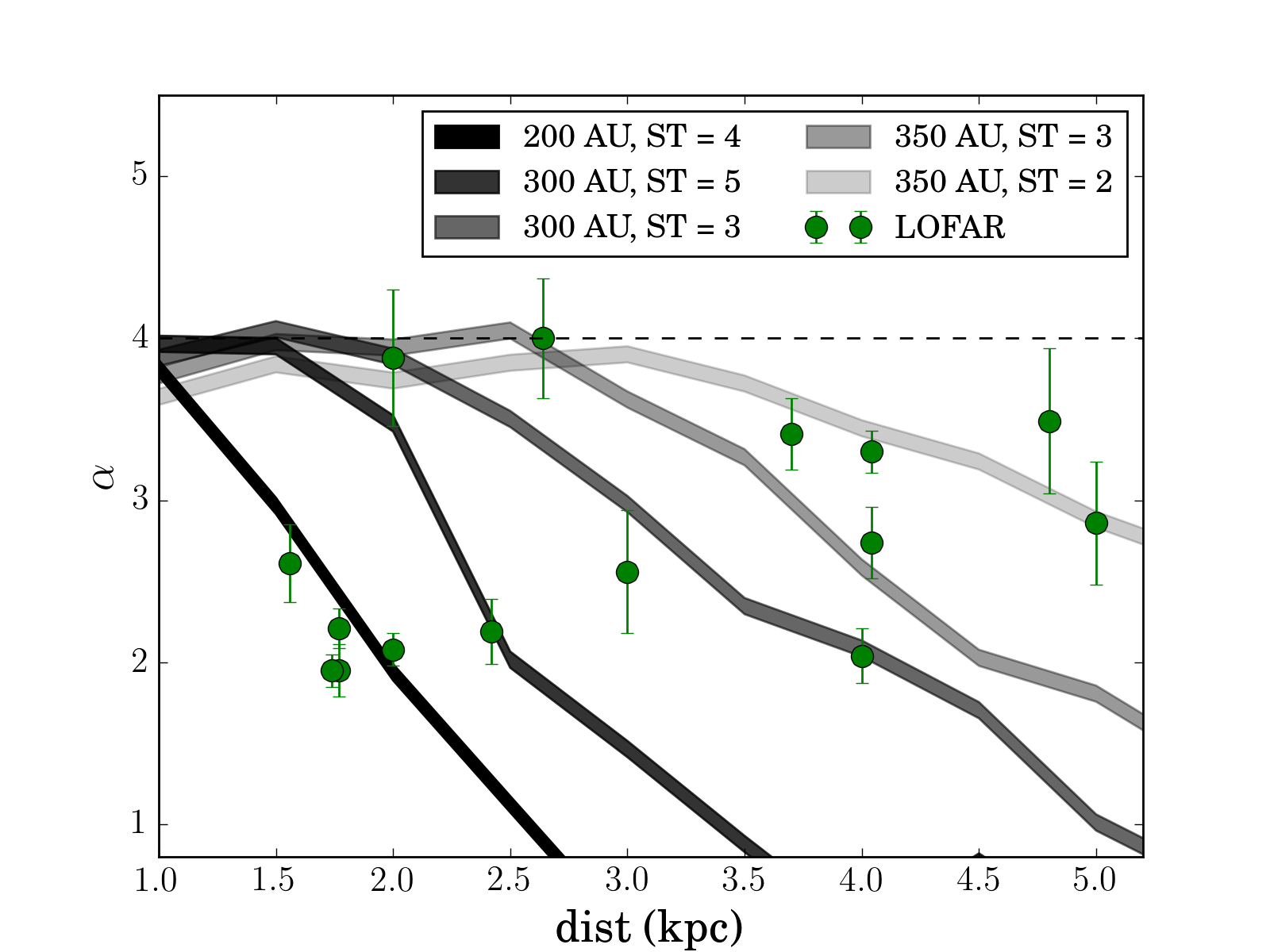}
\caption{The LOFAR data points from this paper (large green circles), along with computed $\alpha$ models, that are based on a simple truncated screen model. Circular screens are placed mid-way along the line of sight, for a set of distances. The screens can differ in size and scattering strength. A simulated pulsar is scattered by the modelled screen and the $\alpha$ value over the HBA frequency range is estimated. The theoretical value of $\alpha = 4$ is shown as a dashed line. The model with the most extremely deviating values of $\alpha$ (darkest line), is based on a screen with radius 200~AU and scattering strength, ST = 4 (see the text for definition). The other model screen sizes and strengths are shown in the legend.}
\label{fig:alfDist}
\end{figure}

The LOFAR data show low $\alpha$ ($<4$) values in the DM range 49 to 217~pc cm$^{-3}$. We investigate whether the distribution of $\alpha$ values with respect to distance (or DM as a proxy for distance) such as seen in Fig. \ref{fig:alfDM} can be produced by a simple picture using truncated scattering screens. As a first step, we simulate the observed scattering of a 0.6 s period pulsar, with a Gaussian shape and a duty cycle of 2.5\% (chosen to represent our set of pulsars) behind a circularly truncated screen. We note that the set of LOFAR pulsars all have distances between 1.5 and 5.0 kpc (excluding PSR~J2305+3100 at 25~kpc), and an $\alpha$ distribution between 1.5 and 4.0. Fig.~\ref{fig:alfDist} shows how this distance and $\alpha$ distribution can be modelled using truncated screens. The setup uses a single truncated screen of varying size and scattering strength. The scattering strength is defined by the standard deviation of the Gaussian distribution in scattering angles, as described in more detail in GK16. Here, we label a scattering strength of $\sigma_{a}  = 3$ mas as $\rm{ST} = 3$. For a range of distance values, a screen is placed mid-way between the pulsar and observer, and the $\alpha$ value over the HBA band calculated. By tweaking the screen size and scattering strength a set of distance-$\alpha$ pairs are obtained. 

In a similar way, low $\alpha$ values, as observed in \citet{Lohmer2001}, for more distant pulsars (between 6.3 and 10.2 kpc, and $ 400 < \rm{DM} < 1100$, in units of pc cm$^{-3}$) can be produced through truncated scattering screens. This, however, would require screens with higher scattering strengths or multiple scattering screens along a given line of sight. It is unlikely that this simplistic picture describes the whole truth. However, evidence for typical sizes and or scattering strengths of screens in the ISM can shed light on the causes of low $\alpha$ values.

\subsection{Correlations between scattering and flux density loss}\label{sec:fluxturn}

For each data set, we calculated an average mean flux density spectrum by summing the intensities of the best fit curve (and dividing it by the number of phase bins across the pulse profile) per frequency channel. This (uncorrected) flux density is corrected for scattering effects through an estimate of the associated raised baseline level (see Section \ref{sec:scatmodels}). Examining the average profiles for each pulsar we find six cases in which profiles with wrap around scattering tails are visible. These are the cases in which we would expect to see an associated flux density loss (see \citet{Geyer2016} for more detail), and therefore a discrepancy between the uncorrected and corrected flux density. The six pulsars are PSRs J0117+5914, J0742$-$2822, J1909+1102, J1917+1353, J1935+1616 and J2257+5909, as discussed in the results section. The flux density spectra for all the sources can be viewed in Appendix \ref{app:flux}. 

From this set of six, four show a correlation between the flux density spectrum turnover and the onset of wrap around scattering tails. This includes PSR~J1909+1102 which has the highest $\tau_{150}/P$ (IM) value of the set, namely 0.15 (see Fig. \ref{fig:alphatauP}) and the joint highest $\tau_{150}$ value of $42 \pm 3$~ms (IM, see Table \ref{table:two}). PSRs J1917+1353 and J1935+1616 show wrap around scattering tails that do not correlate with the associated flux density spectra. In the case of PSR~J1917+1353 there is no clear turnover in its flux density spectrum, and the turnover in the flux density spectrum of PSR~J1935+1616 is at too high a frequency to only be related to long scattering tails. These two pulsars have the lowest $\tau/P$ ratio (0.06) of the six pulsars that show pulse run-in. It is therefore perhaps not unexpected that the flux density spectrum is not dominated by the scattering effects.  As discussed in more detail in Section \ref{sec:J1935} such a turnover could potentially be due to scattering by finite scattering screens.

There are three pulsars for which the spectra appear to turnover without the profiles exhibiting extreme scattering tails. These are PSRs J0614+2229 and J1913$-$0440 for which the flux density spectrum turns over at high frequencies (around 170~MHz), even though $\tau_{150}/P$ is equal to only 0.01, and PSR~J2303+3100 which is only weakly scattered, $\tau_{150}/P = 0.01$, but has a turnover at around 150~MHz.  It therefore seems clear that when the scattering appears to be weak the flux density spectrum shape will be dominated by other effects such as thermal absorption (e.g. \citealt{Rajwade2016}) or intrinsic emission properties, or perhaps some form of anomalous scattering. In the cases for which the scattering is large we do see the expected correlations between scattering and flux density measurements. 

Our method estimates the degree to which flux is lost due to scattering effects. However, for the correction to hold an understanding of the scattering mechanisms is required (i.e. the correct scattering model has to be used). It will therefore be valuable to compare these corrected values to flux density estimates obtained from interferometric pulsar images. The authors are currently analysing joint imagining and beam-formed LOFAR data to conduct such a comparison.

\subsection{Scattering time versus DM and distance}

\begin{figure*}
\includegraphics[width=1.5\columnwidth]{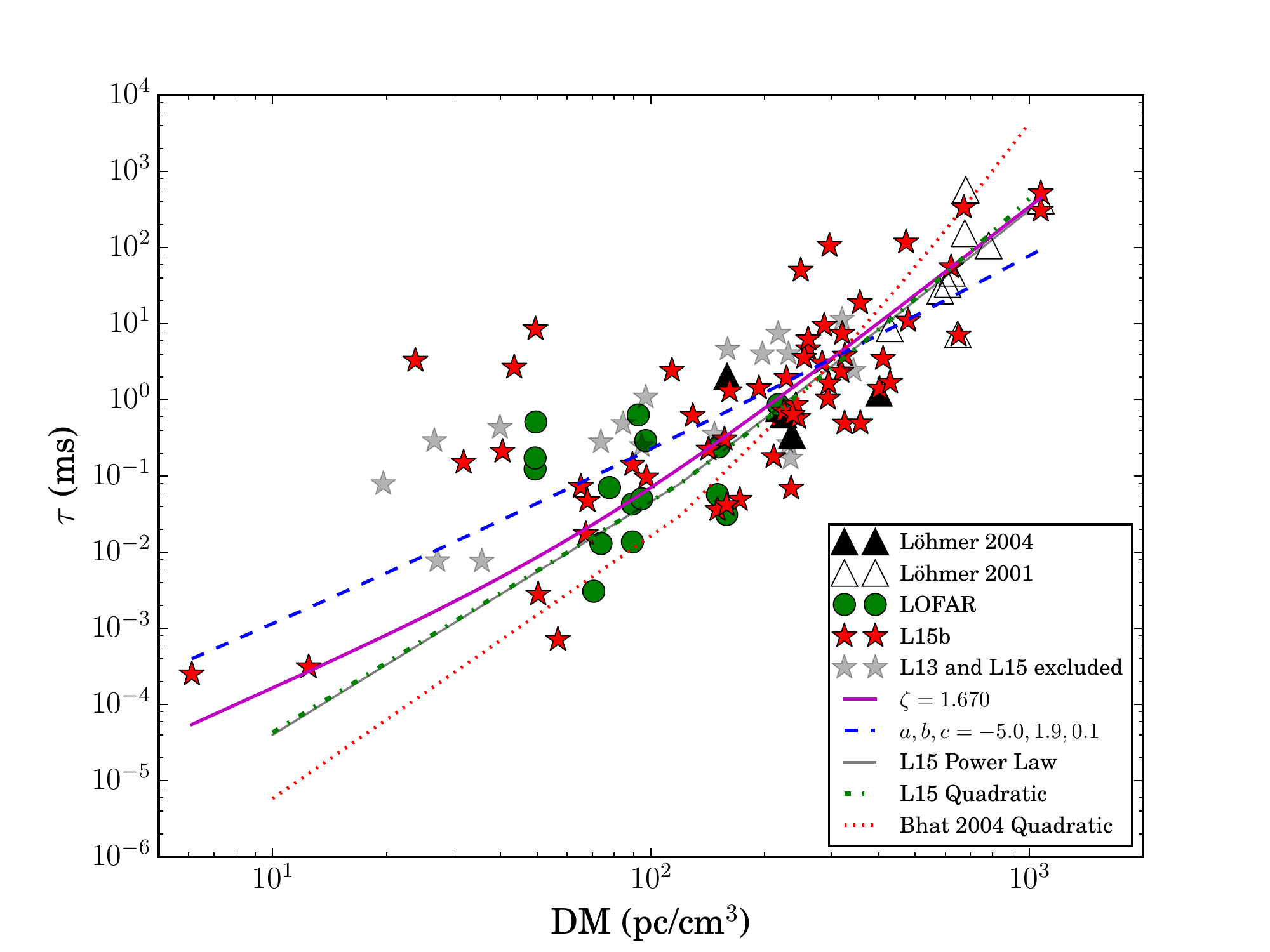}
\caption{Scattering time ($\tau$) at 1~GHz versus DM values, for our data set along with data from \citet{Lohmer2001, Lohmer2004} and L15b. L15b contains $\tau$ values for most sources in L13 and L15. The data points which were considered erroneous in L13 and L15 are also shown (grey stars). The LOFAR data points from this paper are shown as green (dark) circles. To obtain fits all the data, except those marked as \textit{excluded}, were used. The obtained fits are a power-law fit (magenta, solid thick line) and a parabolic fit (blue dashed line). The plot also includes the parabolic fit from \citet{Bhat2004} (red, dotted line), as well as the power-law and parabolic fits from L15 (grey solid, green dash--dotted lines) respectively.}
\label{fig:tauDM}
\end{figure*}

\begin{figure}
\includegraphics[width=\columnwidth]{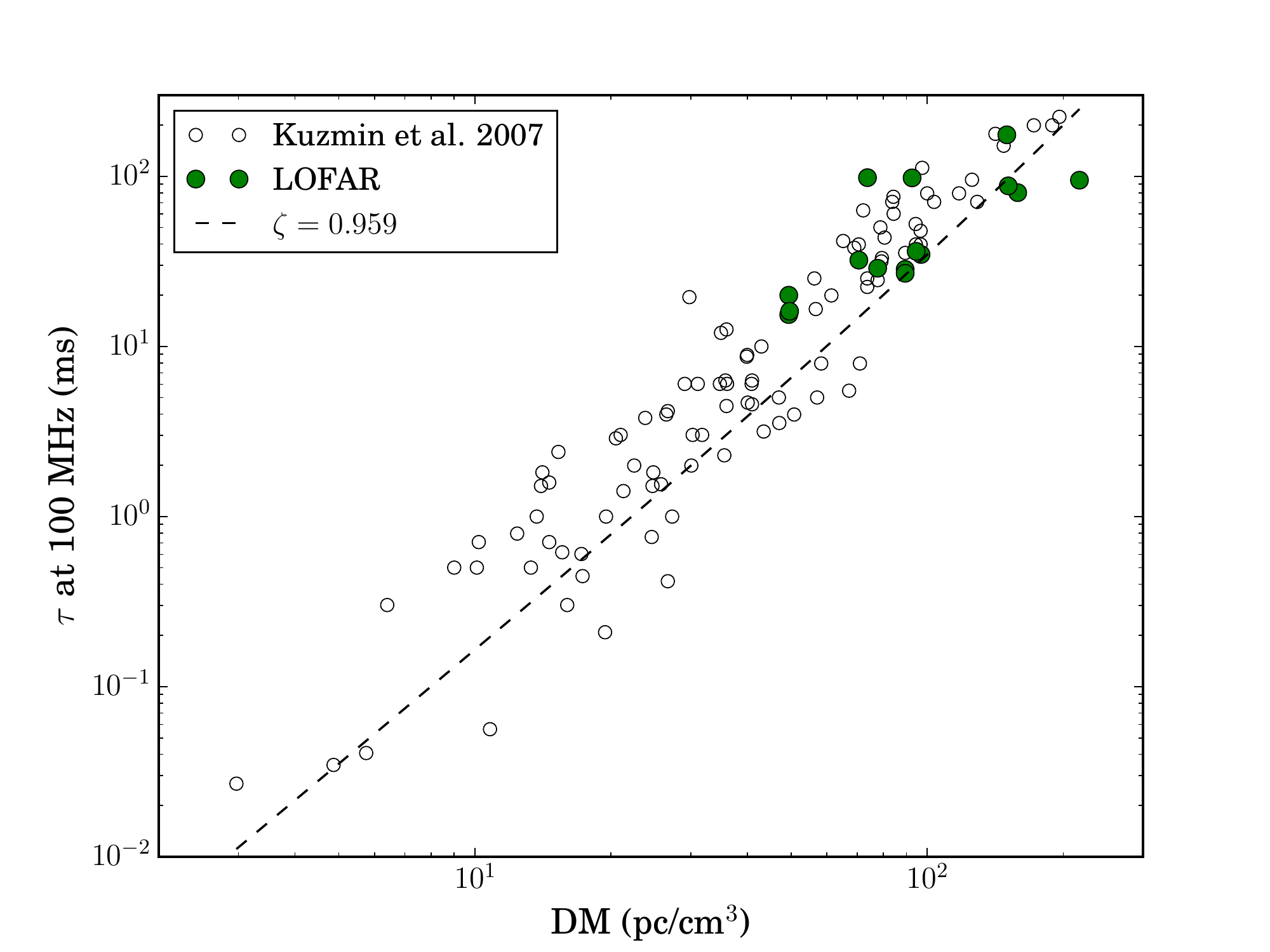}
\caption{The relationship between DM values and  $\tau$ at 100~MHz, as obtained from the \citet{Kuzmin2007} data set (empty circles) and the LOFAR data set in this paper (green circles).}
\label{fig:tau100DM}
\end{figure}

\begin{figure}
\includegraphics[width=\columnwidth]{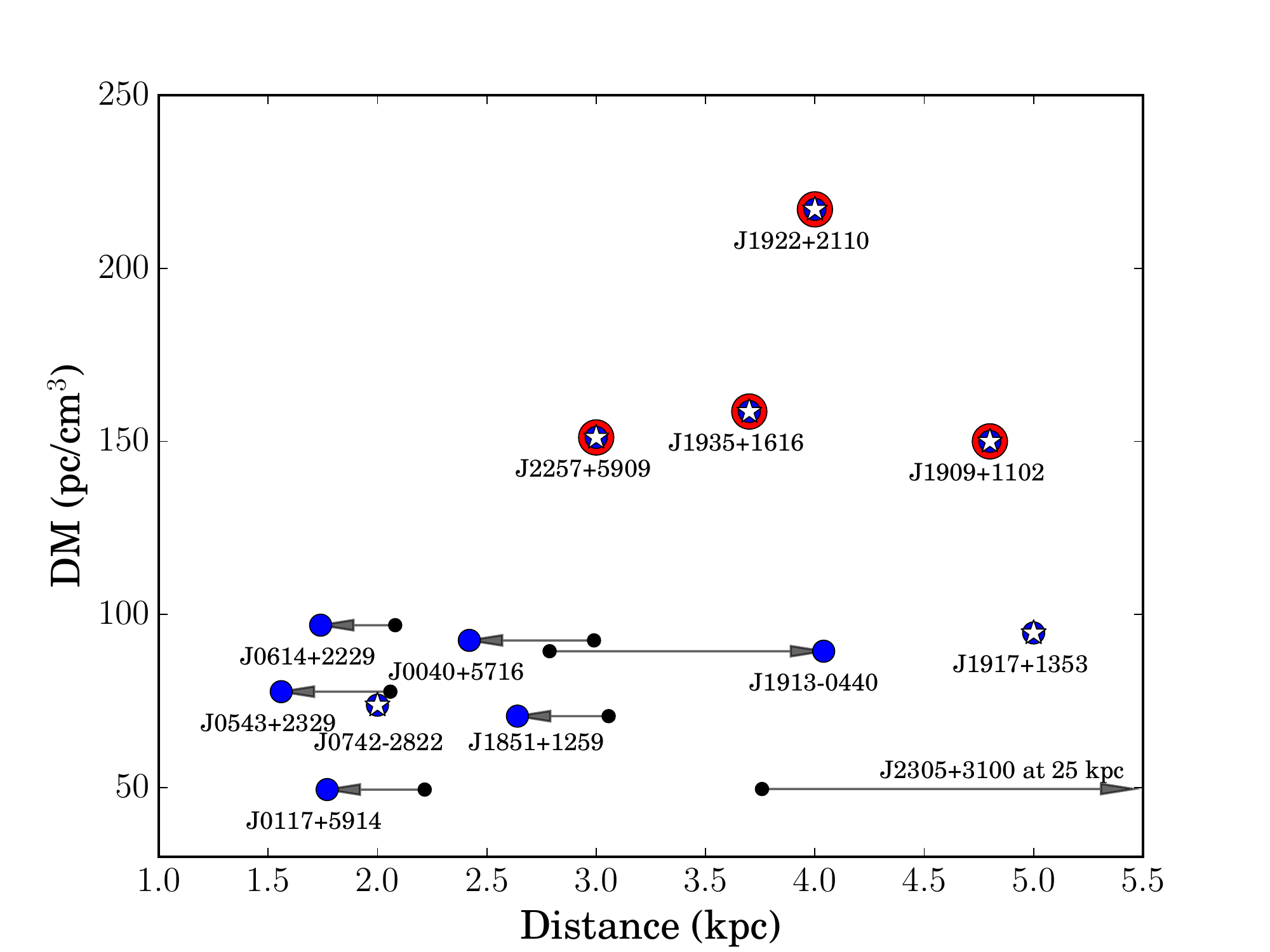}
\caption{The relationship between DM and the distances for our data set. The DM values of the majority of these pulsars are less than 100~pc cm$^{-3}$. The blue (small) circles indicate the distances as estimated by the YMW16 model \citep{Yao2017} or from associated objects (see Table \ref{table:one}). Arrows starting from black dots indicate the change in distance estimates from the NE2001 model \citep{Cordes2002} to the YMW16 model. The data point for PSR~J2303+3100 is excluded as it lies far to the right. There are four clear outliers to the typical distance versus DM trend, shown as large circles with red outlines. These high DM values could indicate a more dense and complex ISM along the lines of sight to these sources. Pulsar for which DM-independent distance measurements exist, are marked with white stars.}
\label{fig:DistDM}
\end{figure}

We now revisit the scattering time ($\tau$) dependence on DM. \cite{Bhat2004} published an empirical relationship between these quantities valid for over 10 orders of magnitude in scattering times, and DM values between 1 and 1000 pc cm$^{-3}$. A parabolic function, $\log\tau = a + b \log \rm{DM} + c\,(\log\rm{DM})^2$, is fitted to their data. Alternatively, power-laws of the form \mbox{$\tau \propto \rm{DM}^{\gamma} (1 + \kappa\,DM^{\zeta})$}, where  $\gamma$ is fixed at 2.2, as determined from a Kolmogorov spectrum, have been used \citep{Ramachandran1997,Lohmer2004}.

Fig. \ref{fig:tauDM} shows our fits to an ensemble of $\tau$ values at 1~GHz versus DM values. Our obtained IM $\alpha$ values have been used to transform $\tau$ values to 1~GHz. The data sets of \citet{Lohmer2001, Lohmer2004} and L15b are also shown, along with obtained fits. The excluded data points from L13 and L15 are shown as grey stars. The LOFAR data points from this paper are shown as green (dark) circles. 

It is clear that the fitted trends obtained by us, L13 and L15 at low DM values, promote higher $\tau$ values than originally proposed by the \citet{Bhat2004} fit. Our fit indicates even larger $\tau$ values at the low DM values than L13 and L15.  Our parameter fits, compared to L15, are $a = -5.0$ (L15, $a = -6.3$), $b = 1.9$ (L15, $b = 1.5$) and $c = 0.1$ (L15, $c = 0.5$). From the power-law fits,  \cite{Ramachandran1997} and \citet{Lohmer2004} have obtained $\zeta = 2.5$ and 2.3, whereas L15 finds 1.74 and we find 1.67.

If indeed the \citet{Bhat2004} relationship can be considered a true reflection of the $\tau$ dependence on DM at higher frequencies (e.g. 1~GHz), then our result shows that in order for the relationship to be upheld, our low spectral index values found at low frequencies, can not persist up to higher frequencies. In order to reach the \citet{Bhat2004} dependence at higher frequencies with our set of pulsars, their spectral index will need to evolve with frequency. e to broad-band data at higher than LOFAR frequencies for these pulsars, where scattering is still measurable, will allow us to investigate whether spectral indices indeed change. 

Fig. \ref{fig:tau100DM} shows a similar plot of $\tau$ versus DM, but for $\tau$ at 100~MHz. Here, we find $\zeta = 0.96$, much lower than in Fig. \ref{fig:tauDM}. The plot shows that our obtained values are in good agreement with the \citet{Kuzmin2007} data set, and again argues for frequency-dependent $\alpha$.

In Fig. \ref{fig:DistDM}, the relationship between the distance and DM for the sources in this paper is shown. Arrows indicate the changes in distance estimates from the NE2001 electron density model \citep{Cordes2002} to the YMW16 model \citep{Yao2017}. The majority of the sources have DM values below 100~pc~cm$^{-3}$. There is no clear increase of DM with distance. Four pulsars are clear outliers on this plot, having higher DM values than the rest of the set. This could point to more complex and dense ISM environments along the lines of sight to these pulsars. In Section \ref{sec:J1935}, we pointed out that the line of sight to PSR~J1935+1616 has been considered anomalous in the literature \citep{Lohmer2004}. 

Three of the four outliers have been discussed in Section \ref{sec:fluxturn} as pulsars with wrap around scattering tails. The fourth pulsar is PSR~J1922+2110, which together with PSR~J1909+1102 has the largest $\tau_{150}$ value of $42 \pm 2$ (IM, Table \ref{table:two}). Its low frequency profiles along with the flux density spectrum reveal that pulses will likely start overlapping at frequencies just below the lowest observed frequency channel of 112.5~MHz.

\section{Conclusions}

In this paper, we show that low frequency pulsar observations lead to measurements of the scattering spectral index $\alpha$ (obtained by isotropic scattering models) ranging between 1.5 and 4.0. This is much lower than expected from simple theoretical models. Our analysis in the temporal domain shows potential evidence for anisotropic scattering in 4 out of 13 sources. The low $\alpha$ value distribution can also be modelled using a set of truncated scattering screens. Studies in the temporal domain do not allow us to fully distinguish between anisotropy and truncated screens. Combining our work with results at higher frequencies, we surmise that the average $\alpha$ across this population is lower at these low frequencies than previous measured values at 1~GHz, suggesting a frequency evolution of $\alpha$. Interferometric imaging, including space-ground experiments, is key in investigating the typical sizes of scattering surfaces, while scintillation results are required for precise scattering measurements at higher frequencies. Both of these techniques should aid the investigation of the frequency dependence of $\alpha$. Furthermore, polarization measurements can serve as a sensitive diagnostic to small amounts of scattering, as demonstrated in \citet{Karastergiou2009}. Such accurate indicators can assist in distinguishing between intrinsic profile evolution and propagation effects. Lastly, the best tests for anisotropic scattering rely on high resolution dynamic spectra, with parabolic arcs in secondary spectra arising as a natural consequence of anisotropy \citep{Stinebring2001}. The pulsars discussed in this paper are good candidates for this list of further investigations.


\section{Acknowledgements}

This paper is based (in part) on data obtained with the International LOFAR Telescope (ILT). LOFAR (van Haarlem et al. 2013) is the Low Frequency Array designed and constructed by ASTRON. It has facilities in several countries, that are owned by various parties (each with their own funding sources), and that are collectively operated by the ILT foundation under a joint scientific policy.
This study also specifically made use of data from the long-term LOFAR core project LT5\_003 (PI:
JPW Verbiest). The authors would like to thank the LOFAR Pulsar Working Group for valuable comments and discussions. 
MG is a Commonwealth Scholar funded by the UK government. VIK, JWTH and DM acknowledge funding from an NWO Vidi fellowship and from the European Research Council under the European Union's Seventh Framework Programme (FP/2007-2013) / ERC Starting Grant agreement no. 337062 (`DRAGNET'). We thank the MNRAS scientific editor and anonymous reviewer for their valuable comments to improve the quality of this manuscript.

\balance
\bibliographystyle{mn2e.bst}
\bibliography{ScatteringBibDesk}

\newpage
\appendix
\onecolumn

\section{Profile fits}\label{app:profiles}

In this section of the Appendix we show the scatter broadened pulse profiles and the corresponding fits for an isotropic (red, solid lines) and extremely anisotropic model (blue, dashed lines). In the interest of saving space, we plot a subset of 8 or fewer average profiles for each pulsar, across the HBA band. In many cases we had split the band into 16 frequency channels, such that we are showing every second channel here.

\noindent
\begin{minipage}{\linewidth}
\makebox[\linewidth]{
  \includegraphics[width=\linewidth]{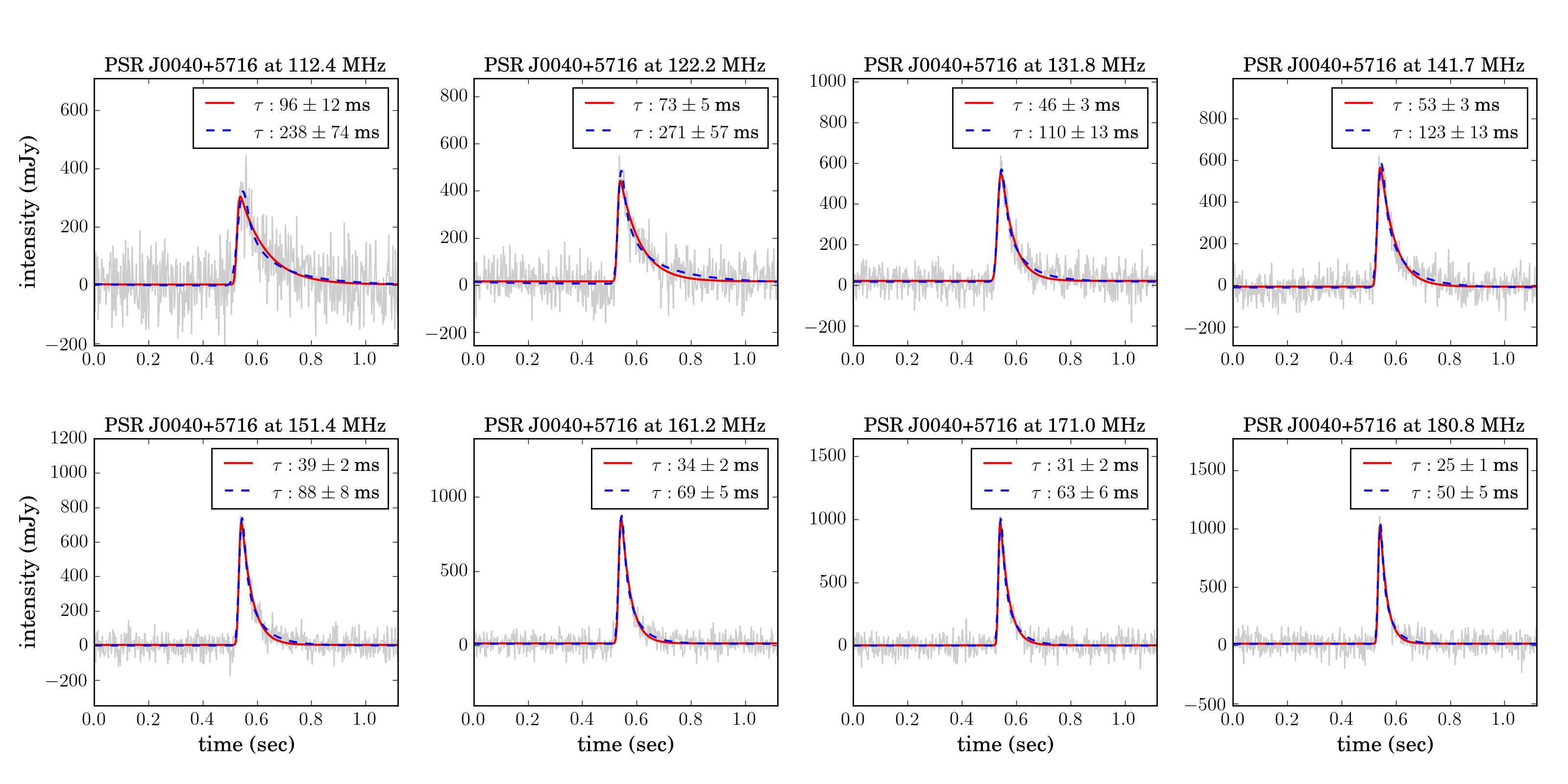}}
\captionof{figure}{PSR~J0040+5914. Fits to 8 (of 16 used) average profiles of Census data.}
\makebox[\linewidth]{
  \includegraphics[width=\linewidth]{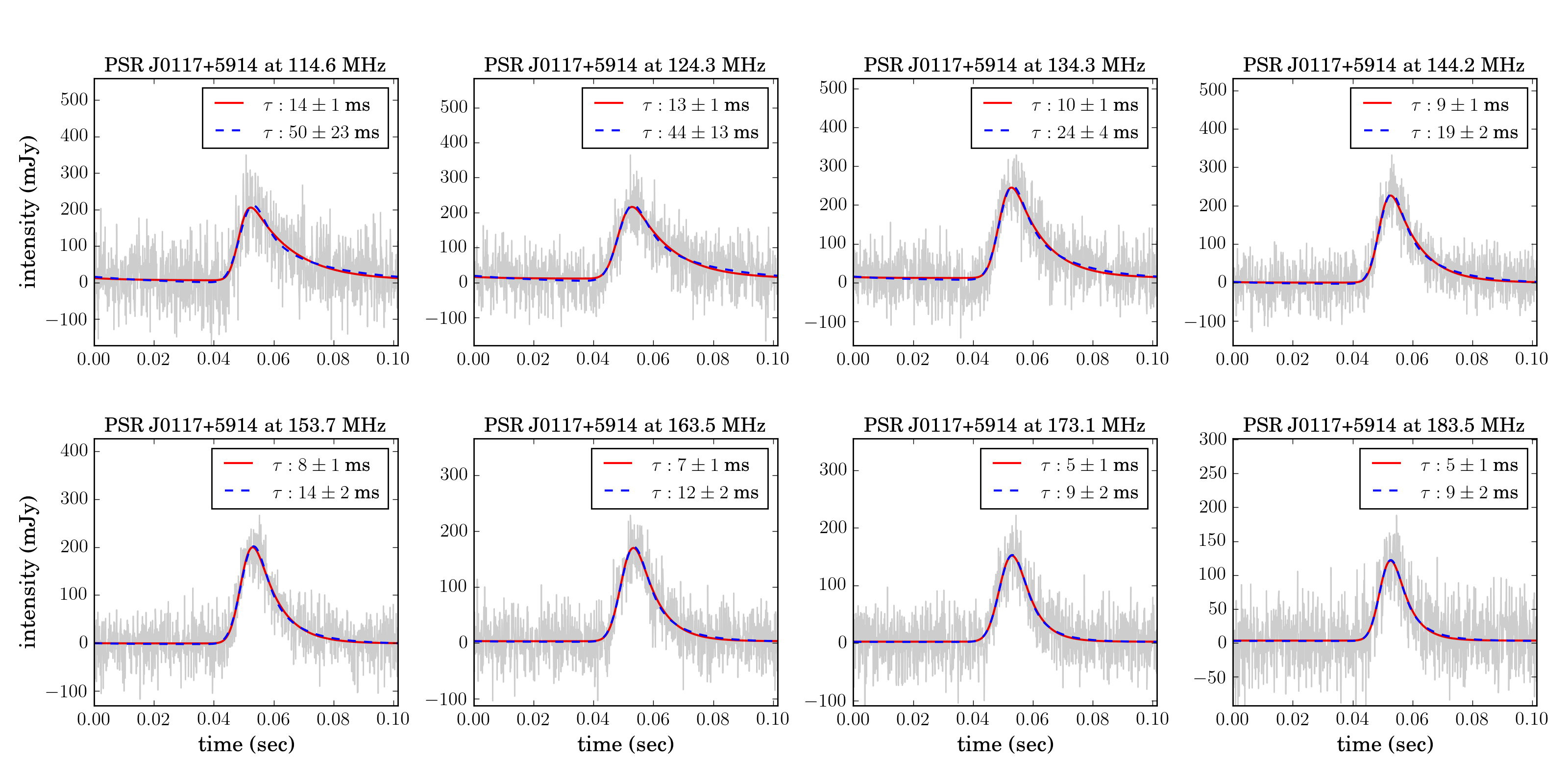}}
\captionof{figure}{PSR~J0117+5716. Profile fits to the 8 average profiles of Commissioning data.}
\end{minipage}

\noindent
\begin{minipage}{\linewidth}
\makebox[\linewidth]{
  \includegraphics[width=\linewidth]{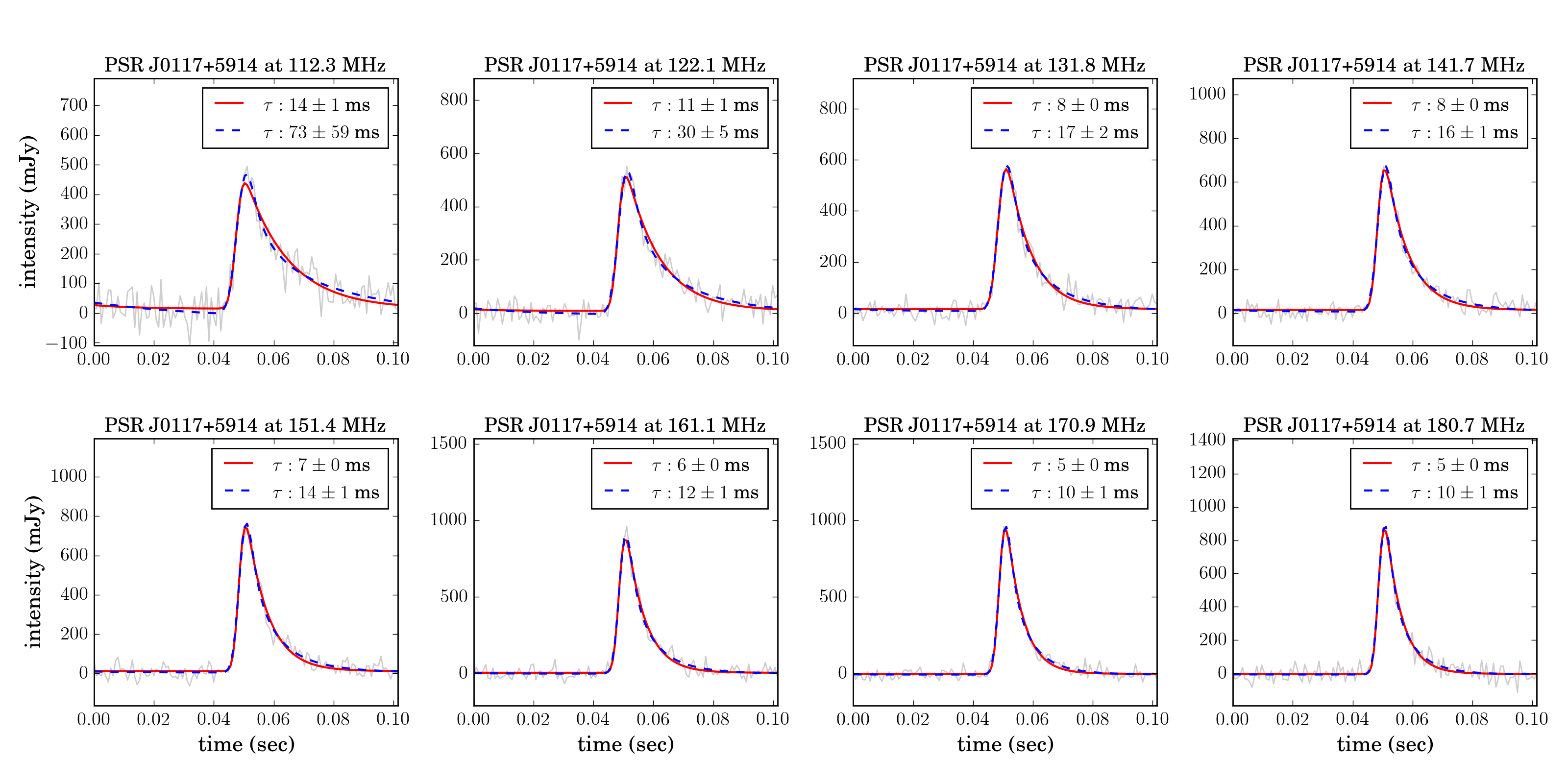}}
\captionof{figure}{PSR~J0117+5716. Profile fits to 8 (of 16 used) average profiles of Census data.}
\makebox[\linewidth]{
  \includegraphics[width=\linewidth]{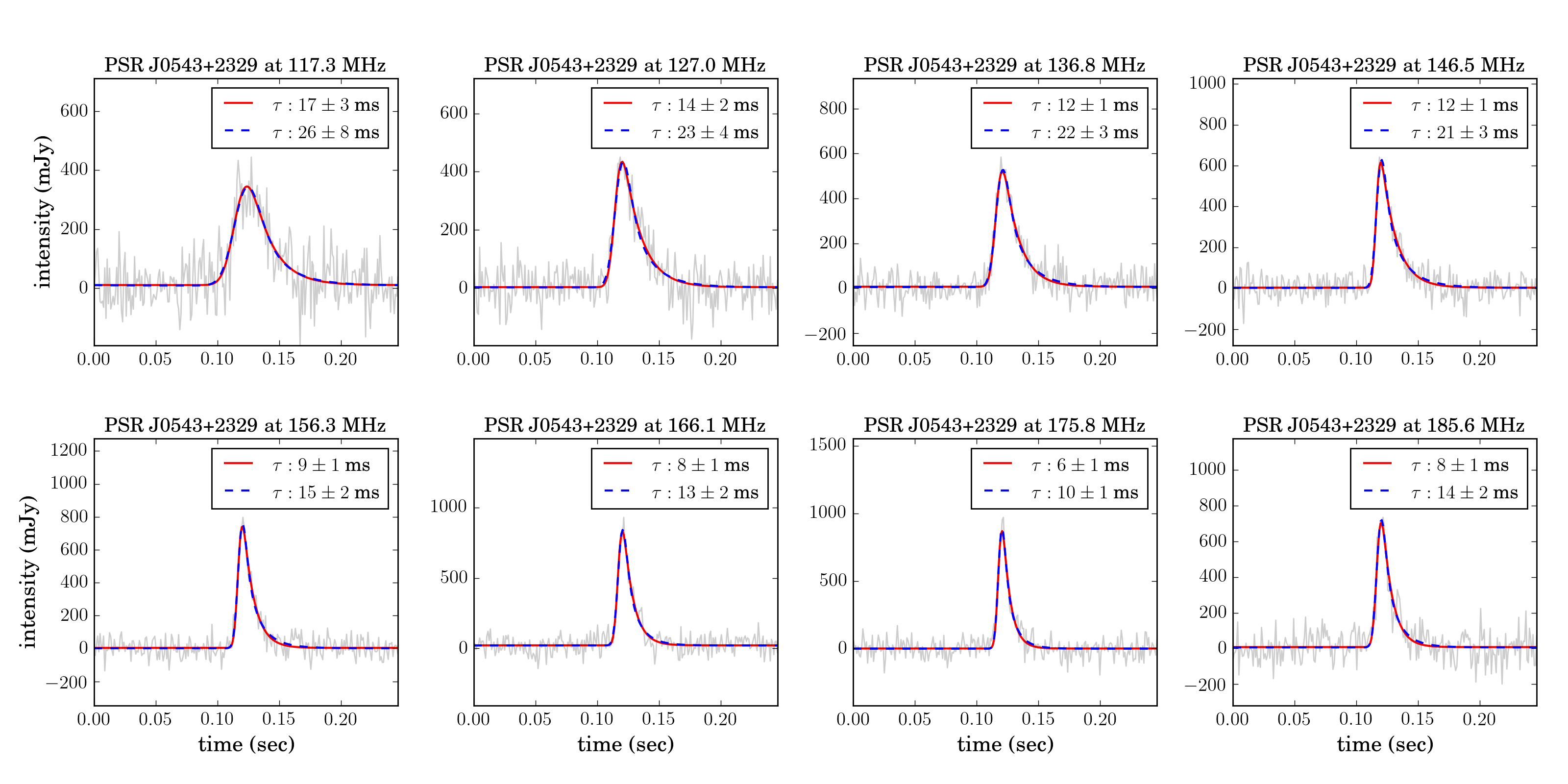}}
\captionof{figure}{PSR~J0543+2329. Profile fits to 8 (of 15 used) average profiles of Commissioning data.}
\end{minipage}

\noindent
\begin{minipage}{\linewidth}
\makebox[\linewidth]{
  \includegraphics[width=\linewidth]{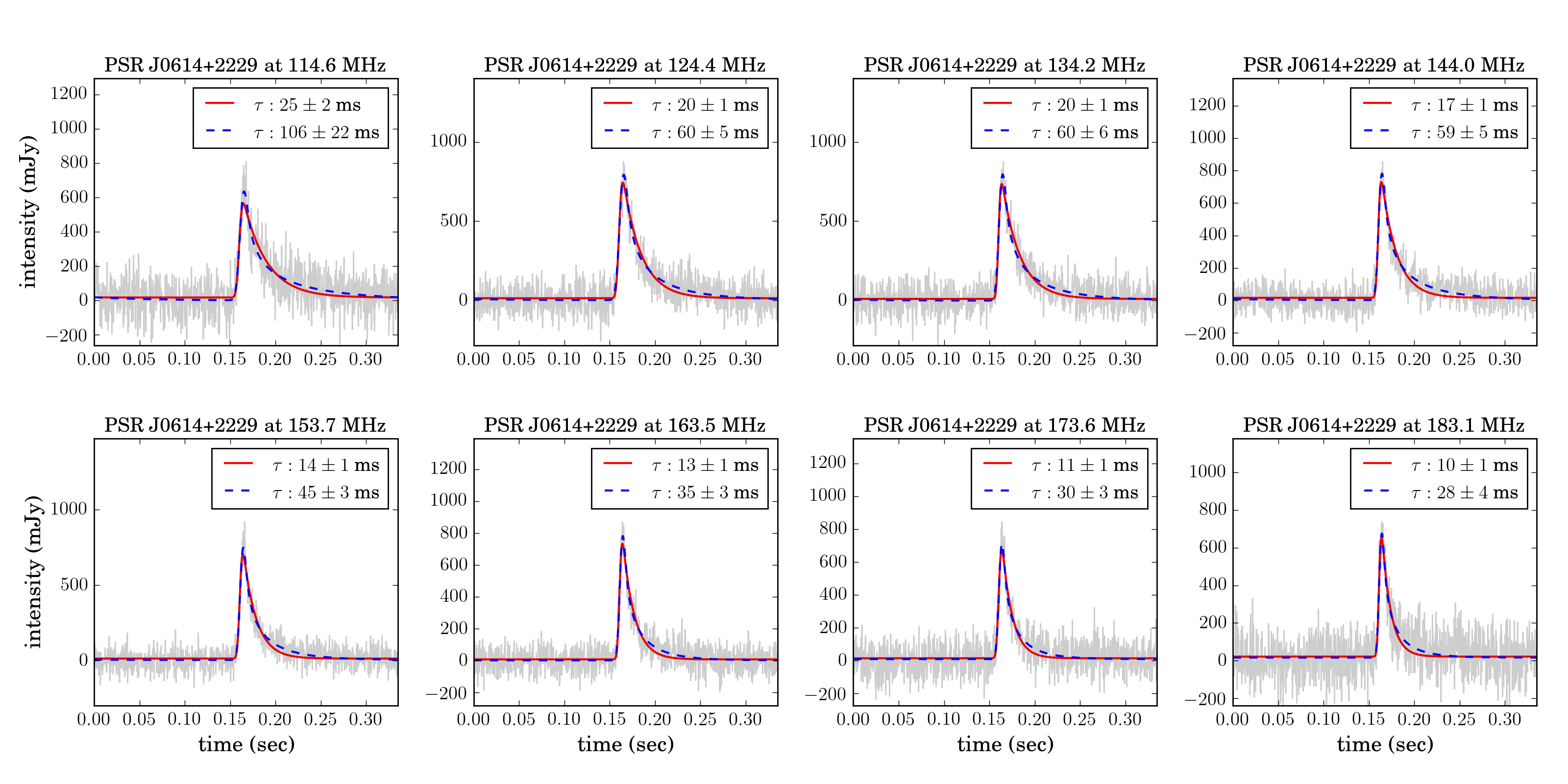}}
\captionof{figure}{PSR~J0614+2229. Profile fits to the 8 average profiles of Commissioning data.}
\makebox[\linewidth]{
  \includegraphics[width=\linewidth]{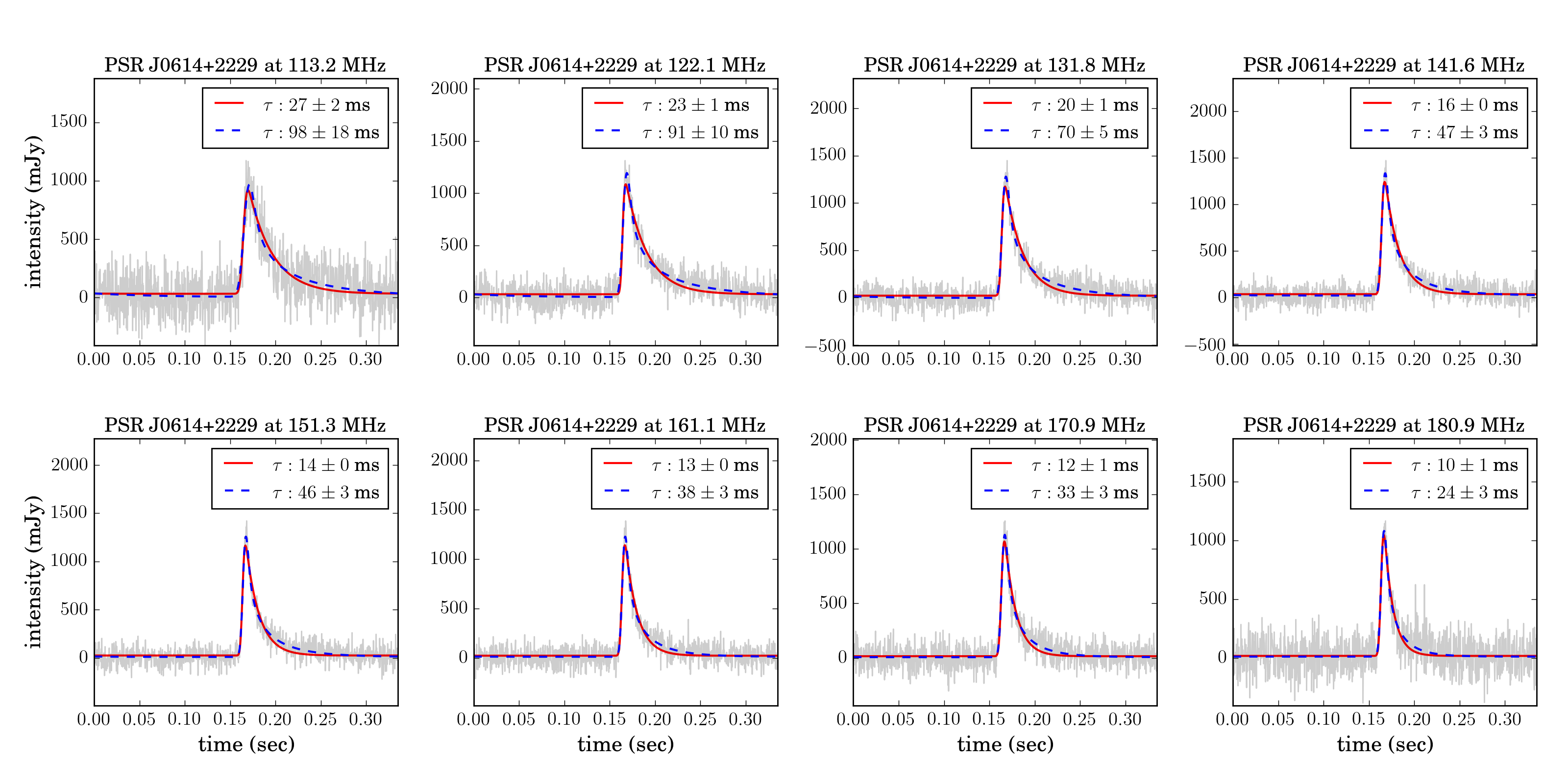}}
\captionof{figure}{PSR~J0614+2229. Profile fits to the 8 (of 16 used) average profiles of Cycle 5 data.}
\end{minipage}

\noindent
\begin{minipage}{\linewidth}
\makebox[\linewidth]{
  \includegraphics[width=\linewidth]{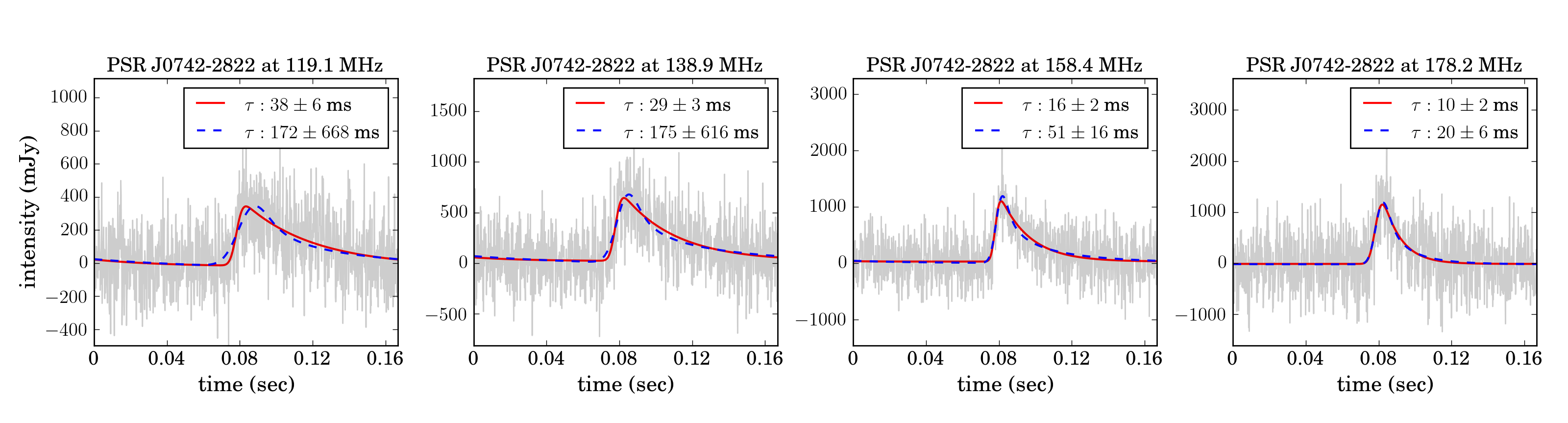}}
\captionof{figure}{PSR~J0742$-$2822. Profile fits to the 4 average profiles of Commissioning data}
\makebox[\linewidth]{
  \includegraphics[width=\linewidth]{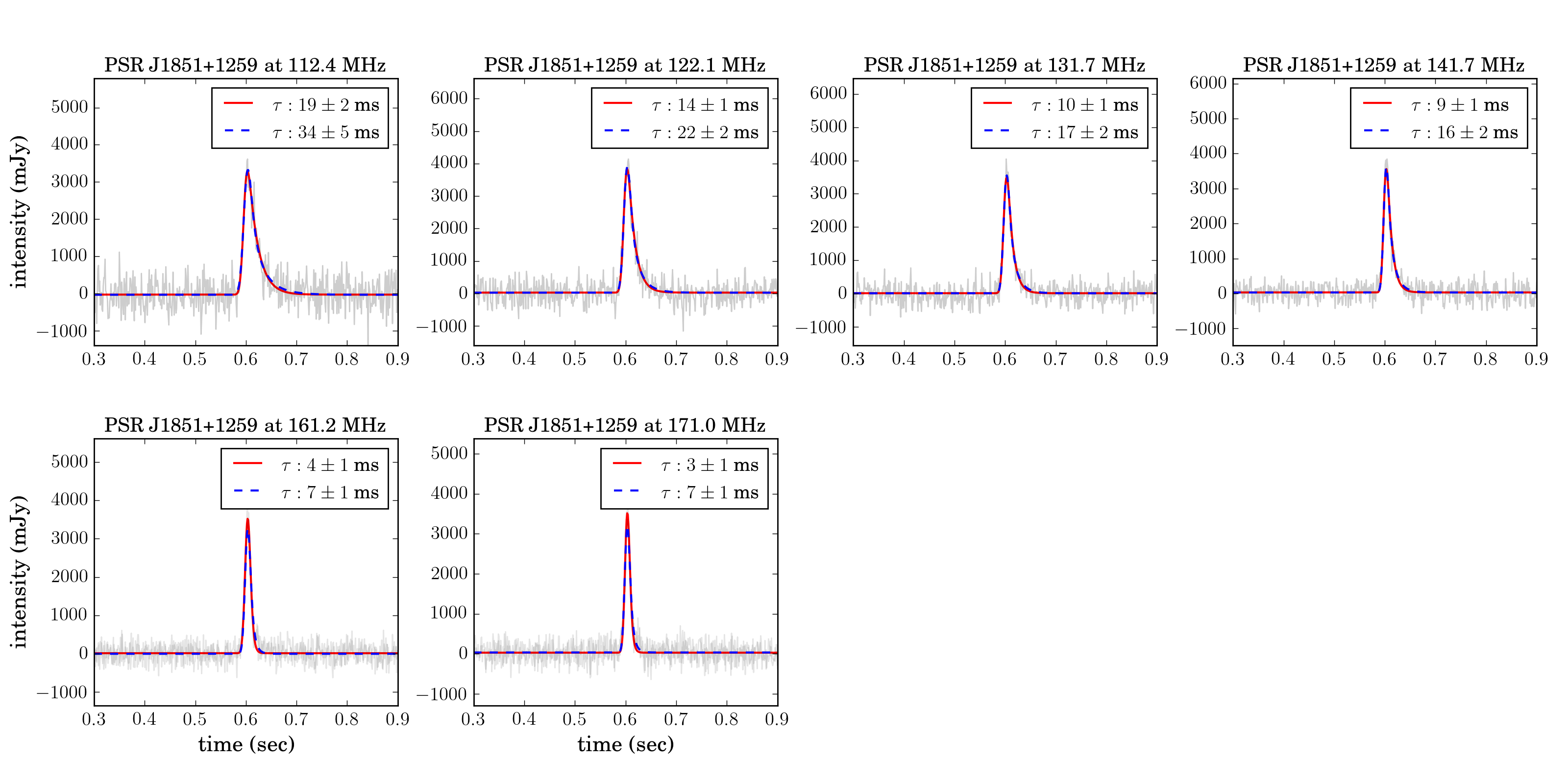}}
\captionof{figure}{PSR~J1851+1259. Profile fits to the 6 (of 12 used) average profiles of Census data. Plots are zoomed to the on-pulse region.}
\end{minipage}

\noindent
\begin{minipage}{\linewidth}
\makebox[\linewidth]{
  \includegraphics[width=\linewidth]{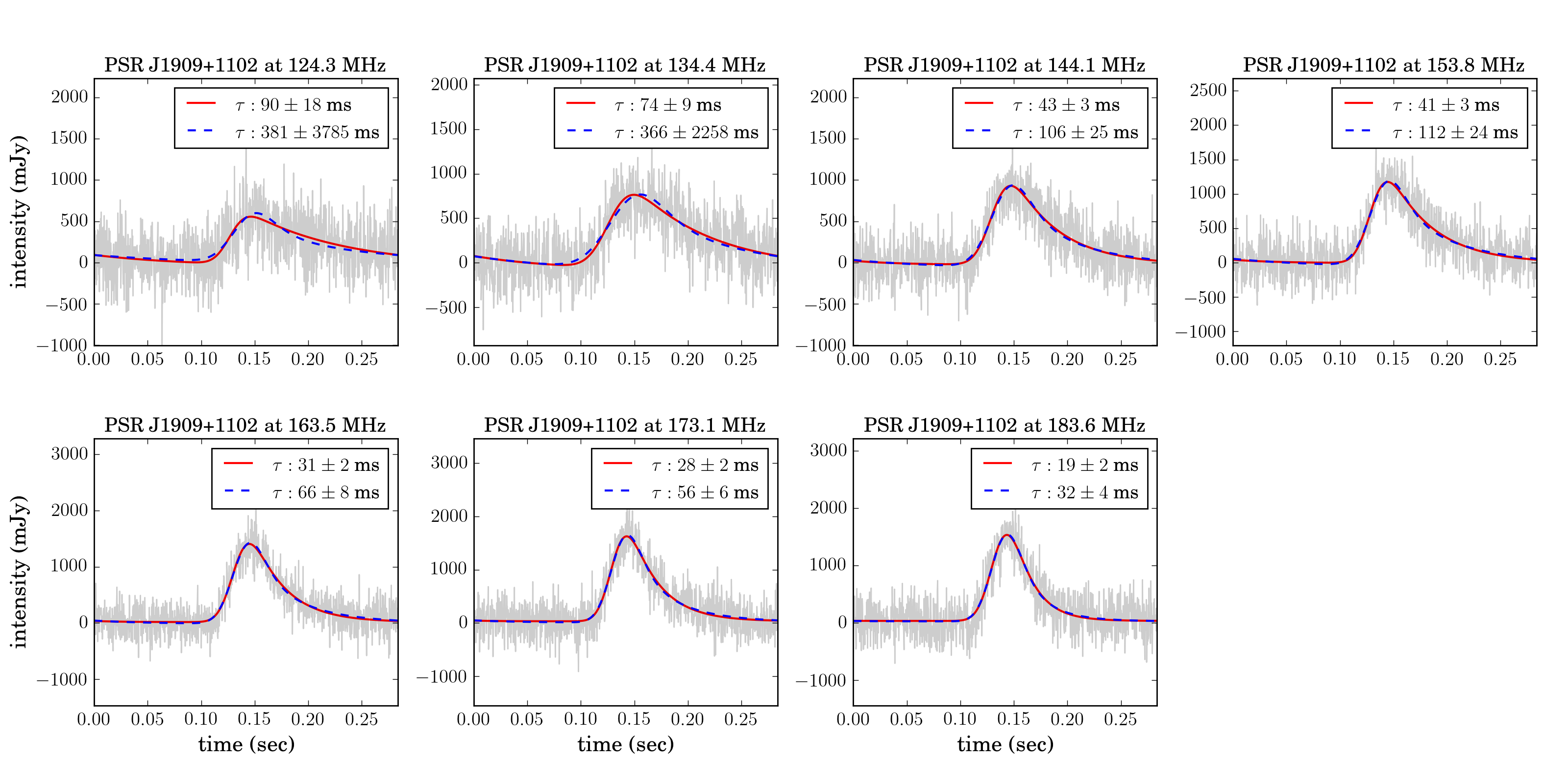}}
\captionof{figure}{PSR~J1909+1102. Profile fits to the 7 average profiles of Commissioning data.}
\makebox[\linewidth]{
  \includegraphics[width=\linewidth]{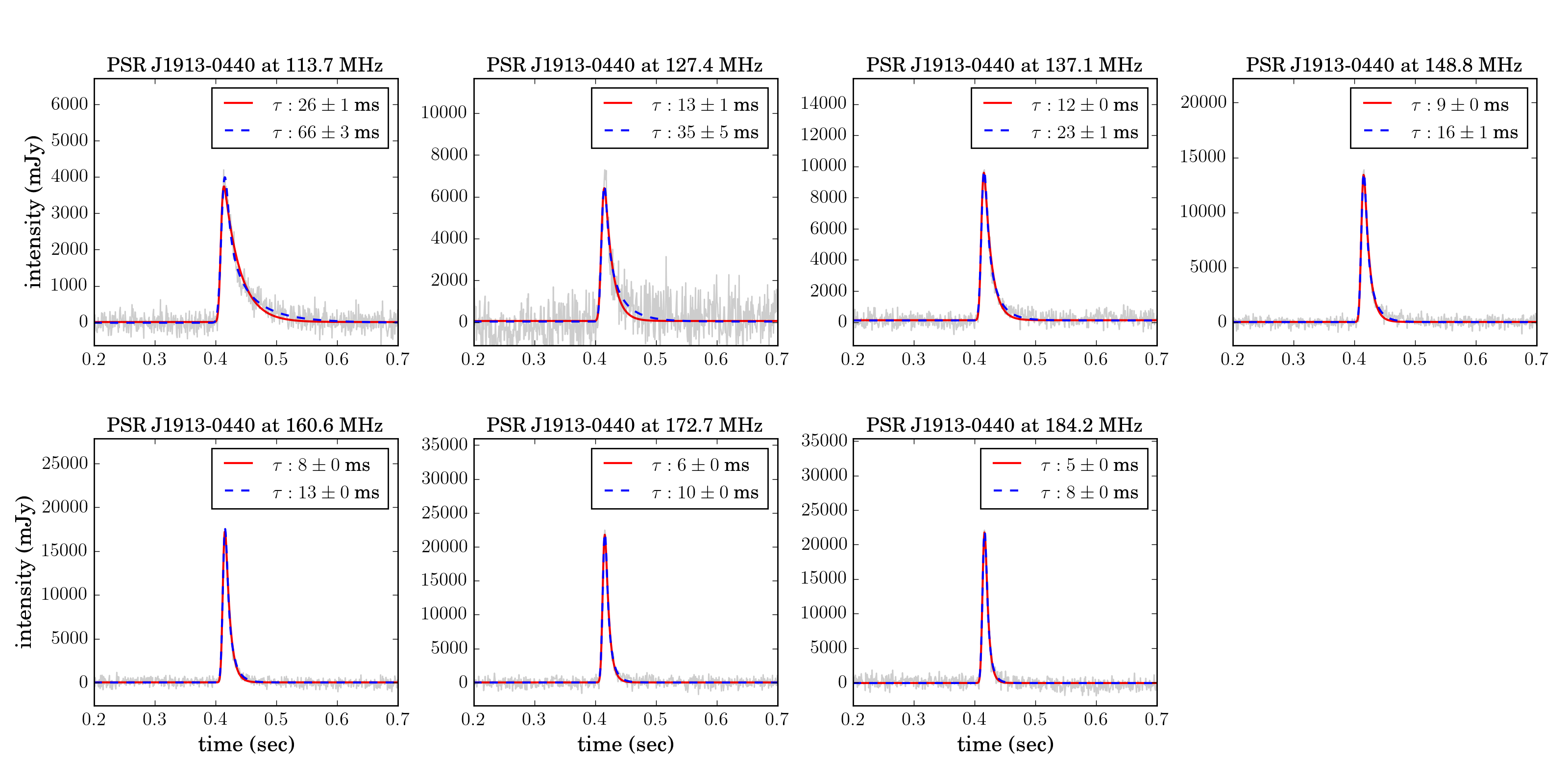}}
\captionof{figure}{PSR~J1913$-$0440. Profile fits to 7 (of 14 used) average profiles of Commissioning data. The fits are zoomed to show 0.5~s of the 0.83~s pulse period. }
\end{minipage}

\noindent
\begin{minipage}{\linewidth}
\makebox[\linewidth]{
  \includegraphics[width=\linewidth]{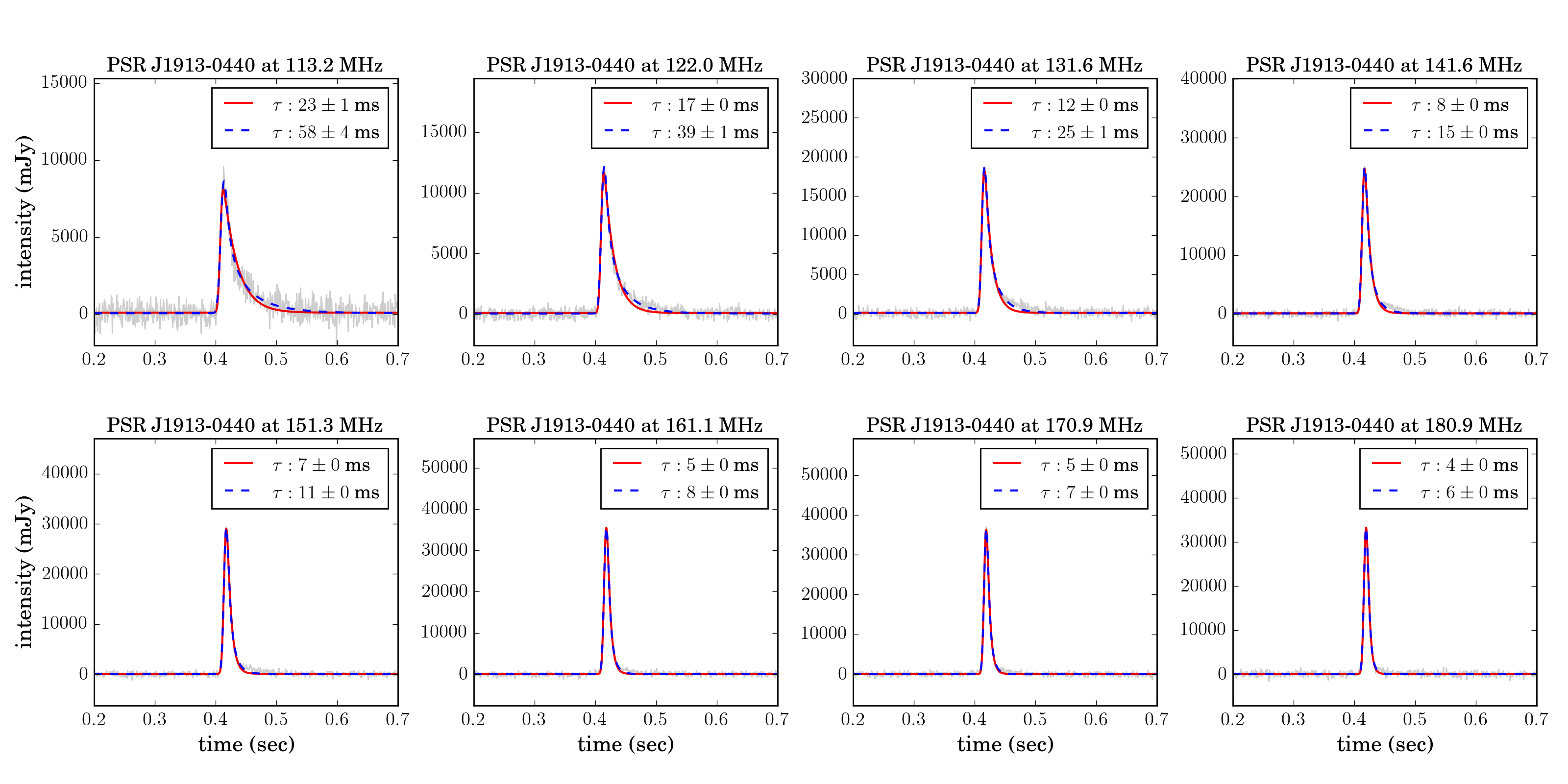}}
\captionof{figure}{PSR~J1913$-$0440. Profile fits to 8 (of 16 used) average profiles of Cycle 5 data. The fits are zoomed to show 0.5~s of the 0.83~s pulse period.}
\makebox[\linewidth]{
  \includegraphics[width=\linewidth]{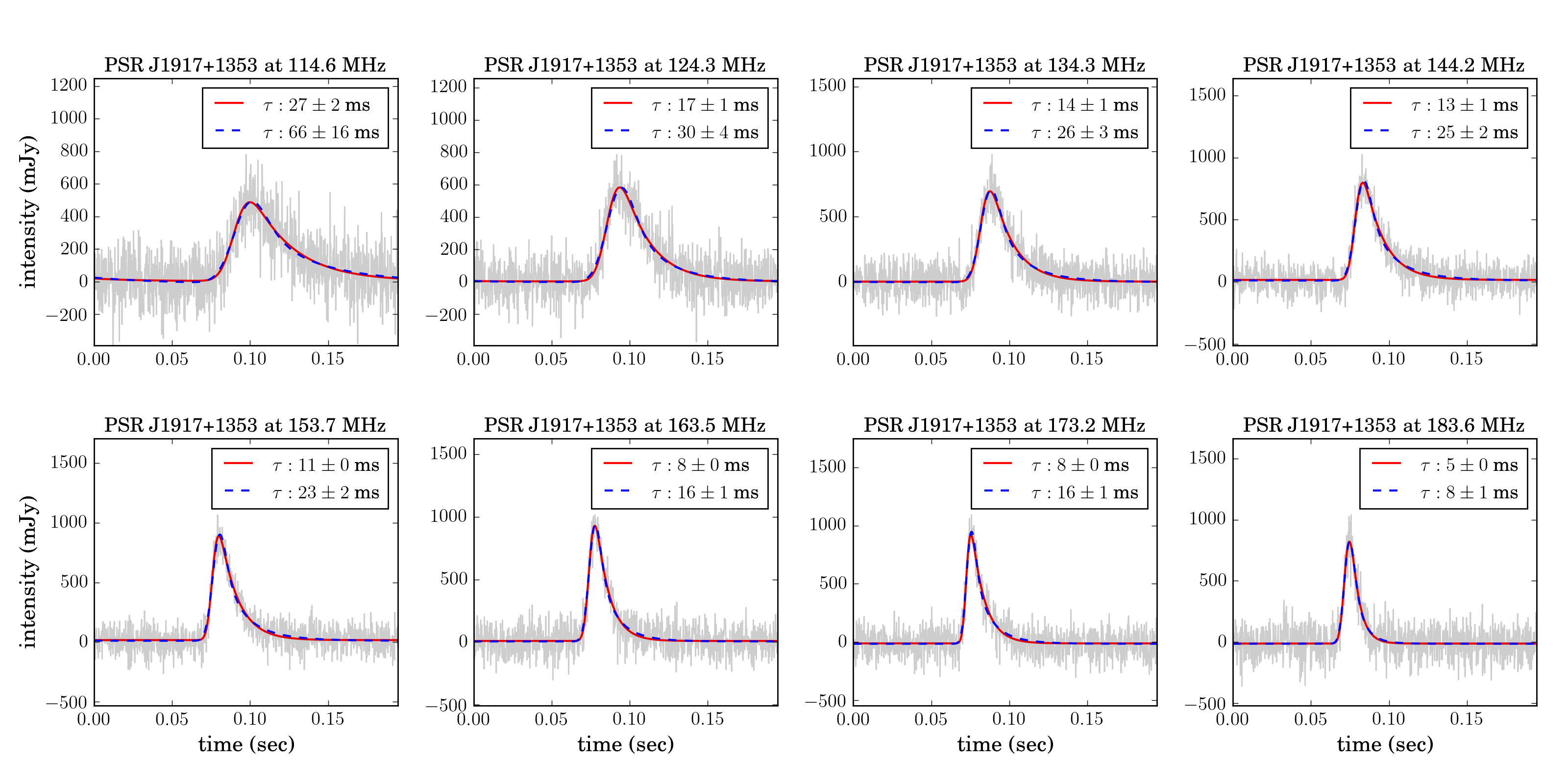}}
\captionof{figure}{PSR~J1917+1353. Profile fits to the 8 average profiles of Commissioning data.}
\end{minipage}

\noindent
\begin{minipage}{\linewidth}
\makebox[\linewidth]{
  \includegraphics[width=\linewidth]{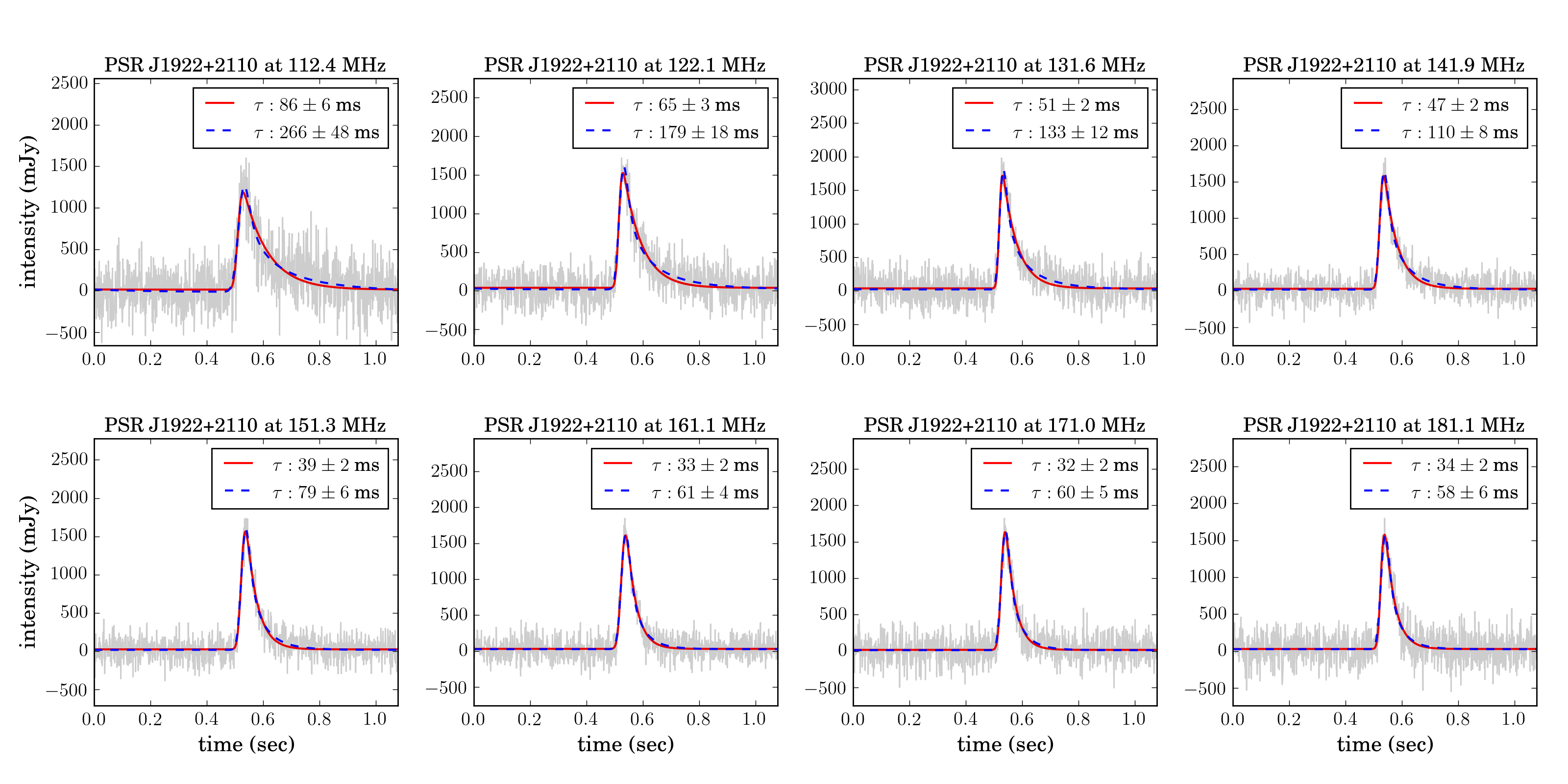}}
\captionof{figure}{PSR~J1922+2110. Profile fits to the 8 (of 16 used) average profiles of Commissioning data.}
\makebox[\linewidth]{
  \includegraphics[width=\linewidth]{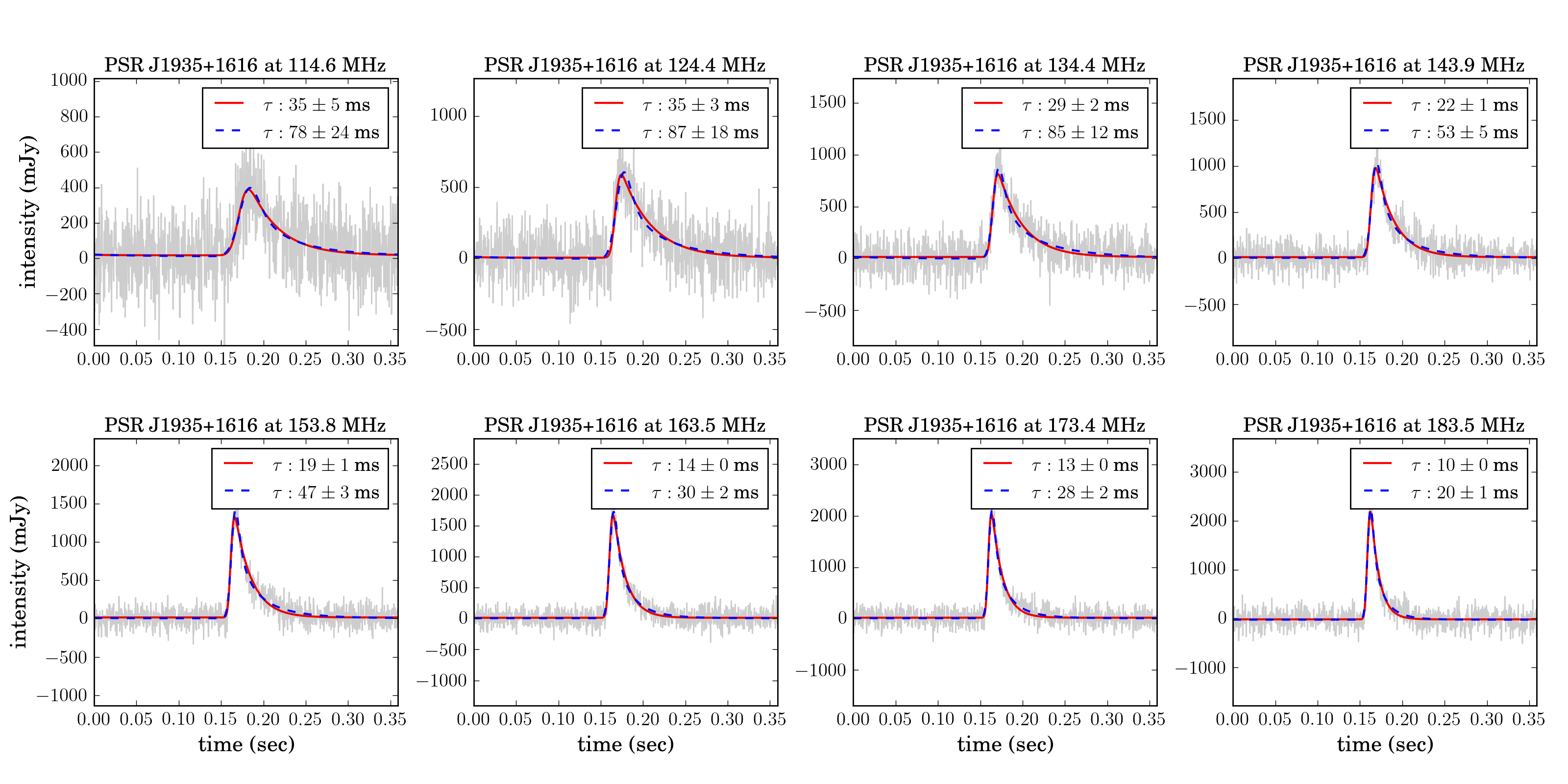}}
\captionof{figure}{PSR~J1935+1616. Profile fits to the 8 average profiles of Commissioning data.}
\end{minipage}

\noindent
\begin{minipage}{\linewidth}
\makebox[\linewidth]{
  \includegraphics[width=\linewidth]{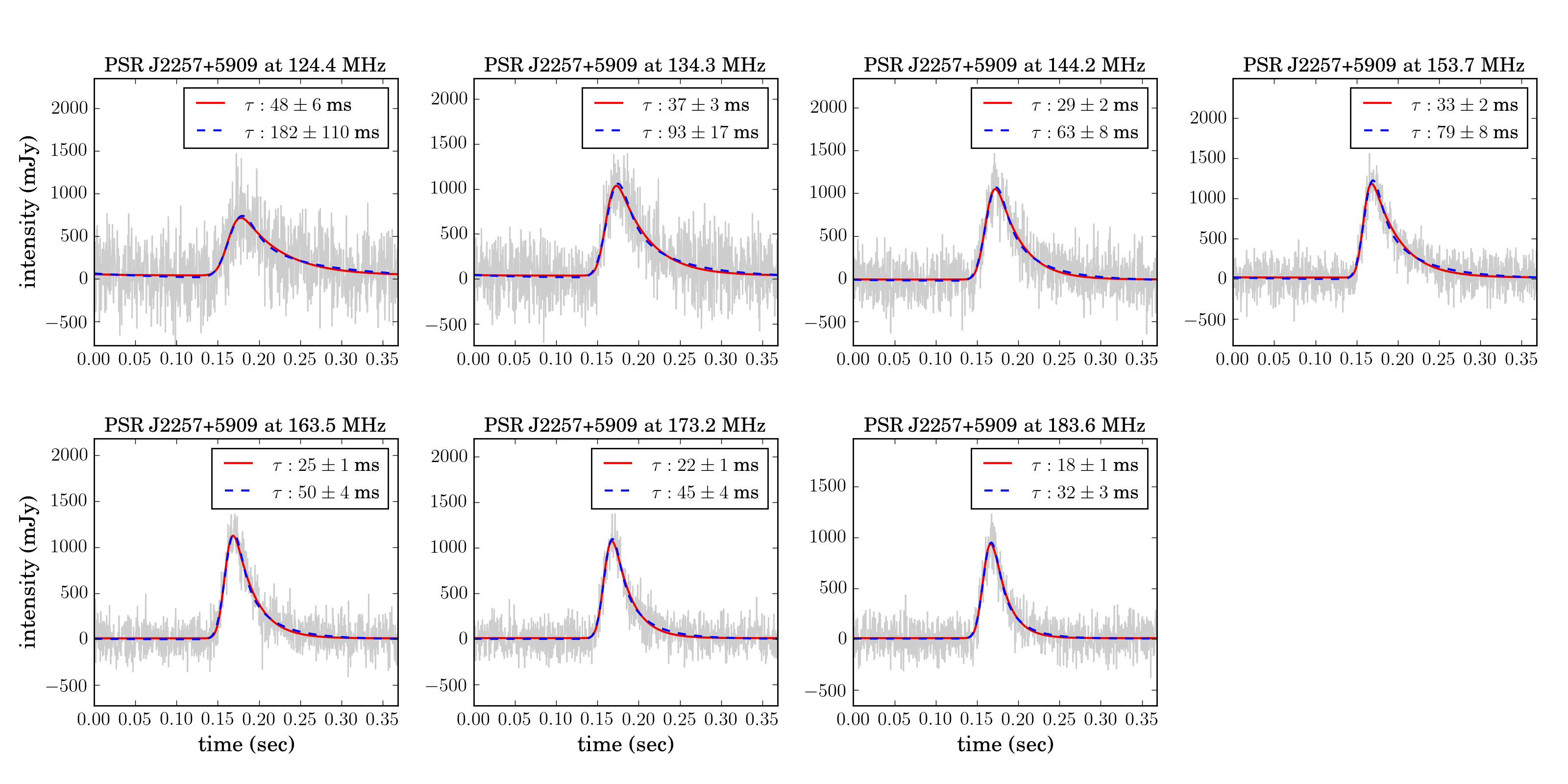}}
\captionof{figure}{PSR~J2257+5909. Profile fits to the 7 average profiles of Commissioning data.}
\makebox[\linewidth]{
  \includegraphics[width=\linewidth]{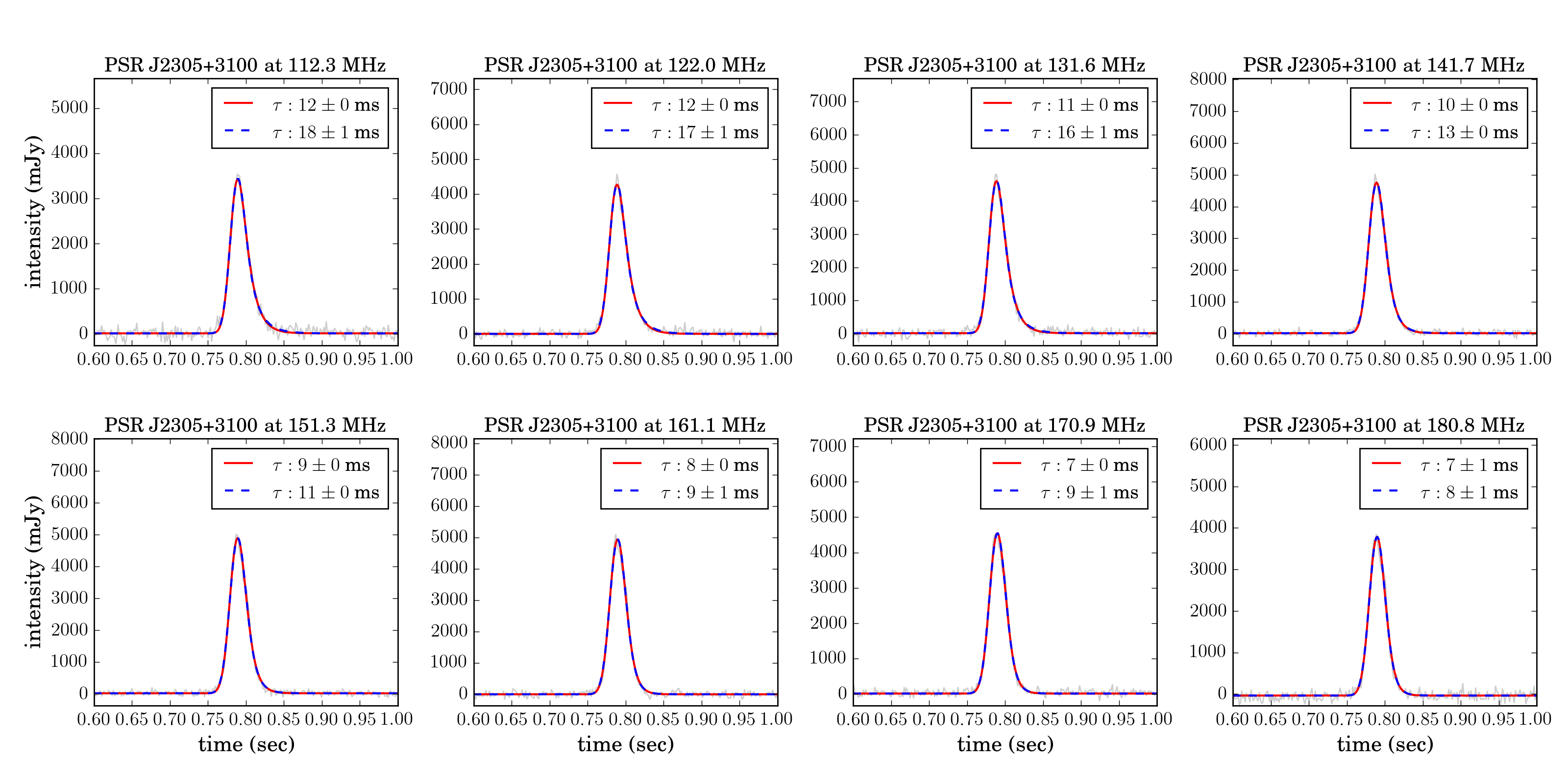}}
\captionof{figure}{PSR~J2305+3100. Profile fits to the 8 (of 16 used) average profiles of Census data. The plots are zoomed to the pulse profile between 0.6 s and 1.0 s. The pulsar has a pulse period of 1.58 s.}
\end{minipage}

\begin{table*}
\centering
\begin{tabular}{p{3.0cm} p{1.7cm} p{1.7cm} p{1.7cm} p{1.5cm}}
\hline
Pulsar &  \multicolumn{2}{>{\centering\arraybackslash}l}{Isotropic Scattering } &  \multicolumn{2}{>{\centering\arraybackslash}l}{Extreme (1D) Anisotropic Scattering} \\
\hline
           &  $\chi^2_{\rm{red}}$  & \mbox{KS p-value} (\%) & $\chi^2_{\rm{red}}$  &\mbox{KS p-value} (\%)\\ [0.5ex] 
\hline
\hline

J0040+5716  &1.73 &  80 &1.74& 76.2\\
J0117+5914 (Co)&  1.88 & 62.2 & 1.87 & 70.8 \\
J0117+5914  (Ce) &  1.43 & 81.2 & 1.33 & 85.6\\
J0543+2329  & 1.69& 74.6 &1.70& 70.8\\
J0614+2229  (Co) &  2.03 & 87.0 & 1.94 & 70.7\\
J0614+2229  (Cy) & 2.05 &80.5 & 1.90& 83.7\\
J0742$-$2822  & 1.9 & 96.7& ... & ...\\
J1851+1259  &1.85 & 82.3 & 1.92 &79.23\\
J1909+1102  & 2.1 & 80.2 & 2.1 & 77.7\\
J1913$-$0440  (Co) &1.95 & 89.8 & 1.90 & 90.0\\
J1913$-$0440 (Cy) & 2.53& 55.8& 2.25&71.1\\
J1917+1353  & 1.92 & 76.9& 1.94& 78.5\\
J1922+2110  &1.9 &87.8& 1.88& 81.9\\
J1935+1616  & 2.13 & 85.2 & 2.11& 85.5\\
J2257+5909  & 1.99 &68.5 & 2.00& 76.4\\
J2305+3100  & 2.24 & 57.8 & 2.26& 56.4 \\

\hline
\hline
\end{tabular}
\caption{Goodness of fits using two different scattering models. The quoted values are averaged over the number of frequency channels studied.}
\label{table:app}
\end{table*}

\begin{table*}
\centering
\begin{tabular}{p{5.0cm}| p{3.0cm} p{3.0cm}| p{3.0cm}}
 PSR~J0117+5914& Comm. data & Census data& \mbox{Difference in datasets} \\
\hline
\hline
Original~DM \mbox{(pc cm$^{-3}$)}& 49.4210 & 49.4207&0.0003\\
\mbox{Isotropic correction} \mbox{(pc cm$^{-3}$)}& $49.4128 \pm 0.0009$ & $49.4143 \pm 0.0006$	&	$-0.0015 \pm 0.0015$\\
Anisotropic correction \mbox{(pc cm$^{-3}$)} &$49.4169 \pm 0.0011$ &	$49.4169 \pm 0.0006$	&	$0.0000 \pm 0.0017$ \\
\hline
 PSR~J0614+2229 & Comm. data & \mbox{Cycle 5} data& Difference in datasets \\
\hline
\hline
Original~DM (pc cm$^{-3}$)& 96.9030 & 96.9100&0.007\\
Isotropic \mbox{correction} (pc cm$^{-3}$)& $96.9000 \pm 0.0007$ & $96.9153 \pm 0.0006$	&	$0.0153 \pm 0.0013$\\
Anisotropic correction (pc cm$^{-3}$) &$96.9063 \pm 0.0005$ &	$96.9209 \pm 0.0008$	&	$0.0146 \pm 0.0013$ \\
\hline
PSR~J1913$-$0440& Comm. data & Cycle 5 data & Difference in datasets \\
\hline
\hline
Original~DM (pc cm$^{-3}$)& 89.3700 & 89.3850&0.0150\\
Isotropic \mbox{correction} (pc cm$^{-3}$)& $89.3460 \pm 0.0009$& $89.3393 \pm 0.0003$	&	$0.0067 \pm 0.0012$\\
Anisotropic correction (pc cm$^{-3}$)& $89.3500\pm 0.0011$ & $89.3469 \pm 0.0003$	& $0.0031 \pm 0.0014$\\
\hline
\end{tabular}
\caption{Original DM values and applied DM corrections between epochs.}
\label{table:app2}
\end{table*}

\newpage

\noindent
\begin{figure}
	\begin{minipage}{\linewidth}
	\section{Width Evolution}\label{app:widths}
	\makebox[\linewidth]{
	 \includegraphics[width=\textwidth]{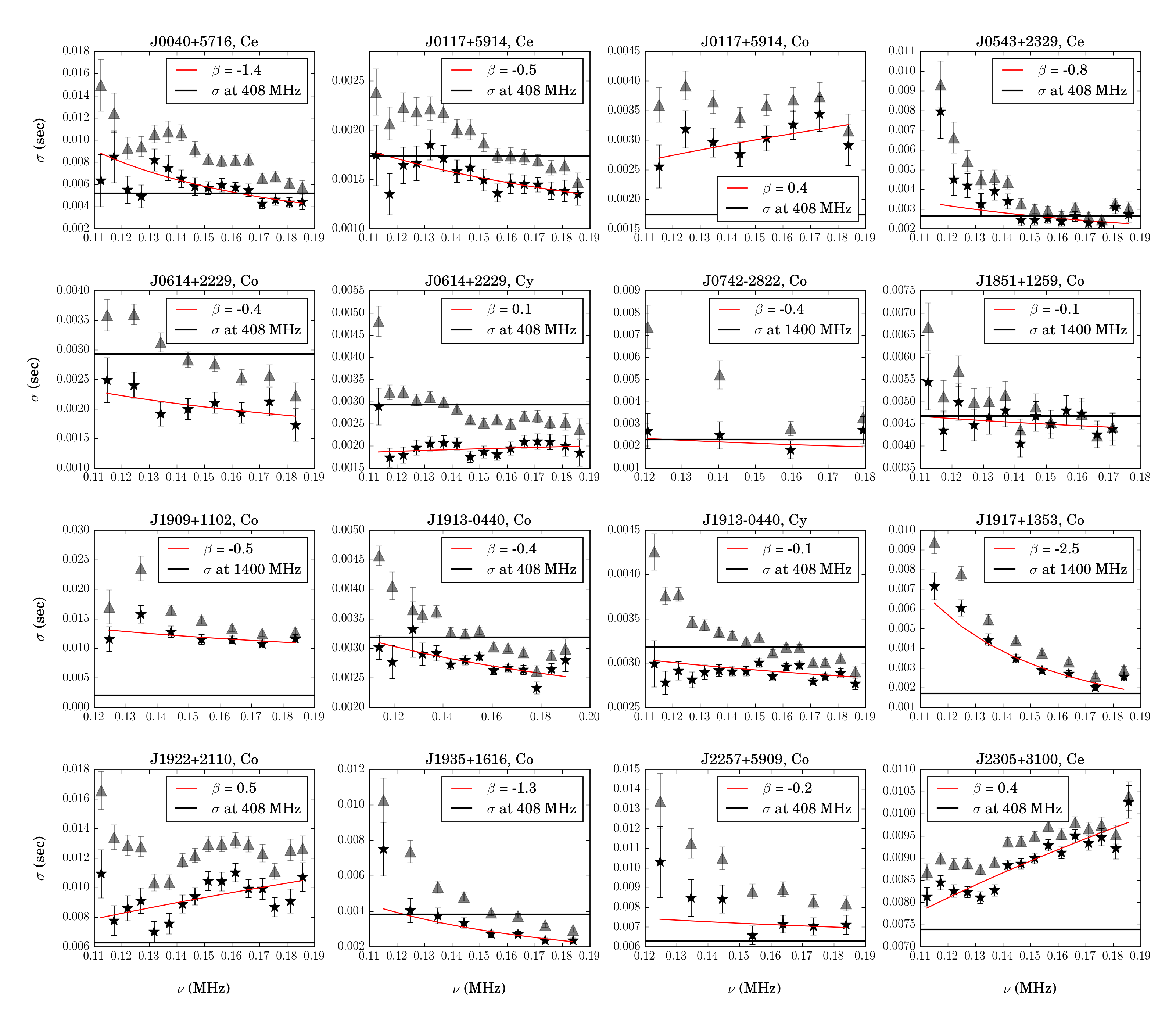}}
	\captionof{figure}{The intrinsic profile width evolution for the pulsars in this dataset. The width is represented by the standard deviation of the intrinsic Gaussian pulse shape ($\sigma$). Isotropic widths and associated error bars are shown in black (stars) and anisotropic in grey (triangles). The spectral indices of a power law fit to the isotropic data ($\sigma \propto \nu^{\beta}$) are shown in the legends. High frequency $\sigma$ values are plotted as black horizontal lines.}
	\end{minipage}
\end{figure}

\newpage

\noindent
\begin{figure}
	\begin{minipage}{\linewidth}
	\section{Flux}\label{app:flux}
	\makebox[\linewidth]{
	 \includegraphics[width=\textwidth]{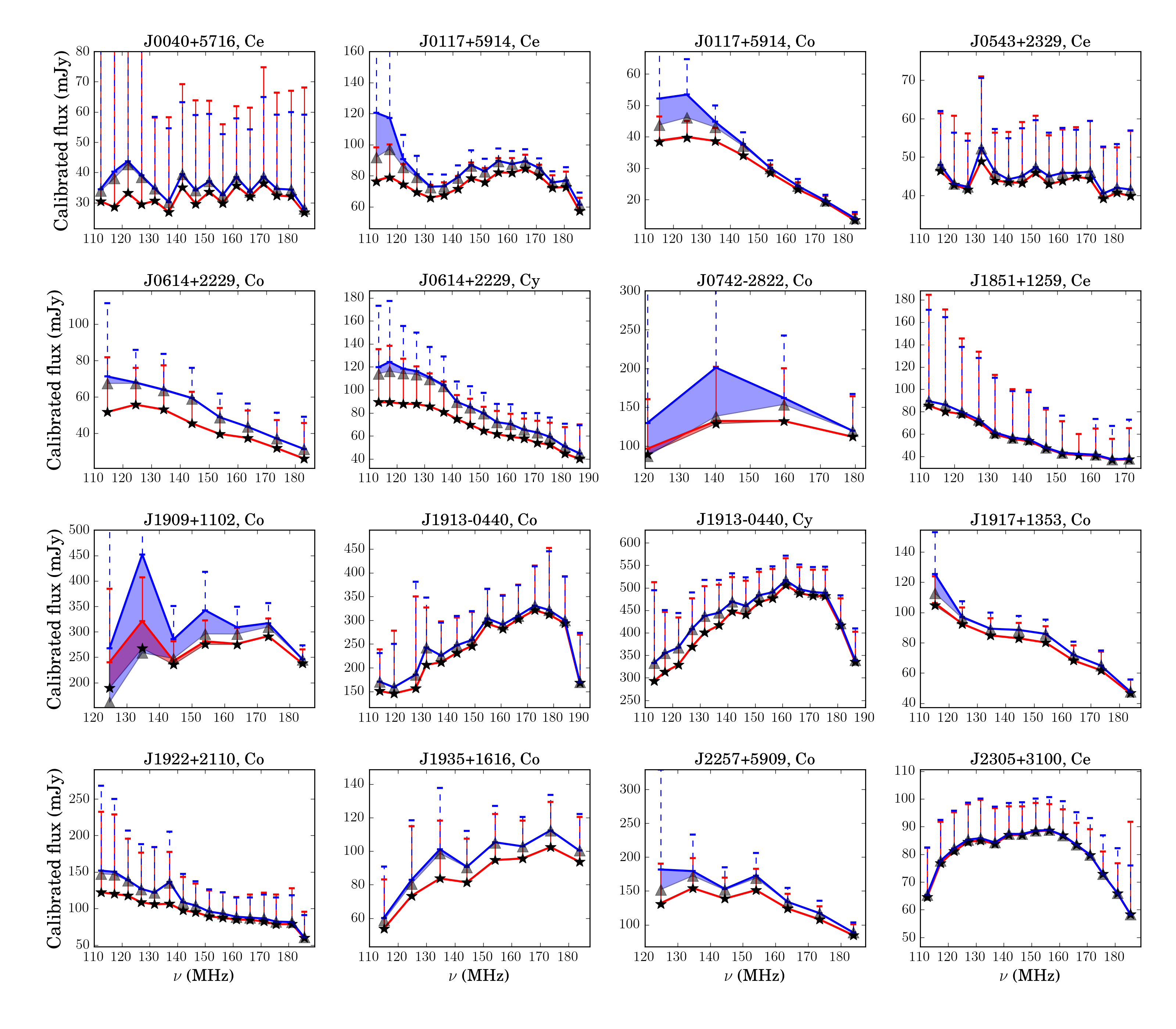}}
	\captionof{figure}{Mean flux density plots for the pulsars in this dataset. The flux densities are calculated from the integrated scattered profiles. Markers and colours are as discussed in the main body of the paper. Indicated error bars are one side of a symmetric error bar.}
	\end{minipage}
\end{figure}

\end{document}